\documentclass[10pt,english,prd,superscriptaddress,nofootinbib,preprintnumbers,showpacs,twocolumn,floatfix]{revtex4-1}
\usepackage[utf8]{inputenc}
\usepackage{amsmath}
\usepackage{amsfonts}
\usepackage{amssymb}
\usepackage{graphicx}
\usepackage{array}
\usepackage{booktabs}
\usepackage{multirow}
\usepackage{mathtools}
\usepackage{subcaption}
\usepackage{makecell}
\usepackage{mathrsfs}

\DeclareMathOperator{\arctanh}{arctanh}
\usepackage[usenames,dvipsnames]{xcolor}
\usepackage{hyperref}   
\hypersetup{colorlinks=true, 
citecolor=MidnightBlue, linkcolor=MidnightBlue, urlcolor=Cyan}
\usepackage{tikz}
\usepackage{verbatim} 
\definecolor{purple}{rgb}{1,0,1}
\definecolor{lime}{HTML}{A6CE39} 

\newcommand{\orcidicon}{%
	\begin{tikzpicture}
	\draw[lime, fill=lime] (0,0) 
		circle [radius=0.16] 
		node[white] {{\fontfamily{qag}\selectfont \tiny ID}};
	\draw[white, fill=white] (-0.0625,0.095) 
		circle [radius=0.007];
	\end{tikzpicture}	\hspace{-2mm}
}
\newcommand\orcidFrancisco{{\href{https://orcid.org/0000-0002-9388-8373}{\orcidicon}}}
\newcommand\orcidManuel{{\href{https://orcid.org/0000-0001-8586-0285}{\orcidicon}}}
\newcommand\orcidTarciso{{\href{https://orcid.org/0009-0007-0450-2672}{\orcidicon}}}
\newcommand\orcidGabriel{{\href{https://orcid.org/0009-0007-3770-8976}{\orcidicon}}}
\begin{document}
\title{Horizons, throats and bounces in hybrid metric-Palatini gravity\\ with a non-zero potential}

\author{Gabriel I. R\'{o}is\orcidGabriel\!\!}
\email{fc54507@alunos.fc.ul.pt}
\affiliation{Instituto de Astrof\'{i}sica e Ci\^{e}ncias do Espa\c{c}o, Faculdade de Ci\^{e}ncias da Universidade de Lisboa, Edifício C8, Campo Grande, P-1749-016 Lisbon, Portugal}
\affiliation{Departamento de F\'{i}sica, Faculdade de Ci\^{e}ncias da Universidade de Lisboa, Edif\'{i}cio C8, Campo Grande, P-1749-016 Lisbon, Portugal}
\author{Jos\'{e} Tarciso S. S. Junior\orcidTarciso\!\!}
\email{tarcisojunior17@gmail.com}
\affiliation{Faculdade de F\'{\i}sica, Programa de P\'{o}s-Gradua\c{c}\~{a}o em 
F\'isica, Universidade Federal do 
 Par\'{a},  66075-110, Bel\'{e}m, Par\'{a}, Brazil}

\author{Francisco S. N. Lobo\orcidFrancisco\!\!} 
\email{fslobo@ciencias.ulisboa.pt}
\affiliation{Instituto de Astrof\'{i}sica e Ci\^{e}ncias do Espa\c{c}o, Faculdade de Ci\^{e}ncias da Universidade de Lisboa, Edifício C8, Campo Grande, P-1749-016 Lisbon, Portugal}
\affiliation{Departamento de F\'{i}sica, Faculdade de Ci\^{e}ncias da Universidade de Lisboa, Edif\'{i}cio C8, Campo Grande, P-1749-016 Lisbon, Portugal}

\author{Manuel E. Rodrigues\orcidManuel\!\!}
\email{esialg@gmail.com}
\affiliation{Faculdade de F\'{\i}sica, Programa de P\'{o}s-Gradua\c{c}\~{a}o em 
F\'isica, Universidade Federal do 
 Par\'{a},  66075-110, Bel\'{e}m, Par\'{a}, Brazil}
\affiliation{Faculdade de Ci\^{e}ncias Exatas e Tecnologia, 
Universidade Federal do Par\'{a}\\
Campus Universit\'{a}rio de Abaetetuba, 68440-000, Abaetetuba, Par\'{a}, 
Brazil}

\date{\LaTeX-ed \today}
\begin{abstract}

This work conducts an in-depth exploration of exact electrically charged solutions, including traversable wormholes, black holes, and black bounces, within the framework of the scalar-tensor representation of hybrid metric-Palatini gravity (HMPG) with a non-zero scalar potential. By integrating principles from both the metric and Palatini formulations, HMPG provides a flexible approach to addressing persistent challenges in General Relativity (GR), such as the late-time cosmic acceleration and the nature of dark matter. Under the assumption of spherical symmetry, we employ an inverse problem technique to derive exact solutions in both the Jordan and Einstein conformal frames. This method naturally leads to configurations involving either canonical or phantom scalar fields. A thorough examination of horizon structures, throat conditions, asymptotic behaviour, and curvature regularity (via the Kretschmann scalar) reveals the intricate causal structures permitted by this theoretical model. The analysis uncovers a diverse range of geometric configurations, with the phantom sector exhibiting a notably richer spectrum of solutions than the canonical case. These solutions encompass traversable wormholes, black universe models, where the interior of a black hole evolves into an expanding cosmological phase rather than a singularity, as well as black bounce structures and multi-horizon black holes. The results demonstrate that introducing a non-zero scalar potential within HMPG significantly expands the array of possible gravitational solutions, yielding complex causal and curvature properties that go beyond standard GR. Consequently, HMPG stands out as a powerful theoretical framework for modelling extreme astrophysical environments, where deviations from classical gravity are expected to play a crucial role. Future research will focus on evaluating the stability of these solutions and investigating potential observational signatures, such as gravitational lensing effects and gravitational wave emissions.

\end{abstract}
\maketitle
\def\HMS{{\scriptscriptstyle{\rm HMS}}}

\section{Introduction}

Black holes, once considered mere theoretical curiosities, have emerged as central objects in modern astrophysics and gravitational research. These enigmatic regions of spacetime, characterized by an event horizon from which neither matter nor light can escape, have been instrumental in testing the predictions of Einstein’s General Relativity (GR) in the strong-field regime. Observational breakthroughs, such as the detection of gravitational waves by the LIGO/Virgo collaborations \cite{LIGOScientific:2016aoc,LIGOScientific:2016vlm} and the imaging of black hole shadows by the Event Horizon Telescope (EHT) Collaboration \cite{EventHorizonTelescope:2021bee,EventHorizonTelescope:2021srq,EventHorizonTelescope:2023gtd,EventHorizonTelescope:2022wkp,EventHorizonTelescope:2022apq}, have provided compelling evidence for the existence of black holes and opened new avenues for exploring their properties.
Indeed, the first direct detection of gravitational waves, GW150914, marked a historic achievement \cite{LIGOScientific:2016aoc}. The observed signal was attributed to the merger of two stellar-mass black holes, confirming not only the existence of such objects but also their dynamic behavior in binary systems. Subsequent detections have reinforced these findings, establishing gravitational wave astronomy as a powerful tool for probing the universe.
Complementing these advancements, the EHT provided the first image of a supermassive black hole’s shadow in the galaxy M87 \cite{EventHorizonTelescope:2021bee,EventHorizonTelescope:2021srq}. This striking observation revealed a bright emission ring surrounding a dark central region, consistent with theoretical predictions of the black hole shadow. Further analysis of the Sagittarius A* black hole at the center of the Milky Way \cite{EventHorizonTelescope:2022wkp,EventHorizonTelescope:2022apq} demonstrated the universality of these phenomena across different mass scales and environments. Polarization studies have also shed light on the magnetic field structures near event horizons \cite{EventHorizonTelescope:2021bee,EventHorizonTelescope:2021srq,EventHorizonTelescope:2023gtd}, offering insights into accretion processes and jet formation.

While classical GR predicts the existence of singularities at the cores of black holes, these points of infinite density pose challenges to the consistency of the theory \cite{Penrose:1969pc}. These singularities challenge the completeness of GR and motivate a deeper understanding of the nature of spacetime \cite{Visser:2008cjw, Cardoso:2019rvt}.
Regular black holes, characterized by the absence of singularities, and black bounces, offer promising avenues for addressing these challenges \cite{Dymnikova:2015hka, Simpson:2018tsi}. The exploration of such alternatives has been motivated within the context of nonlinear electrodynamics, scalar-tensor theories, and quantum gravity \cite{Bronnikov:2000vy, Carballo-Rubio:2018pmi, Lobo:2020kxn}.
Regular black holes modify classical solutions by incorporating structures that prevent singularities \cite{Rodrigues:2017yry,Rodrigues:2018bdc,Rodrigues:2020pem}. These models often rely on nonlinear electrodynamics or modifications to the stress-energy tensor to ensure a finite curvature at the core \cite{Dymnikova:2015hka, Bronnikov:2005gm}. Notable examples include solutions proposed by Bronnikov \cite{Bronnikov:2000vy} and Fan and Wang \cite{Fan:2016hvf}, which demonstrate the viability of such metrics under specific energy conditions \cite{Zaslavskii:2010qz}. These solutions are further supported by studies exploring their thermodynamic stability and causal structures \cite{Carballo-Rubio:2019fnb}.
Black bounces provide a dynamic framework where spacetime may smoothly transition for instance between a black hole and a traversable wormhole \cite{Simpson:2018tsi, Lobo:2020kxn}. These geometries circumvent singularities by replacing them with a throat that connects two asymptotic regions 
\cite{Rodrigues:2022mdm,Rodrigues:2023vtm,Pereira:2023lck,Alencar:2024yvh,Pereira:2024gsl}. Recent research \cite{Simpson:2018tsi} has shown how such models can be embedded within the Vaidya spacetime \cite{Simpson:2019cer}. The dynamic nature of black bounces also offers insights into the evolution of spacetime under collapsing scenarios \cite{Bolokhov:2012kn}.

The ongoing study of regular black holes \cite{Junior:2015fya,Rodrigues:2015ayd,Rodrigues:2016fym,deSousaSilva:2018kkt,Rodrigues:2019xrc,Junior:2023ixh,Junior:2024xmm} and black bounces \cite{Junior:2022zxo,Junior:2023qaq,Fabris:2023opv,Junior:2024vrv,Junior:2024cbb} within the context of modified theories of gravity is of fundamental importance for advancing our understanding of the spacetime structure and the nature of strong gravitational fields. These studies not only aim to resolve singularities, but also provide crucial insights into the viability of alternative gravity models in describing astrophysical black holes. By investigating deviations from the standard Kerr and Schwarzschild solutions, one may explore how modified gravity can influence event horizon dynamics, causal structures, and the potential observational signatures of non-singular black hole candidates.
Beyond the singularity issue, the stability properties of black hole spacetimes within scalar-tensor theories have been a subject of extensive scrutiny \cite{Bronnikov:1999wh,Bronnikov:2024uyb}. Stability analyses play a key role in determining whether these black holes are physically viable and whether they can form dynamically from realistic initial conditions. These investigations further elucidate how modifications to the gravitational action impact the intricate interplay between spacetime curvature and matter fields \cite{Lobo:2020ffi,Rois:2024qzm}, offering a deeper theoretical framework for understanding deviations from GR.
More specifically, hybrid metric-Palatini gravity (HMPG), an extended framework that incorporates both metric and Palatini variational principles \cite{Harko:2011nh,Capozziello:2012ny,Carloni:2015bua,Capozziello:2015lza,Harko:2020ibn,Lobo:2023ddb,Capozziello:2012qt,Capozziello:2013uya,Capozziello:2013yha,Rosa:2018jwp,Rosa:2021yym,Rosa:2020uoi}, has emerged as a compelling approach for investigating non-singular black hole solutions. This formulation naturally modifies the Einstein field equations by introducing new dynamical terms that arise from the simultaneous variation of the metric and connection, leading to enriched phenomenology in the strong-field regime. 

In fact, research by Bronnikov and collaborators \cite{Bronnikov:2019ugl,Bronnikov:2020vgg,Bronnikov:2021tie} has demonstrated that this hybrid framework admits a broad class of spherically symmetric black holes and wormholes, often exhibiting novel causal structures that distinguish them from the standard GR solutions. These results suggest that HMPG may provide a theoretically consistent and observationally testable setting for exploring deviations from classical black hole physics.
Indeed, recently a systematic study of exact solutions for electrically charged wormholes, black holes, and black bounces within the HMPG framework, was undertaken \cite{Rois:2024qzm}. However, the  focus was on configurations with a vanishing scalar potential under spherical symmetry and derive solutions in both the Jordan and Einstein conformal frames. The analysis revealed a diverse array of solutions, including traversable wormholes, black holes with extremal horizons, and ``black universe'' models where the spacetime beyond the horizon transitions to an expanding cosmological solution rather than a singularity. These configurations were classified based on the properties of the scalar field, with detailed examinations of their horizon and throat structures, asymptotic behaviours, and singularity profiles.
The results highlight the adaptability of HMPG in modelling intricate gravitational phenomena, significantly broadening the theoretical framework's applicability to various astrophysical scenarios. In the present work, we extend the analysis carried out in \cite{Rois:2024qzm}, in the presence of a non-zero scalar potential under spherical symmetry, which provides an extremely richer causal structure.

This paper is organised in the following manner: In Sec. \ref{chap 1}, we briefly present the formalism of hybrid metric-Palatini gravity, namely, the actions in the Jordan and Einstein frames, and the respective field equations. In Sec. \ref{chap 2}, we present the static and spherically symmetric metric, and the methodology we adopt to analyse the regularity of the solution, via the Kretschmann scalar, and to examine and characterise the existence of horizons, throats and bounces. In Secs. \ref{example1_section} and \ref{example2_section}, we thoroughly analyse the canonical and phantom sectors, respectively, and explore the rich and intricate causal structures through the study of horizons, throat geometries, asymptotic limits, and curvature properties characterized by the Kretschmann scalar. Finally, in Sec. \ref{section:conclusion}, we summarize and discuss our results.

\section{Hybrid Metric-Palatini Gravity: Formalism}\label{chap 1}

\subsection{Action: Jordan frame}

The action for the HMPG theory is given by \cite{Harko:2011nh,Capozziello:2012ny}:
\begin{equation}\label{2}
    S=\int d^4x\sqrt{-g}[R+f(\mathcal{R})]+S_m\,\,,
\end{equation}
where $S_m$ is the matter action, which, in the general case, is defined as $S_m=\int d^4x \sqrt{-g} \mathcal{L}_m$, where $\mathcal{L}_m$ is the matter Lagrangean. The term $R$ stands for the metric Einstein-Hilbert term, and $\mathcal{R} \equiv g^{\mu\nu}\mathcal{R}_{\mu\nu} $ is the Palatini curvature. The Palatini Ricci curvature tensor $\mathcal{R}_{\mu\nu}$ is defined in terms of an independent connection $\hat{\Gamma}^\alpha{}_{\mu\nu}$ as:
\begin{equation}
\mathcal{R}_{\mu\nu} \equiv \hat{\Gamma}^\alpha_{\mu\nu ,\alpha} -
\hat{\Gamma}^\alpha_{\mu\alpha , \nu} +
\hat{\Gamma}^\alpha_{\alpha\lambda}\hat{\Gamma}^\lambda_{\mu\nu}
-\hat{\Gamma}^\alpha_{\mu\lambda}\hat{\Gamma}^\lambda_{\alpha\nu}.
\end{equation}

Now, by introducing a new auxiliary field $E$, the hybrid metric-Palatini action (\ref{2}) can be reformulated in the equivalent form of a scalar-tensor theory, given by
\begin{equation}\label{eq2sst}
S=\int\mathrm{d}^{4}x\sqrt{-g}[R+f(E)+f^{\prime }(E)(%
\mathcal{R}-E)]+S_m,
\end{equation}
where for $E=\mathcal{R}$, the action (\ref{eq2sst}) reduces to action (\ref{2}).
By introducing the following definitions:
\begin{equation}\label{3}
    \phi\equiv f^\prime(E)\,, \qquad  V(\phi)=Ef^\prime(E)-f(E)\,\,,
\end{equation}
it is possible to reformulate Eq. (\ref{2}) into a scalar-tensor theory with the following action \cite{Harko:2011nh}:
\begin{equation}\label{4}
    S=\int d^4x\sqrt{-g}[R+\phi\mathcal{R}-V(\phi)]+S_m\,\,.
\end{equation}
Then, by varying the action (\ref{4}) with respect to the metric $g_{\mu\nu}$, the scalar field $\phi$ and the independent connection $\hat{\Gamma}^\alpha_{\mu\nu}$, respectively, we obtain the field equations
\begin{align}
    R_{\mu\nu} + \phi \mathcal{R}_{\mu\nu} - \frac{1}{2} (R + \phi \mathcal{R} - V) g_{\mu\nu} &= \kappa^2 T_{\mu\nu}\;, \label{metricFEq} \\
       \mathcal{R} - V^{\prime}(\phi) &= 0 \;, \label{RVphi}\\
       \hat{\nabla}_\alpha \big( \sqrt{-g}\phi g^{\mu\nu} \big) &= 0\; \label{FieldEqphi},
\end{align}
where $V^{\prime}(\phi)\equiv dV(\phi)/d\phi$. Moreover, it is possible to relate $\mathcal{R}_{\mu\nu}$ and $R_{\mu\nu}$ using
\begin{equation} \label{eq:conformal_Rmn}
\mathcal{R}_{\mu\nu}=R_{\mu\nu}+\frac{3}{2\phi^2}\partial_\mu \phi \partial_\nu \phi-\frac{1}{\phi}\left(\nabla_\mu
\nabla_\nu \phi+\frac{1}{2}g_{\mu\nu}\Box\phi\right) \,,
\end{equation}
and rewrite the action (\ref{2}) in an equivalent scalar-tensor representation given as (we refer the reader to  Refs. \cite{Harko:2011nh,Capozziello:2012ny,Danila:2018xya} for more details):
\begin{equation}\label{5_}
    S=\int d^4x\sqrt{-g}\left[(1+\phi)R+\frac{3}{2\phi}(\partial\phi)^2 -V(\phi)\right]+S_m\,\,.
\end{equation}
In fact, it is interesting to note that this action is a special case of the Bergmann-Wagoner-Nordtvedt scalar-tensor theories, given by \cite{Bronnikov:2020vgg,Bergmann:1968ve,Wagoner:1970vr,Nordtvedt:1970uv}
\begin{equation}\label{6}
    S=\int d^4x\sqrt{-g}\left[f(\phi)R+g(\phi)(\partial\phi)^2 -V(\phi)\right]+S_m\,\,,
\end{equation}
where $f(\phi)$, $g(\phi)$ and $V(\phi)$ are arbitrary functions of $\phi$, which, in the case of the HMPG theory, by comparison with action (\ref{5_}), are given by:
\begin{equation}\label{7}
     f(\phi)=1+\phi\,,\qquad g(\phi)=\frac{3}{2\phi}\,,
\end{equation}
and $V(\phi)$ is the scalar field potential, represented by the same function.

\subsection{Action: Einstein frame}

Indeed, linking our theory to the Bergmann-Wagoner-Nordtvedt framework is of considerable significance, as it enables the application of a transformation between the Jordan conformal frame, represented by Eq.\;(\ref{5_}), and the Einstein conformal frame. 
The transition to this conformal frame is particularly useful, as it offers a more convenient and structured framework for analysing and solving the field equations. In this frame, the equations are typically simplified, making the overall analysis more tractable. 

More specifically, for both the general case and the particular instance of our theory, as outlined by the expressions in Eq.\;(\ref{7}), this transformation can be expressed as follows, according to the formulation in \cite{Bronnikov:2020vgg}, namely:
\begin{eqnarray}
\bar{g}_{\mu\nu} &=& f(\phi) g_{\mu\nu}\,, 
	\nonumber \\
\frac{d\phi}{d\bar{\phi}} &=& f(\phi)\left|f(\phi)g(\phi)-\frac{3}{2}\left(\frac{df}{d\phi}\right)^2\right|^{-1/2}, \notag 
\end{eqnarray} \label{8a}
which implies
\begin{equation} \label{8}
    \begin{minipage}{0.1\textwidth}
    \vspace{-0.5cm}
        \begin{align}
            &\bar{g}_{\mu\nu}=(1+\phi)g_{\mu\nu}\,\,, \notag \\
            &\phi=-\tanh^2\frac{\bar{\phi}}{\sqrt{6}}\,\,,\,\, \text{if  }\,\, -1<\phi<0\,\,, \notag \\
            &\phi=\tan^2\frac{\bar{\phi}}{\sqrt{6}}\,\,,\,\, \text{if  }\,\, \phi>0\,\,, \notag
        \end{align}
    \end{minipage}
\end{equation}
respectively.

By applying this transformation, we are able to derive the general form of the HMPG action within the Einstein frame. In this formulation, the quantities denoted by bars correspond to the transformed variables, as outlined in \cite{Bronnikov:2020vgg}. This approach provides a clearer and more manageable expression of the action in the context of the Einstein frame, which is given by:
\begin{eqnarray}\label{9_}
	S_E=\int d^4x\sqrt{-\bar{g}}\Big[\bar{R}-n\bar{g}^{\mu\nu}\bar{\phi}_{,\mu}\bar{\phi}_{,\nu} 
		\nonumber \\
	- ~U(\bar{\phi}) + \frac{\bar{\mathcal{L}}_m}{(1+\phi)^2} \Bigg].
\end{eqnarray}

In this context, $n$  is determined by the sign of the kinetic term of the scalar field in action (\ref{5_}), which is influenced by the sign of the scalar field itself. Specifically, we have $n=-\text{sign }(\frac{3}{2\phi})=-\text{sign }\phi$; if $n=+1$, $\phi$ is a canonical scalar field, and if $n=-1$, $\phi$ is a phantom scalar field. Additionally, we define the Einstein-frame potential as $U(\bar{\phi})\coloneqq V(\phi)/(1+\phi)^2$, with $\phi=\phi(\bar{\phi})$. This dependence also applies in the matter term, on the right, which is obtained by transforming $S_m$. Here, $\bar{\mathcal{L}}_m$ represents the transformed matter Lagrangian, whose form depends on an expression that remains to be determined.

\subsection{Field Equations for $V(\phi)\not\equiv 0$}

In this study, our goal is to explore and examine novel static, spherically symmetric, and electrically charged solutions within the framework of HMPG theory. This theory is generally described by the action (\ref{5_}), under the assumption that $V(\phi)\not\equiv 0$. Accordingly, our analysis will proceed within the Jordan conformal frame, given by the following action:
\begin{eqnarray}
	S=\int d^4x\sqrt{-g}\Bigg[(1+\phi)R+\frac{3}{2\phi}(\partial\phi)^2
		\nonumber \\
	-V(\phi) +F_{\mu\nu}F^{\mu\nu}\Big]\,,
\end{eqnarray}
which corresponds to considering $\mathcal{L}_m= F_{\mu\nu}F^{\mu\nu}$ in Eq. (\ref{5_}), which is a pure Maxwell term, where $F_{\mu\nu}\equiv \nabla_\mu A_\nu -\nabla_\nu A_\mu$ is the electromagnetic field tensor or Maxwell tensor, and $A_\mu$ is the electromagnetic 4-potential.

It is important to emphasize that, for the scope of this analysis, we will restrict the electromagnetic field to being purely electric. This assumption is crucial for simplifying the derivation and solution of the corresponding field equations. To facilitate this process, we will apply a transformation to the Einstein conformal frame, as specified in Eq. (\ref{8}). This approach is consistent with similar transformations that have been used in previous research on related topics. As aforementioned, it plays a crucial role in simplifying the process of solving the field equations. Once the transformation is applied, the resulting expression governing the system takes the following form:
\begin{equation}\label{S Vdiff0}
	S_E=\int d^4x\sqrt{-\bar{g}}\Big[\bar{R}-n\bar{g}^{\mu\nu}\bar{\phi}_{,\mu}\bar{\phi}_{,\nu}
	- ~U(\bar{\phi})+F_{\mu\nu}F^{\mu\nu}\Big].
\end{equation}
By varying this action with respect to the metric $\bar{g}^{\mu\nu}$, the scalar field $\bar{\phi}$ and the electromagnetic 4-potential $A_\alpha$, we obtain the Einstein conformal frame field equations, given by:
\begin{eqnarray}
   && \bar{G}_{\mu\nu}-n\bar{\phi}_{,\mu}\bar{\phi}_{,\nu}+\frac{1}{2} \bar{g}_{\mu\nu}\left(\bar{g}^{\alpha\beta}\bar{\phi}_{,\alpha}\bar{\phi}_{,\beta}\right)+\frac{1}{2}\bar{g}_{\mu\nu}U(\bar{\phi})
   	\nonumber \\
   && \qquad \qquad -\left(\frac{1}{2}\bar{g}_{\mu\nu} F_{\alpha\beta}F^{\alpha\beta} - 2 F{_\mu}^{\alpha} F_{\nu\alpha}\right)=0\,, \label{eq:G}\\
    && 2n \bar{\Box}\bar{\phi} - \frac{d}{d\bar{\phi}} U(\bar{\phi})=0\,, \label{eq: phi}\\
    && \bar{\nabla}_\mu F^{\mu\alpha}=0\,,\label{eq: F}
\end{eqnarray}
respectively, where $\bar{G}_{\mu\nu}$ is the transformed Einstein tensor, defined as $\bar{G}_{\mu\nu}=\bar{R}_{\mu\nu}-\frac{1}{2}\bar{g}_{\mu\nu}R$.

\section{Static and spherically symmetric metric: Regularity and causal structure}\label{chap 2}

\subsection{Metric}

To obtain static, spherically symmetric solutions to the Einstein-frame field equations, we consider the line element
\begin{equation}\label{dsE}
    ds_E^2=A(x)dt^2-\frac{1}{A(x)}dx^2-r^2(x)d\Omega^2 \,,
\end{equation}
where $A(x)$ and $r(x)$, the spherical radius function in this case, are functions to be determined of the radial coordinate $x$, and $d\Omega^2=(d\theta+\sin^2\theta d\varphi^2)$. This line element will be adopted because it inherently satisfies the ``quasiglobal" coordinate condition, under which $g_{00} = -g^{11}$ is satisfied ($x$ is known as a ``quasiglobal" radial coordinate). By satisfying this condition, the process of integrating the field equations and deriving solutions becomes significantly more straightforward. 

One of the advantages of satisfying this coordinate condition, and thus adopting this line element, is that it facilitates the study of the metric's asymptotic behaviour. In particular, if we analyse the line element at a spatial infinity -- a regular point of $x$ where $r(x)\to\infty$ -- and find that both $g_{00}$ and $g_{11}$ are constant and finite non-zero, then we have an asymptotically flat spacetime. In the particular case where $g_{00}=-g_{11}=1$, the spacetime is asymptotically Minkowskian, which is flat nonetheless. 
Apart from these cases, it is also possible that $g_{00}\sim -\Lambda y^2$ at infinity, where $y$ is a function of $x$ that diverges there, and $\Lambda$ is the cosmological constant. If $\Lambda<0$, then, at that limit, $g_{00}\to\infty$ and spacetime is asymptotically Anti-de Sitter (AdS). Conversely, if $\Lambda>0$, then $g_{00}\to-\infty$ and spacetime is asymptotically de Sitter (dS). Both cases are asymptotically non-flat, with negative and positive space-time curvature at infinity, respectively. Note that, as we will see in section \ref{example2_section}, in this work we always have $y=x$.  
Finally, if $g_{00}=0$ at infinity, then there are several possibilities, for which we need to analyse curvature invariants, in particular the Kretschmann scalar. A detailed discussion about this scalar will be addressed shortly (see the explanation following Eq. (\ref{K})). In general, at infinity, if it diverges, a singularity exists there; if it is non-zero finite (as when $g_{00}\sim -\Lambda y^2$), then spacetime is asymptotically non-flat there, but neither AdS nor dS; if it is null (as is the case when $g_{00}$ is finite non-zero), spacetime is asymptotically flat at that limit.

In fact, there may be metrics in which multiple points exhibit the features of a spatial infinity (radius function diverging at a regular $x$), and we could consider infinity to be located at any one of them. However, generally, only one spatial infinity exists in a given spacetime. Therefore, we must select only one of those suitable points to represent spatial infinity and begin our analysis from there.
There are cases, however, in which spacetime presents two infinities, corresponding to two adjacent suitable points. In such cases, we refer to them as the ``first'' and ``second'' spatial infinities, and begin our analysis from the former. 
Note that if a given metric does not feature an infinity, we must discard it, since, according to our physical knowledge, such a feature is required.

At flat spatial infinities we are able to determine the global mass of the configuration, also known as the Schwarzschild mass. On the other hand, at non-flat infinities, it is only possible to compute a quasi-local mass. However, we will not delve into this further. To determine the Schwarzschild mass, we compare the asymptotic behaviour of the metric, at infinity, with the Schwarzschild solution, shown in \cite{dInverno:1992gxs}, through a series expansion of $g_{00}$. However, to obtain it, we will adopt an expression adapted from the general expression used in \cite{Bronnikov:2020vgg}, which was modified to match the line element defined in Eq. (\ref{dsE}). This expression, derived from the previous comparison, is given by
\begin{equation}\label{massa}
  m=\lim_{x\to x_\infty} |r(x)| \frac{ d\left[\log{\sqrt{A(x)}}\right]/dx}{ d\left[\log{r(x)}\right]/dx}\,\,,
\end{equation}
where $x_\infty$ represents the value of $x$ where spatial infinity is located. We will see later that in this work we always have either $x_\infty=\infty$ or $x_\infty=-\infty$, or both. The limit $x\to x_\infty$ in the expression above will be used in our analysis, however, it can be adapted to other coordinates, as long as it corresponds to a spatial infinity. When there are two spatial infinities, the mass can be determined at each. In particular, evaluating the mass at $x\to\infty$ or $x\to-\infty$ will lead to different results. Finally, we will impose a positive mass, $m>0$, whenever it is determined at the first spatial infinity.

In the subsequent sections, we will derive particular solutions to Eqs.~(\ref{eq:G}) and (\ref{eq: phi}), while assuming the line element specified in Eq.~(\ref{dsE}). By doing so, we will obtain the Einstein-frame solutions. However, prior to proceeding, it is necessary to solve Eq.~(\ref{eq: F}) to determine the explicit form of the Maxwell tensor, $F^{\mu\nu}$. Since we are considering a static, spherically symmetric spacetime, the electromagnetic field source generates a purely radial and static field. In this scenario, and due to the anti-symmetric property of the Maxwell tensor, $F^{\alpha\beta} = -F^{\beta\alpha}$, there exist only two independent non-zero components, namely $F^{10}$ and $F^{23}$. 
Nevertheless, as previously mentioned, we restrict our analysis to a source that is solely electrically charged (with vanishing magnetic charge). Under this assumption, only the component $F^{10}$ remains non-zero. Imposing these conditions, we solve Eq.~(\ref{eq: F}), leading to the following result:
\begin{equation}\label{}
    F^{10}(x)=\frac{q}{r(x)^2}\,, \label{F determined}
\end{equation}
where $q$ is an integration constant that may be interpreted as the charge of the source. 

After obtaining the Einstein-frame solutions, and before proceeding with the analysis, we will also first need to obtain the metric back in the Jordan frame, since that is the one we are interested in. Inverting the metric transformation equation and using the transformation of $\phi$, for both types of scalar field, from Eq.\;(\ref{8}), we obtain the following:
\begin{align}
      g_{\mu\nu}=\cosh^2(\bar{\phi}/\sqrt{6})\bar{g}_{\mu\nu}\,,\quad &\text{if}\quad -1<\phi<0\,\,,\,n=+1\,,\label{g munu can}\\
      g_{\mu\nu}=\cos^2(\bar{\phi}/\sqrt{6})\bar{g}_{\mu\nu}\,,\quad &\text{if}\quad \phi>0\,\,,\,n=-1\,.\label{g munu pha}
\end{align}

This way, in terms of the respective line elements, we obtain, by defining $\psi\coloneqq\bar{\phi}/\sqrt{6}$, the following relations:
\begin{align}
        ds_J^2=\cosh^2\psi\, ds_E^2\,,\quad &\text{if}\quad -1<\phi<0\,\,,\,n=+1\,, \label{sJ can}\\
        ds_J^2=\cos^2\psi\, ds_E^2\,,\quad &\text{if}\quad \phi>0\,\,,\,n=-1\,,\label{sJ pha}
\end{align}
where the subscripts $J$ and $E$ are relative to the Jordan and Einstein frames, respectively. Note that the ``quasiglobal" coordinate condition is no longer satisfied by these line elements. However, as long as $ds_E^2$ satisfies it, the analysis of the asymptotic behaviour discussed previously remains applicable.

In this work, we will analyse two distinct examples of general solutions derived within this framework, each corresponding to a different potential $V(\phi)$. To obtain each of them, we employ the inverse problem method. This approach involves postulating a suitable expression for $r(x)$, a separate choice for each example, rather than directly assuming a form for the potential. 
Through this methodology, three unknown functions emerge that must be determined by integrating the Einstein-frame field equations: $A(x)$, $\bar{\phi}(x)$, and $U(\bar{\phi})$. Note that, via the expression $U(\bar{\phi})= V(\phi)/(1+\phi)^2$, we can subsequently determine $V(\phi)$.

Before delving into those two examples, we now present the systematic approach that will be used to analyse the regularity, as well as the horizon and throat structures, of each solution.

\subsection{Regularity of the metric and the Kretschmann scalar}\label{Regularity_Kscalar}

Upon completing the preceding analysis of the asymptotic behaviour, the next step is to examine the regularity of the metric. To assess this, we will focus on evaluating the Kretschmann scalar, $K$, which serves as a key indicator of potential singularities. The Kretschmann scalar can be expressed in the following form:
\begin{equation}\label{K}
	K=4K_1^2+8K_2^2+8K_3^2+4K_4^2\,\,,
\end{equation}
where the components are given by:
\begin{eqnarray} \label{Ki}
	K_1=g^{11}R^0_{101}\,,\qquad  K_2=g^{22}R^0_{202}\,,
		\nonumber \\
	 K_3=g^{22}R^1_{212}\,,\qquad K_4=g^{33}R^2_{323}\,\,,
\end{eqnarray}
respectively.

To analyse the scalar, we examine each of its components separately, however, in general, all components must be considered collectively. A singularity in the spacetime corresponds to a point in $x$ where the Kretschmann scalar $K$ diverges, which occurs if any one of the four components diverges. Furthermore, if $K$ diverges at multiple points, the singularity is located at the first such point from spatial infinity, as spacetime terminates at this location. Apart from this, if at a given point in $x$ none of the four components in Eq.\;(\ref{Ki}) diverges, then the spacetime is regular at that point.

The first divergence of $K$ will only be classified as a singularity if there is no regular point of $x$ before it where $r\to\infty$ or $r\to 0$ without being a minimum (or maximum).
In these cases, there is a second spatial infinity or a regular centre, respectively, and no singularity exists, even if $K$ diverges. This is because the range of $x$ corresponding to the spacetime does not even reach the point of divergence, being always regular. 
Let us denote a singularity by $x_s$ and the last point of the radial coordinate, with the (first) spatial infinity as the starting point, by $x_{\text{end}}$.
If the first divergence of $K$ actually corresponds to a singularity, there we define $x_{\text{end}}=x_s$. When a regular point like the ones just discussed occurs before it, we define there $x_{\text{end}}$. 

In general, the locations of these ``end" points  depend on the specific form of the metric. As we will see in the analysis of each general solution, it is possible that they coincide with the end of the validity interval opposite to the (first) spatial infinity, or some other point preceding that. Thus, their actual locations are only determined after a thorough analysis of the metric.

In the presence of a singularity, it is possible to determine whether it exhibits an attractive or repulsive nature. This classification arises from the behaviour of the metric in an asymptotically flat spacetime. As $r(x) \to \infty$, the metric component $g_{00}$ behaves as $g_{00} = 1 + U(x)$, where $U(x)$ represents the Newtonian gravitational potential in the weak-field limit. For $r(x) \geq 0$, a generalized Newtonian potential can be considered, where $g_{00}$ serves as an analogue to this potential. Consequently, the derivative of $g_{00}$ with respect to $x$ refers to the gravitational force. This allows us to classify the singularity as follows: if $g_{00}' > 0$ as $x \to x_s^+$, or if $g_{00}' < 0$ as $x \to x_s^-$, the singularity is classified as attractive, as the gravitational force in its vicinity is attractive. Conversely, if $g_{00}' < 0$ as $x \to x_s^+$, or if $g_{00}' > 0$ as $x \to x_s^-$, the singularity is classified as repulsive, as the force becomes repulsive. 
To better understand this classification, one can think of $g_{00}$ as a potential function, where a minimum corresponds to an attractive potential, and a maximum indicates a repulsive potential. 

It is important to note that when $x \to \infty$ corresponds to spatial infinity, which is one of the possibilities in this work, as will be seen later, $x$ always approaches $x_s$ from above. However, when spatial infinity is located at $x\to -\infty$, $x$ always approaches $x_s$ from below.
Alternatively, based on the aforementioned explanation, the singularity can be analysed without directly considering the derivative of $g_{00}$. Instead, if $g_{00} \to 0^+$ (or $0^-$) at $x_s$, it indicates an attractive (or repulsive) nature. If $g_{00} \to \infty$ (or $-\infty$) at that point, the singularity is classified as repulsive (or attractive). This method, though applicable only for these specific values, is often preferred due to its greater precision and absence of numerical errors.

In addition to the aforementioned classification, singularities can also be classified as time-like or space-like (apart from light-like ones, which will be discussed in the next subsection), in a manner analogous to the singularities found in the Reissner-Nordström and Schwarzschild solutions (see \cite{dInverno:1992gxs}), respectively. A time-like singularity arises when, in its vicinity, the metric signature is $(+--\,-)$, whereas a space-like singularity occurs when the roles of the time and radial coordinates are interchanged, resulting in a metric signature of $(-+-\,-)$. Note that, as we are considering a Lorentzian spacetime, the metric functions $g_{00}$ and $g_{11}$ must always have opposite signs. This means that, to identify the metric signature of a given spacetime region, we may only evaluate the sign of $g_{00}$. Accordingly, the singularity is time-like if $g_{00}>0$, and space-like if $g_{00}<0$, at its vicinity.

\subsection{Horizons, throats and bounces}

Subsequent to this analysis, we proceed to examine the existence of horizons within the metric. Horizons are characterized as regular points of $x$, which are neither minima nor maxima of the radius function, occurring prior to a singularity, at which $g_{00} = 0$. We will denote them, in general, by $x_H$ (other designations will be used when referring to particular types of horizons). A horizon corresponding to a simple root of $g_{00}$, which we will denote as a simple horizon, is classified as either an event (EH), internal (IH), or Cauchy (CH) horizon. This classification depends on the total number of horizons. If there is only one, then it is an EH; if there are two, the external one is an EH and the internal one is a CH; if there are three or more, the most external one is an EH and the remaining ones are IHs. Note that the metric signature always changes beyond any of these horizons.

Apart from this, when two horizons exist in a solution, and a choice of parameters causes them to coincide, an extremal horizon emerges. A particular case occurs when this happens at a double root of $g_{00}$, resulting in a degenerate or double horizon. Note that the metric signature always remains unchanged beyond such horizons. Furthermore, in this work, all extremal horizons are degenerate, but we will always refer to them simply as extremal horizons.

The classification of each horizon, presented before, suffers modifications when at least one of them is extremal. When the outermost horizon, or the only one, is extremal, it is classified as an extremal EH. In this case, all remaining horizons, regardless of their number, are IHs or extremal IHs. If an EH exists and there is an extremal horizon inside it, the latter is classified as an extremal IH (rather than CH); if there are two or more horizons inside the EH, all of them are IHs or extremal IHs, similar to the previous classification.

When considering the relationship between horizons and singularities, it follows that if at least one horizon exists, the solution is classified as a black hole solution. Conversely, if no horizon is present, the solution is considered to correspond to a naked singularity. For an illustrative example of such behaviour, one may refer to the Reissner-Nordström solution, which can exhibit 0 horizons, 1 extremal horizon, or 2 simple horizons, depending on the relationship between the electric charge $q$ and the mass $m$ \cite{dInverno:1992gxs}. 
Additionally, if $g_{00} = 0$ occurs at a point of singularity, the singularity is classified as light-like, or null, and is also referred to as a singular horizon \cite{Clement:2009ai}.

Turning to the spherical radius function, it is defined as $R(x)\equiv \sqrt{-g_{22}}$. Note that in the Einstein frame that is simply the $r(x)$ present in the line element of Eq.\;(\ref{dsE}), whereas in the Jordan frame we also have to consider the conformal factor. Regarding this function, we observe the following: if at a given point of $x$, $R(x)  \to \infty$, it implies the presence of a surface located at infinity, with infinite area. Conversely, when $R(x)$ is non-zero finite, we have a sphere with non-zero finite area. Additionally, if $R(x) \to 0$ at a singular point or at a regular one that is not a minimum (or maximum) of $R(x)$, it corresponds to a central region. Note that, a spatial infinity occurs at a regular point of $x$ where $R(x)\to\infty$ (we previously defined it in terms of $r(x)$).

Finally, we also aim to investigate the presence of throats and bounces within the metric. To do so, it is necessary to closely examine the radius function, particularly its derivative. Both structures occur at regular local minima of $R(x)$, which correspond to zeros of its derivative, always prior to a singularity point.
A throat occurs at a minimum located in a spacetime region where the metric signature is $(+--\,-)$, thus, where $g_{00}>0$. On the other hand, if the signature is $(-+-\,-)$, thus if $g_{00}<0$, a bounce occurs at that point \cite{Simpson:2018tsi}. Furthermore, if at a minimum of $R(x)$ we also have $g_{00}=0$, a throat arises there as well, but it is now referred to as an extremal null throat \cite{Simpson:2018tsi}.

Additionally, when analysing the radius function, if it exhibits a regular local maximum (which is also a zero of the derivative) and $g_{00} > 0$, such a structure is termed an anti-throat \cite{Rodrigues:2022mdm}.

The presence of a throat, in the absence of an anti-throat, any horizon and a singularity, is generally linked to the formation of a wormhole.  In this scenario, the spacetime remains entirely regular and features a throat, with a second spatial infinity located beyond this structure. Such geometries can be classified in two distinct categories: if $g_{00}\not=0$ at any point, then the minimum of $R(x)$ is a throat and the configuration corresponds to a two-way traversable wormhole; whereas if $g_{00}=0$ at the minimum of $R(x)$, then it is an extremal null throat, as aforementioned, and the geometry is classified as a one-way traversable wormhole \cite{Simpson:2018tsi}.

However, it is also possible for a throat to not be associated with a wormhole geometry. This occurs, for example, in cases where an anti-throat is present \cite{Rodrigues:2022mdm}, as such configurations do not support a second spatial infinity. Another notable example occurs whenever the minimum of $R(x)$, independently of the sign of $g_{00}$ of the region where it is located, is inside one or more horizons, whichever they are. This geometry is known as a ``black bounce spacetime", or simply ``black bounce" \cite{Simpson:2018tsi}. Essentially, it represents a type of regular black hole in which the central singularity is removed and replaced by a regular minimum of $R(x)$, beyond which the interior evolves into an expanding cosmological solution. Although the term suggests the existence of a bounce, as defined before, it is used to refer to any spacetime featuring a minimum of $R(x)$ in such a configuration. When one of the infinities is flat or AdS and the other is dS, this geometry is also known as ``black universe" \cite{Bolokhov:2012kn, Bronnikov:2006fu}.

A regular minimum of $R(x)$, in the context of a wormhole or a ``black universe" solution, represents a region of spacetime that serves to connect two distinct and separate regions. These regions may either belong entirely to our own Universe, or one region may be part of our Universe while the other pertains to a parallel universe, in the case of an asymmetric metric, or a copy of our universe, in the case of a symmetric metric. In this work we consider that beyond the minimum of $R(x)$ lies a distinct universe. 

This notion of symmetry is relative to the specific location of this minimum.
To verify it, we must examine whether all metric functions remain invariant under the transformation $x \to -x$ when the structure is located at $x = 0$. In the case that the structure resides at a different point, we can perform a translation of the coordinate system so that the structure is positioned at $x = 0$, and then check for the symmetry. If all functions exhibit the considered symmetry, then it follows that the metric itself, and by extension the entire spacetime, must also exhibit such a symmetry. An initial approach to analysing the potential asymmetry of the metric is to consider the spherical radius function, which we already use in identifying these structures. If the radius function is asymmetric with respect to the minimum of $R(x)$, this immediately implies that the metric is likewise asymmetric. 
In such cases, we may encounter phenomena such as horizons on one side of the minimum of $R(x)$, but absent on the other, or a differing evolution of the spherical radius on each side. 
 Beyond the conventional scenarios explored thus far, more specific and unique cases like these may emerge during the subsequent analyses, where they will be thoroughly examined and discussed.

With the completion of this analysis, we are now in a position to construct the Penrose diagrams for each of the spacetime geometries under consideration. The key structures that require particular attention when constructing these diagrams include spatial infinities, horizons, throats, bounces, and singularities.

At this point, we have established the requisite conditions to proceed with the detailed analysis of each of the aforementioned examples of general solutions.

\section{Example 1: Canonical Sector}\label{example1_section}

\subsection{Metric} 

In this first example, in order to solve the Einstein-frame field equations (\ref{eq:G}) and (\ref{eq: phi}), we start by assuming, for the spherical radius the following function (as is done in Refs.~\cite{Bronnikov:2020vgg, Zloshchastiev:2004ny}):
\begin{equation}\label{r can}
    r(x)=\sqrt{x(x+1)}\,.
\end{equation}
The function $r(x)$, as defined, diverges as $x \to \pm \infty$ and vanishes at $x = 0$ and $x = -1$. In this context, the constant $1$ is interpreted as a characteristic length scale, which could, in principle, take any value, but for simplicity, it has been normalized to unity. 
Referring to the analysis outlined in \cite{Rois:2024qzm}, for the case $V(\phi) \equiv 0$, the Einstein-frame solution obtained was compatible with both canonical and phantom scalar fields. This flexibility arose because both the parameter $n$ and the scalar field itself could assume either sign (as discussed below Eq. (\ref{9_})), with no constraints to restrict these possibilities. However, in the present case, the assumption of this specific expression for $r(x)$ intrinsically enforces the scalar field to be canonical, without the possibility of being phantom. Consequently, this solution is naturally associated with the canonical sector. 
This conclusion is a direct outcome of the equations of motion derived from Eq. (\ref{eq:G}), where the only non-zero components correspond to $\mu = \nu$. By subtracting the equation for $\mu = \nu = 1$ from the one for $\mu = \nu = 0$, we obtain:
\begin{equation}
    \frac{1}{2}A(x)\left(\frac{1}{x^2(1+x)^2}-2n\bar{\phi}'\,^2(x)\right)=0 \,,
\end{equation}
which implies
\begin{equation}
	 \frac{1}{x^2(1+x)^2}=2n\bar{\phi}'\,^2(x)\,,
\end{equation}
where we assume $A(x)\not=0$, which holds everywhere, except at a horizon or a light-like singularity, which does not invalidate that the remaining term has to be null everywhere else. By analysing the second equation, we see that we must have $n=+1$ so that the equation is real-valued, which means the scalar field is canonical. 

After adopting the outlined assumptions and requirements, substituting Eq.\;(\ref{F determined}) into Eq.\;(\ref{eq:G}), determining the non-zero components of all tensors present in Eqs. (\ref{eq:G}) and (\ref{eq: phi}), and combining these equations appropriately, we ultimately derive a general solution that depends on free parameters, with each specific choice of parameters yielding a distinct particular solution. However, to determine the function $A(x)$ as fully as possible, including appropriately fixing certain integration constants, we impose the condition that the general solution be asymptotically flat, specifically Minkowskian, as $x \to \infty$. This assumption is justified because, at this limit, the spacetime is regular and $r \to \infty$, as previously noted, thereby corresponding to a spatial infinity. Furthermore, empirical and theoretical evidence suggests that the universe is asymptotically flat. The same condition could have been imposed for the limit $x \to -\infty$.
To enforce asymptotic flatness, we match the series expansion of $A(x)$ to the expansion of $g_{00}$ in the Schwarzschild metric, expressed as:
\[
1 - \frac{2M}{x} + \mathcal{O}\left(\frac{1}{x^2}\right).
\]
Here, $M$ represents the mass of the configuration, which corresponds to the Schwarzschild mass of our Einstein-frame solution. Based on the preceding discussion, we have the following solution in the Einstein frame:
\begin{equation}
	\bar{\phi}(x) = \bar{\phi}_0+\frac{1}{\sqrt{2}}\log \left(\frac{x}{x+1}\right)\,,\label{phi exp}
\end{equation}
\begin{widetext}
\begin{eqnarray}
    U(\bar{\phi}(x)) &=& \frac{1}{x^2 (x+1)^2}\left\{-2 \left[18 M x (x+1) (2 x+1)+q^2 (12 x (x+1)+1)\right]-4 x (x+1) \log \left(\frac{x}{x+1}\right)\right. \label{potential} 
    \nonumber \\
    && \qquad \times \left.\left[3 \left(6 M x (x+1)+M+2 q^2 (2 x+1)\right)+q^2(6 x (x+1)+1) \log \left(\frac{x}{x+1}\right)\right]\right\}\,, 
    \\
    A(x) &=&1-6 M (2 x+1)-4 q^2-4 \log \left(\frac{x}{x+1}\right) \left[3 M x (x+1)+q^2 x (x+1) \log \left(\frac{x}{x+1}\right)+q^2 (2 x+1)\right]\,,
    	\label{A(x) can}
\end{eqnarray}
\end{widetext}
where $\bar{\phi}_0 \coloneqq \bar{\phi}(\infty)$ is an integration constant that can take any arbitrary value. It is worth noting that the solution for the particular case $q=0$ has been obtained and discussed in Refs.~\cite{Bronnikov:2020vgg, Zloshchastiev:2004ny}.

From these equations, it is evident that all the functions are well-defined within the domain $x \in ]-\infty, -1[ \;\cup\; ]0, +\infty[$, and they diverge at $x = -1$ and $x = 0$. These divergences indicate potential singular points. 
Additionally, the scalar field potential is negative throughout this domain, tending to $-\infty$ as $x$ approaches either $x = -1$ or $x = 0$, irrespective of the values of the constants. Conversely, the behaviour of $A(x)$, particularly whether it possesses zeros and the number of such zeros, will depend on the relationship between $M$ and $q$, as will be demonstrated in the following analysis.

Before delving into a more thorough analysis, we may first transform the solution into the Jordan conformal frame, particularly focusing on its line element and potential, as this is the solution of primary interest. Although a detailed examination of the Einstein-frame solution is feasible, it is not the central objective of our study. To perform the transformation, we begin by defining:
\begin{equation}
    \psi(x)\coloneqq\bar{\phi}(x)/\sqrt{6}=\frac{1}{2\sqrt{3}}\log \left(\frac{x}{x+1}\right)+\psi_0\,\,,
\end{equation}
where $\psi_0=\bar{\phi}_0/\sqrt{6}$. Thus, since the scalar field is canonical, by using Eq. (\ref{sJ can}) in combination with Eqs. (\ref{dsE}) and (\ref{r can}), we obtain, in the present case, the following line element in the Jordan conformal frame:
\begin{eqnarray}\label{dsJ can}
    ds_J^2&=&\cosh^2\left[\frac{1}{2\sqrt{3}}\log \left(\frac{x}{x+1}\right)+\psi_0\right]\times
    	\nonumber \\
    &&\times \left[A(x)dt^2-\frac{1}{A(x)}dx^2-x(x+1)d\Omega^2\right],
\end{eqnarray}
where the expression of $A(x)$, given by Eq. (\ref{A(x) can}), was not substituted as it is too lengthy. Apart from this, we may also transform $U(\bar{\phi})$ back to the Jordan frame, by using the transformation of $\phi$ of Eq. (\ref{8}), so that we have the entire solution in that frame. By doing so, we obtain the scalar field potential $V(\phi)$:
\begin{equation}\label{V_canonical}
    U(\bar{\phi})=\frac{V(\phi)}{(1+\phi)^2} \, \quad \Leftrightarrow\, \quad V(\phi)=\cosh^{-4}\psi \,\,U(\bar{\phi})\,.
\end{equation}

Note that the particular case $q=0$ of this solution has also been briefly discussed in Ref.~\cite{Bronnikov:2020vgg}.

Now, we finally have the conditions necessary to analyse the Jordan-frame general solution. First of all, by analysing Eq. (\ref{dsJ can}), we see that, as was discussed before, the largest domain of validity of the coordinate $x$ for which the line element takes real values is $x\in\,]-\infty, -1[ \;\cup\; ]0, +\infty[$.  This way, as the metric has to be defined in a continuous interval, there are two distinct intervals of validity: $x\in\,]-\infty, -1[$ or $x\in\,]0, +\infty[$. Note that these are two possible ranges of $x$ that are not simultaneously valid. In fact, by analysing the metric, we find that both show similar behaviour, due to the freedom of choice of the constants, thus, we only analyse the positive one. 

In this case, as the limit $x\to\infty$ corresponds to spatial infinity (as in the Einstein-frame metric), $x$ is restricted to the range $0\leq x_\text{min} < x<\infty$, where $x_\text{min}$, as the name suggests, is the minimum value, allowed by the metric behaviour, that $x$ can have (it is analogous to the previously defined $x_\text{end}$). Accordingly, if there is a singularity, we define $x_\text{min}=x_s$.
Conversely, if the spacetime is regular and either extends to a second spatial infinity or presents a regular centre at a given point, the analysis ends at that location and $x_\text{min}$ is accordingly defined there.

Now, by analysing the line element in Eq.\;(\ref{dsJ can}) as $x\to\infty$, we find $g_{00}=-g_{11}=\cosh^2\psi_0$, thus, this is an asymptotically flat spacetime, being Minkowskian if $\psi_0=0$. By using Eq.\;(\ref{massa}), the Schwarzschild mass of this Jordan frame solution is given by:
\begin{equation}\label{massa ex 1}
    m=\frac{1}{6}\cosh(\psi_0) [6 \,M\cosh (\psi_0) + \sqrt{3} \sinh (\psi_0) ]\,.
\end{equation}
By imposing $m>0$, we are able to derive a constraint on $M$ and $\psi_0$, namely $M>-\tanh (\psi_0)/2\sqrt{3}$ (note that we are only imposing $m>0$, not $M>0$, thus, it may have any sign). The range of this expression is, approximately, $[-0.2887,\,0.2887]$, where the positive values are related to $\psi_0<0$ and the negative ones to $\psi_0>0$. From this, we are able to obtain two straightforward relations, namely, this shows us that positive values of both constants, $M$ and $\psi_0$, always lead to a positive mass and negative values of both are never allowed. We also see that any value of $M$ above $0.2887$ always leads to a positive $m$, and below $-0.2887$ is never allowed, independently of $\psi_0$. Apart from these relations, a proper use of the expression is required. In the following analysis we always assume combinations of constants that lead to $m>0$. 

\subsection{Regularity conditions} 

Now, to test the regularity of the metric, we are interested in analysing the Kretschmann scalar. Accordingly, we start by analysing the component $K_1$, which is given by:
\begin{widetext}
\begin{eqnarray}
      K_1&=&-\frac{1}{12 x^2 (x+1)^2}\text{sech}^2\left(\frac{1}{2 \sqrt{3}}\log \left(\frac{x}{x+1}\right)+\psi_0\right) \left\{2 \sqrt{3} \left[-6 M-4 q^2 \log \left(\frac{x}{x+1}\right)+2 x+1\right] \right.
      	\nonumber \\
     &&\;\; \times \left.\tanh\left(\frac{1}{2 \sqrt{3}}\log \left(\frac{x}{x+1}\right)+\psi_0\right)+\left[4 \log \left(\frac{x}{x+1}\right) \left[3 M x (x+1)+q^2 (2 x+1)+q^2 x (x+1) \log \left(\frac{x}{x+1}\right)\right]\right.\right.
     	\nonumber \\
     &&\left. +6 M (2 x+1)+4 q^2-1\biggl] \text{sech}^2\left(\frac{1}{2 \sqrt{3}}\log \left(\frac{x}{x+1}\right)+\psi_0\right)+24 \left[2 x (x+1) \log\left(\frac{x}{x+1}\right) \biggl[3 M x (x+1) \right.\right.
     	\nonumber \\
     &&\left.\left.\left.+q^2 (2 x+1)+q^2 x (x+1) \log \left(\frac{x}{x+1}\right)\right]+3 M x (x+1) (2 x+1)+q^2 (2 x (x+1)-1)\right]\right\}\,.
\end{eqnarray}
\end{widetext}

By analysing this expression, at the limit $x\to\infty$, we find it is zero, as well as the remaining terms of $K$, which supports our previous description of the spatial infinity. Furthermore, we find that the logarithmic function $\log \left(\frac{x}{x+1}\right)$ and the term $1/(x^2(x+1)^2)$ always cause $K_1$ to diverge at $x=0$ (and $x=-1$, if we were to consider the negative interval of $x$). Apart from that, independently of the value of any constants ($\psi_0$, $M$ and $q$), there is no divergence in $K_1$ at any positive $x$. Additionally, by analysing the remaining terms of $K$ we find the same results, which means, since there is neither a second spatial infinity (the spherical radius, $R(x)$, given by a combination of Eq. (\ref{r can}) and the conformal factor, only goes to infinity at $x\to\infty$) nor a regular centre, a singularity occurs at $x_\text{min}=x_s=0$.

\subsection{Characterization of the horizons} 

The existence of horizons, identified as regular points of \( x \) where \( g_{00} = 0 \) -- or equivalently $A(x)=0$, since the conformal factor never vanishes -- is determined by the relationship between the electric charge \( q \) and the Einstein-frame mass parameter \( M \), while remaining independent of \( \psi_0 \). Depending on their values, the system may exhibit 0, 1, or 2 horizons. The number of horizons is directly associated with critical values of \( M \) or \( q \), denoted as \( M_c \) and \( q_c \), respectively. Importantly, these critical values are not unique and depend on the value of the other parameter.
To calculate the critical values, we require that \( g_{00} \) and its derivative vanish simultaneously. However, this condition could not be resolved analytically, necessitating the fixing of \( q \) (or \( M \)) to specific values, resulting in numerical solutions. Consequently, obtaining distinct critical values requires repeating this approach. For example, when \( q = 5 \), we find the critical value \( |M_c| \approx 5.00389 \) (the same is obtained when $q=-5$), and for \( M = 5 \), we find \( |q_c| \approx 4.99611 \) (we find the same for $M=-5$). Because fixing one parameter yields two symmetric critical values for the other, absolute values are used. Additionally, since the metric always involves \( q^2 \), only the magnitude of $q$ is relevant, in any case, as its sign does not affect the results. 
Consequently, we may always represent $q$ by $|q|$ (and $q_c$ by $|q_c|$). On the other hand, this is not the case for $M$, as is discussed below.

\begin{figure*}[ht!]
    \centering
    \includegraphics[width=0.45\linewidth]{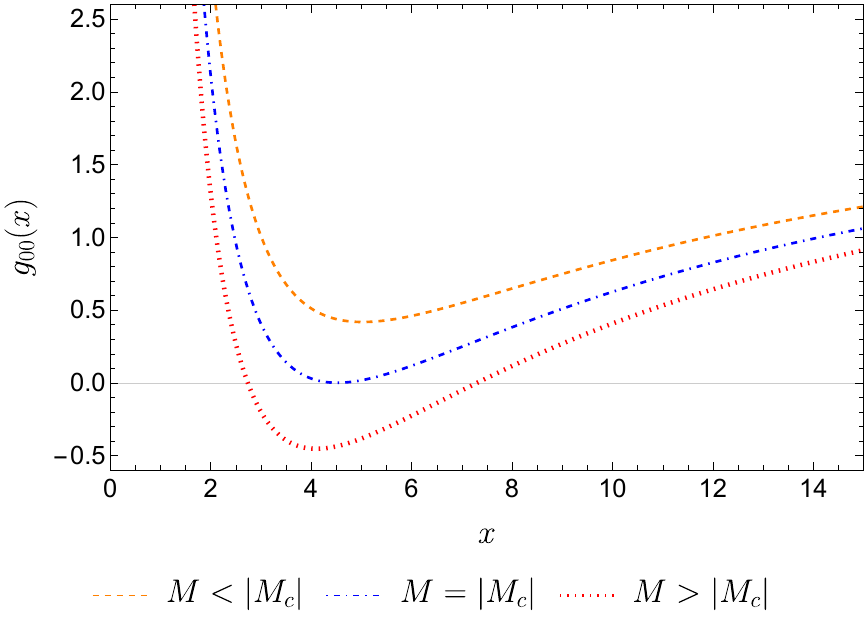}
    \hspace{0.6cm}
    \includegraphics[width=0.45\linewidth]{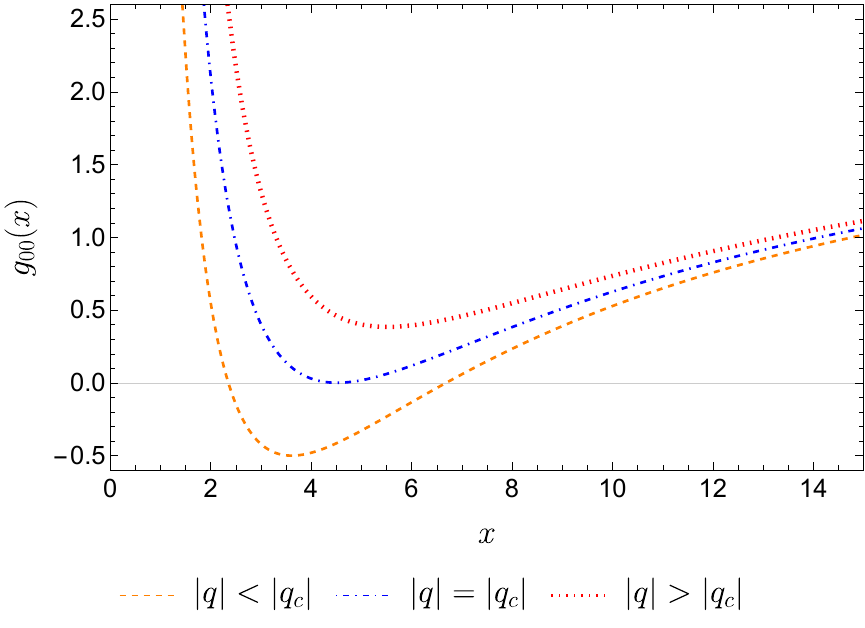}
    \caption{Left plot: Plot of the metric function $g_{00}(x)=\cosh^2(\psi)~A(x)$ (with $A(x)$ given by Eq.~(\ref{A(x) can})) for $q=5$ and $|M_c|= 5.00389$. Three different mass scenarios associated with this critical mass are shown. The cases $M<|M_c|$ and $M>|M_c|$ correspond to $M=|M_c|-0.5$ and $M=|M_c|+0.5$, respectively. By extending these relations to $|M|$, similar results are obtained in the range $x<-1$ if we consider $M<0$ and the negative counterpart $M_c= -5.00389$. Right plot:  Plot of the metric function $g_{00}(x)$ for $M=5$ and $|q_c|= 4.99611$. Three different charge scenarios associated with this critical charge are shown. The cases $|q|<|q_c|$ and $|q|>|q_c|$ correspond to $|q|=|q_c|-0.5$ and $|q|=|q_c|+0.5$, respectively. As is clear from the use of absolute values, both signs of $q$ lead to the same results.}
    \label{critical_can}
\end{figure*}

The number of horizons depends on the relative magnitude of \( M \) to \( |M_c| \), or $|q|$ to $|q_c|$, as can be seen from Fig. \ref{critical_can}. The plot on the left shows the behaviour of $g_{00}$, with respect to $x$, in different mass scenarios, considering $|q|=5$ and $M_c=5.00389$, and the one on the right shows that behaviour in different charge scenarios, with $M=5$ and $|q_c|=4.99611$. Note that we are only considering the positive range of $x$.

By analysing the left plot, we find:
\begin{itemize}
    \item If \( M > |M_c| \), the system has two horizons: an external event horizon at \( x_{EH} \) and an internal Cauchy horizon at \( x_{CH} \).
    \item If \( M = |M_c| \), there exists an extremal event horizon at \( x_{eH} \).
    \item If \( M < |M_c| \), no horizons are present.
\end{itemize}

Then, by analysing the right plot we find that for \( |q| \) the scenario is inverted. 

Note that the above relations hold for $|M|$ as well, but with the horizons occurring in the negative range of $x$ when $M<0$. In fact, unlike what was previously discussed for $q$, the sign of $M$ actually plays a crucial role: symmetric values of $M$ produce curves that are approximately mirror-symmetric, with the horizons, when present, always lying on the same side of the \( x \)-axis as \( M \). Thus, for \( M < 0 \), no horizons are found in the interval of interest (\( x > 0 \)); the reverse would apply if we were to consider the negative interval of $x$ (\( x < -1 \)). Consequently, only positive \( M_c \) values are physically meaningful for this study. Henceforth, we consider only $M_c>0$.

It is also notable that the locations of horizons are independent of \( \psi_0 \). Additionally, as illustrated in Fig. \ref{critical_can}, in all cases, as \( x \to x_s \), \( g_{00} \to \infty \), and the metric signature near the singularity is \( (+--\,-) \).

Furthermore, upon analysing the spherical radius, we find that it vanishes at \( x = x_s \), which implies the presence of a time-like, repulsive, central singularity at \( x = x_s \). Additionally, by examining the derivative of the spherical radius, we observe that there are no extrema points, and thus, no throats are present in any case.
Accordingly, for a fixed value of \( |q| \), we classify the solution as follows:
\begin{itemize}
    \item If \( M < M_c \), a naked singularity exists at \( x = x_s \);
    \item If \( M = M_c \), the configuration corresponds to a black hole with an extremal event horizon;
    \item If \( M > M_c \), the configuration represents a black hole with both an event horizon and a Cauchy horizon, with \( x_{CH} < x_{EH} \).
\end{itemize}

In Appendix~\ref{apA}, we present the detailed derivation of the $f(\mathcal{R})$ function from action (\ref{2}) -- which is ultimately obtained in parametric form -- corresponding to a solution in the $M>M_c$ scenario, in particular with $q=5$, $M=5.50389$ ($M=|M_c|+0.5$) and $\psi_0=0$.

Considering the results above, we conclude that this solution is analogous to the Reissner-Nordström solution. Consequently, the Penrose diagrams of the spacetime geometries analysed, shown in Figs. \ref{PenroseCan0} and \ref{PenroseCan}, are also analogous to those of that solution. Figure \ref{PenroseCan0} refers to the case $M<M_c$, the plots from left to right in Fig. \ref{PenroseCan} correspond to the diagrams of the cases  $M=M_c$ and $M>M_c$, respectively. More specifically:

\begin{itemize}
	\item
Figure \ref{PenroseCan0} depicts a Penrose diagram for a naked, central, time-like singularity solution. 
In this type of diagram, null geodesics are drawn at a $45^\circ$ angle. On the right of the plot, the asymptotically flat infinity is depicted, where $i^0$ is the flat spatial infinity, the upper and lower diagonal lines represent the future and past null infinities, denoted by $\mathscr{I^+}$ and $\mathscr{I^-}$, respectively, and $i^+$ and $i^-$ are the future and past time-like infinities. This representation and notation is used for any asymptotically flat infinity in this work. The time-like singularity appears as a vertical double line, on the left, at $x=x_s$.

\begin{figure}[ht!]
	\centering
	\includegraphics[scale=0.75]{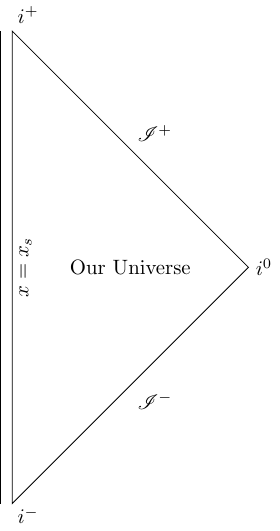}
	\caption{Penrose diagram for a naked, central, time-like singularity solution. We refer the reader to the text for more details.}
	\label{PenroseCan0}
\end{figure}

\begin{figure*}[ht!]
	\centering
	\includegraphics[scale=0.525]{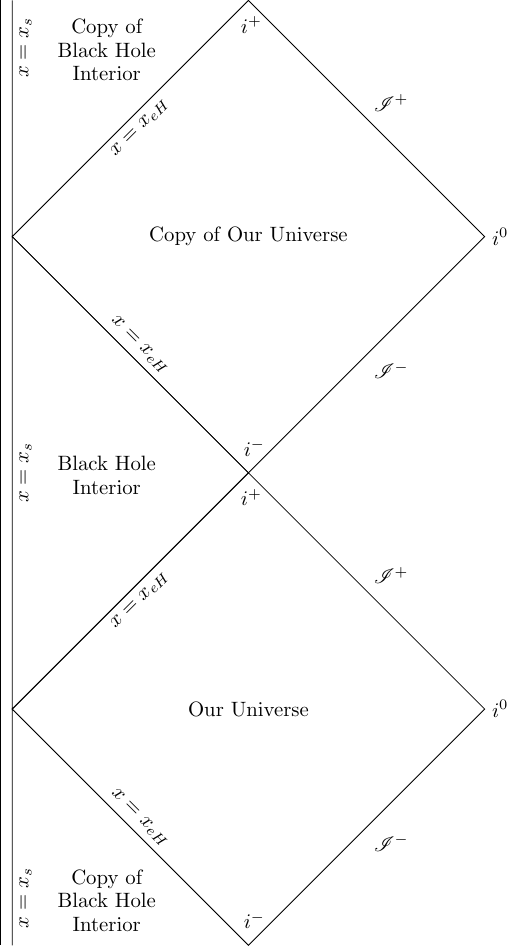}
	\hspace{1cm}
	\includegraphics[scale=0.442]{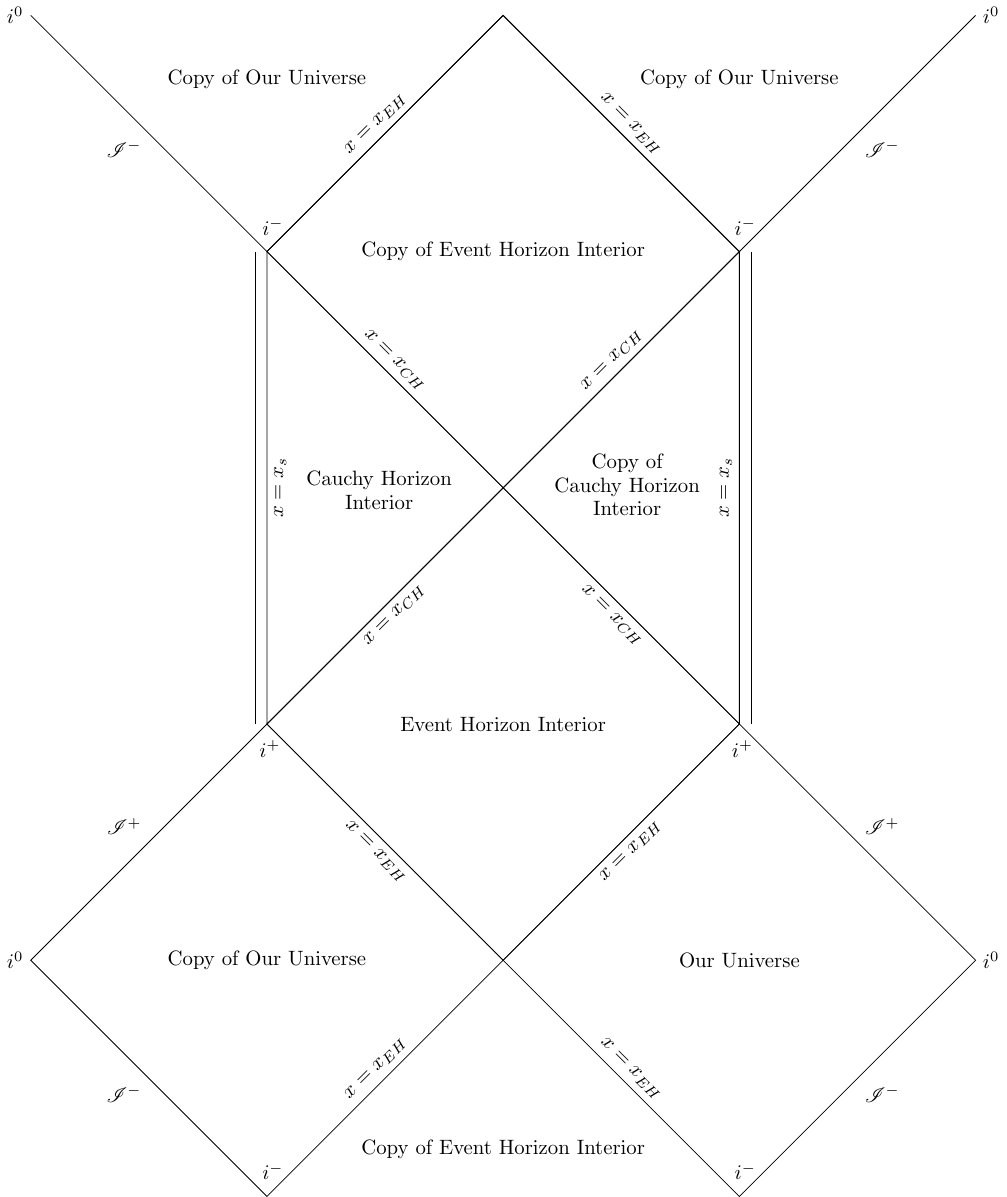}
	\caption{Left plot: Penrose diagram for a black hole solution with an extremal event horizon and a central, time-like singularity. Right plot: Penrose diagram for a black hole solution with an event horizon, a Cauchy horizon, and a central, time-like singularity. We refer the reader to the text for more details.}
	\label{PenroseCan}
\end{figure*}

\item
The left plot of Fig. \ref{PenroseCan} denotes the Penrose diagram for a black hole solution with an extremal event horizon and a central, time-like singularity. In the region labelled ``Our Universe", the infinity appears on the right, while on the left the upper and lower diagonal lines at $x=x_{eH}$ depict future and past branches of the extremal event horizon, respectively. This horizon does not change the metric signature (which holds for any extremal horizon), and within the future branch (the region labelled ``Black Hole Interior'') there are future and past branches of that horizon, on the right, and a time-like singularity, on the left. Note that copies of each region exist, accessed by traversing certain horizon branches. Moreover, the entire diagram actually extends infinitely upwards and downwards, from both regions labelled ``Copy of Black Hole Interior", repeating the same structure depicted here in this portion. 

\item
The right plot of Fig. \ref{PenroseCan} depicts a Penrose diagram for a black hole solution with an event horizon, a Cauchy horizon, and a central, time-like singularity. Now, the diagonal lines at $x=x_{EH}$ in ``Our Universe" correspond to future and past branches of the event horizon. This horizon changes the metric signature (which holds for any simple horizon) to $(- +-\,-)$, and within the future branch (``Event Horizon Interior"), there are two past branches of that horizon, and two future branches of the Cauchy horizon, which change the metric signature back to $(+--\,-)$, depicted as the diagonal lines at $x=x_{CH}$. Within the left one of these (``Cauchy Horizon Interior"), there are future and past branches of that horizon, on the right, and a time-like singularity, on the left. Moreover, copies of each region exist and the entire diagram actually extends infinitely upwards and downwards, as before. The ``Copy of Our Universe" lying to the left of the original is connected to it by an Einstein-Rosen bridge.

\end{itemize}

\section{Example 2: Phantom Sector}\label{example2_section}

\subsection{Metric} 

In this second example, in order to solve the Einstein-frame field equations (\ref{eq:G}) and (\ref{eq: phi}), we assume a different expression for the spherical radius, given by (as is done in Refs.~\cite{Bolokhov:2012kn, Bronnikov:2020vgg}):
\begin{equation}\label{r pha}
    r(x)=\sqrt{x^2+1}\;\;,
\end{equation}
which diverges as $x \to \pm\infty$ and is never null. For simplicity, the length scale is once again chosen to be $1$. In contrast to the previous example, where the scalar field was assumed to be canonical, the choice of the expression for $r(x)$ in this case yields a phantom field. As previously explained, this conclusion follows from the equations of motion derived from Eq.~(\ref{eq:G}). As before, we may subtract the equation for $\mu = \nu = 1$ from the one for $\mu = \nu = 0$, yielding:
\begin{equation}
    A(x)\left(-\frac{2}{(x^2+1)^2}-n\bar{\phi}'\,^2(x)\right)=0 \,,
\end{equation}
which implies
\begin{equation}
	  -\frac{2}{(x^2+1)^2}=n\bar{\phi}'\,^2(x)\;,
\end{equation}
where, as explained before, we are assuming $A(x)\not=0$ in order to obtain the equation on the right. By analysing it, we find that, for it to be real-valued, we must have $n=-1$, which implies that we are now dealing with a phantom scalar field. 

By considering these assumptions and requirements, and following the same procedure that allowed us to obtain Eqs. (\ref{phi exp})--(\ref{A(x) can}), we get the following general Einstein-frame solution for the current case:
\begin{widetext}
\begin{eqnarray}
 \bar{\phi}(x) &=& \bar{\phi}_0+\sqrt{2}\arctan(x)\,,\label{phi ph} \\
    U(\bar{\phi}(x)) &=& \frac{1}{2 \left(x^2+1\right)^2}\left\{4 \left(x^2+1\right) \arctan(x) \left[M \left(9 x^2+3\right)-q^2 \left(3 \pi  x^2-6 x+\pi \right)+q^2 \left(3 x^2+1\right) \arctan(x)\right]\right.
    \nonumber \\
    && \qquad\left.-6 M \left(x^2+1\right) \left(3 \pi  x^2-6 x+\pi \right)+q^2 \left[-12 \pi  \left(x^3+x\right)+12 x^2+\pi ^2 \left(3 x^4+4x^2+1\right)+8\right]\right\} \,,
    \label{potential ph} \\
    A(x) &=& \frac{1}{4} \left\{4 \arctan(x) \left[3 M \left(x^2+1\right)-\left[q^2 \left(\pi  x^2-2 x+\pi \right)\right]+q^2 \left(x^2+1\right) \arctan(x)\right]-6 M\left(\pi  x^2-2 x+\pi \right)\right.
    	\nonumber \\
    &&\qquad\left.+q^2 \left[\pi  \left(\pi  x^2-4 x+\pi \right)+4\right]+4\right\}\,, 
    \label{A(x) ph}
\end{eqnarray}
\end{widetext}
where $\bar{\phi}_0 \coloneqq \bar{\phi}(0)$ is an integration constant, and $M$ represents the mass parameter of the Schwarzschild metric. Notably, $M$ also corresponds to the Schwarzschild mass of the present solution in the Einstein frame. It is worth noting that this solution has already been obtained and discussed in Ref.~\cite{Bolokhov:2012kn}, as has the particular case $q=0$ in Ref.~\cite{Bronnikov:2020vgg}.

Upon analysing these functions, we observe that they are well-defined for all real numbers, remaining regular throughout the domain $x \in ]-\infty, +\infty[$, as no divergences occur. 
Furthermore, the characteristics of the scalar field potential, particularly its sign and the presence of extrema, as well as the behaviour of $A(x)$, including the presence and number of zeros, will depend on the relationship between $M$ and $q$. These dependencies will become evident in the subsequent analysis.

Before proceeding with a more detailed analysis, however, we will first derive the solution's line element in the Jordan conformal frame, as this is the frame of interest. To achieve this, we will define, as previously:
\begin{equation}
    \psi(x)\coloneqq\bar{\phi}(x)/\sqrt{6}=\frac{1}{\sqrt{3}}\arctan(x)+\psi_0\,\,,
\end{equation}
where $\psi_0=\bar{\phi}_0/\sqrt{6}$. This way, as we are dealing with a phantom scalar field, by using Eq.\;(\ref{sJ pha}) in combination with Eqs.\;(\ref{dsE}) and (\ref{r pha}), we get the following line element in the Jordan conformal frame:
\begin{eqnarray}\label{dsJ pha}
    ds_J^2&=&\cos^2\left(\frac{\arctan(x)}{\sqrt{3}}+\psi_0\right)
    \times 
    	\nonumber \\
    && \times \left[A(x)dt^2-\frac{1}{A(x)}dx^2-(x^2+1)d\Omega^2\right],
\end{eqnarray}
(once again, the expression of $A(x)$, from Eq. (\ref{A(x) ph}), was not substituted as it is too extensive). Similarly, we may also transform $U(\bar{\phi})$ back to the Jordan frame, by using the transformation of $\phi$ of Eq. (\ref{8}), obtaining the scalar field potential $V(\phi)$ of this solution:
\begin{equation}\label{V_phantom}
    U(\bar{\phi})=\frac{V(\phi)}{(1+\phi)^2} \quad \Leftrightarrow \quad V(\phi)=\cos^{-4}\psi \,\,U(\bar{\phi})\;\;.
\end{equation}

Note that the particular case $q=0$ of this solution has also been obtained and discussed in Ref.~\cite{Bronnikov:2020vgg}.

Following the same type of analysis carried out in the previous example, we now start analysing this general Jordan-frame solution. Firstly, by considering Eq. (\ref{dsJ pha}), we find that the largest interval of validity of the coordinate $x$, for which the metric takes real values, is $x\in\, ]-\infty,+\infty[$. In general, at both endpoints the metric is regular and $R(x)\to\infty$ (the spherical radius of Eq. (\ref{dsJ pha}), which involves the conformal factor), being these the only points that may correspond to a spatial infinity. There are, however, certain scenarios in which the conformal factor prevents $R(x)$ from approaching infinity, thereby also preventing a spatial infinity from arising at both limits simultaneously, as we will see. 

If we treat $x\to\infty$ to be the (first) spatial infinity, 
when it can be considered as such, then $x$ is restricted to the range $-\infty \leq x_\text{min} < x < \infty$, where $x_\text{min}$ represents the same as before (even though the Einstein-frame metric is everywhere regular and its radius never null, this may be a singular or regular central point, due to the conformal factor, as we will see). On the other hand, if we treat $x\to-\infty$ to be the (first) spatial infinity, $x$ is then restricted to the range $-\infty < x < x_\text{max}\leq\infty$, where $x_\text{max}$ is analogous to $x_\text{min}$. 

In general, either limit may be chosen as the (first) spatial infinity. However, based on our current physical understanding, our Universe is asymptotically flat, which means, in our analysis, if a flat spatial infinity is a possibility, it will always be chosen as the (first) one. Furthermore, when one of the limits cannot be considered as an infinity, the other, which is the only alternative, must always be chosen, since a spacetime must have at least one spatial infinity to be physically meaningful, even if it is non-flat.

By analysing the line element of Eq.\,(\ref{dsJ pha}) as $x\to\infty$, we find $g_{00}=-g_{11}=\cos^2(\frac{\pi}{2\sqrt{3}}+\psi_0)$. Note that:
\begin{itemize}
	\item If $\psi_0\neq\frac{\pi}{2}-\frac{\pi}{2\sqrt{3}}+c_1\pi$, where $c_1$ is an integer, it is an asymptotically flat spacetime, in particular, it is Minkowskian if $\psi_0=-\frac{\pi}{2\sqrt{3}}+c_1\pi$.
	\item Conversely, if $\psi_0=\frac{\pi}{2}-\frac{\pi}{2\sqrt{3}}+c_1\pi$, then, at that limit, both $g_{00}$ and $g_{11}$ are null and $R(x)\to 1/\sqrt{3}$, thus, it is not a spatial infinity. In this case, we must consider $x\to-\infty$ instead. 
\end{itemize}

Now, analysing the line element in the limit $x\to-\infty$, we must also consider two different cases of $\psi_0$: 
\begin{itemize}
	\item If $\psi_0\neq\frac{\pi}{2}+\frac{\pi}{2\sqrt{3}}+c_1\pi$ ($\psi_0$ may be equal to the aforementioned values), we may consider two cases.
	
	\begin{itemize}
		\item
	When $M\neq \pi q^2/3$, at this limit we have $g_{00}\sim (-3M+\pi q^2)\cos^2(\frac{\pi}{2\sqrt{3}}-\psi_0)x^2$ and $g_{11}\sim 1/[(-3M+\pi q^2)\cos^2(\frac{\pi}{2\sqrt{3}}-\psi_0)x^2]$, and so, spacetime is asymptotically dS if $M>\pi q^2/3$ ($g_{00}\to -\infty$) and AdS if $M<\pi q^2/3$ ($g_{00}\to \infty$). In the former case the spatial infinity is asymptotically dS, and in the latter, it is asymptotically AdS. Note that these should not be interpreted as conventional spatial infinities due to the peculiar asymptotic structures of these spacetimes; rather, as defined earlier, we adopt the term ``spatial infinity" to refer to the regular point in $x$ where $R(x)\to\infty$. Furthermore, on a separate note, in these cases (dS and AdS), $x\to-\infty$ will only be considered as the (first) spatial infinity when $x\to\infty$ cannot be considered as such. 
	\item On the other hand, when $M=\pi q^2/3$, we find $g_{00}=-g_{11}=\cos^2(\frac{\pi}{2\sqrt{3}}-\psi_0)$ (similar to the case obtained for $x\to\infty$), thus, spacetime is asymptotically flat, being Minkowskian when $\psi_0=\frac{\pi}{2\sqrt{3}}+c_1\pi$. In this case, this limit may always be considered as the (first) spatial infinity. 
	\end{itemize}

\item Alternatively, if $\psi_0=\frac{\pi}{2}+\frac{\pi}{2\sqrt{3}}+c_1\pi$, then, at $x\to-\infty$, for any $M$ and $q$, we have $R(x)\to 1/\sqrt{3}$. Thus, in this situation, this is not a spatial infinity and we must instead consider $x\to\infty$. Note that each of these values of $\psi_0$ is symmetric to one that prevents $x\to\infty$ from being an infinity.
\end{itemize}

Since the conformal factor is always positive and finite when the endpoints are infinities, in these cases, the asymptotic behaviours, at both limits, of the term $A(x)$, shown in Eq. (\ref{A(x) ph}), which is the Einstein-frame $\tilde{g}_{00}$ metric function, is the same as that of $g_{00}$, aside from the distinction between flat and Minkowskian spacetimes (in the Einstein frame these are always Minkowskian).

Now, by using Eq. (\ref{massa}), the Schwarzschild mass of the Jordan-frame solution at $x\to\infty$ and $x\to-\infty$, when they are flat spatial infinities ($M=\pi q^2/3$ in the latter limit), is given by: 
\begin{equation}\label{m ex 2-1}
    m=\frac{1}{6} [3\,M+3 \,M\cos (\pi/\sqrt{3}\pm2\psi_0) - \sqrt{3} \sin (\pi/\sqrt{3}\pm\psi_0) ]\,,
\end{equation}
where the ``$+$" sign is relative to the limit $x\to\infty$, whereas the ``$-$" sign is relative to $x\to-\infty$ (this correspondence will be used again in the following discussion).
When this last limit is a dS or AdS spatial infinity, it is not possible to determine this, which reflects the global mass of the configuration, being only possible to determine the quasi-local mass, as aforementioned.

At both infinities, when spacetime is asymptotically Minkowskian, which occurs when  $\psi_0=\pm\frac{\pi}{2\sqrt{3}}+c_1\pi$ (each sign is relative to each infinity, as before), as aforementioned, we find $m=M$, and so, in this case, the Jordan-frame's mass is equal to the Einstein-frame's one. By imposing $m>0$, we obtain, in this case, that $M$ must also be positive. On the other hand, in general, with that imposition -- note that we are not imposing $M>0$, thus, it may have any sign --, we are able to obtain a constraint on both $M$ and $\psi_0$, given by
\begin{equation}\label{Mphant}
    M>\frac{\sin(\pi/\sqrt{3}\pm2\psi_0)}{\sqrt{3}[1+\cos(\pi/\sqrt{3}\pm2\psi_0)]}\,\,.
\end{equation}

Due to this constraint, we verify that the full range of allowed values of $\psi_0$ increases with $M$ and decreases as it gets smaller. Furthermore, we are also able to find that for the limit $x\to\infty$, when $M=0$ (which can be seen as a threshold) the entire range $\psi_0\in]\frac{\pi}{2}-\frac{\pi}{2\sqrt{3}}+c_1\pi,-\frac{\pi}{2\sqrt{3}}+(c_1+1)\pi[$ is allowed, whereas $\psi_0\in]-\frac{\pi}{2\sqrt{3}}+c_1\pi,\frac{\pi}{2}-\frac{\pi}{2\sqrt{3}}+c_1\pi[$ is not. As $M$ increases, the former range remains always allowed, while the latter becomes increasingly more allowed, expanding from the lower endpoint. On the other hand, as $M$ decreases from $0$, the latter range remains never allowed, while the former becomes increasingly less allowed, contracting from the higher endpoint. For the limit $x\to-\infty$,
when $M=0$, the entire range $\psi_0\in]\frac{\pi}{2\sqrt{3}}+c_1\pi,\frac{\pi}{2}+\frac{\pi}{2\sqrt{3}}+c_1\pi[$ is allowed, whereas $\psi_0\in]\frac{\pi}{2}+\frac{\pi}{2\sqrt{3}}+c_1\pi,\frac{\pi}{2\sqrt{3}}+(c_1+1)\pi[$ is not. We find that these, as well as the way they expand or contract with $M$, are mirror-symmetric to those for the previous limit, respectively.

\subsection{Regularity conditions} 

Now, as before, we are interested in analysing the Kretschmann scalar to test the regularity of the metric. Thus, we start with the component $K_1$, given by
\begin{widetext}
\begin{eqnarray}
      K_1&=&\frac{1}{12\left(x^2+1\right)^2}\sec ^2\left(\frac{\arctan(x)}{\sqrt{3}}+\psi_0\right) \biggl\{4 \arctan(x) \biggl[3 \left(x^2+1\right)\left[3 M \left(x^2+1\right)-q^2 \left(\pi  x^2-2 x+\pi \right)\right]
      \nonumber \\
      &&\left.\left. - 4 \sqrt{3} q^2 \tan \left(\frac{\arctan(x)}{\sqrt{3}}+\psi_0\right)\right]-\biggl[4 \arctan(x) \left[3 M \left(x^2+1\right)-q^2 \left(\pi x^2-2 x+\pi \right)+q^2 \left(x^2+1\right) \arctan(x)\right]\right.
      \nonumber \\
      &&\left.-6 M \left(\pi  x^2-2 x+\pi \right)+q^2 \left[\pi \left(\pi  x^2-4 x+\pi \right)+4\right]\biggl] \sec ^2\left(\frac{\arctan(x)}{\sqrt{3}}+\psi_0\right)+8\sqrt{3} \left(-3 M+\pi  q^2+x\right) \right.
      \nonumber \\
      &&\qquad \left. \times \tan \left(\frac{\arctan(x)}{\sqrt{3}}+\psi_0\right)-18 M\left(x^2+1\right) \left(\pi  x^2-2 x+\pi \right)+3 q^2 \left[\left(\pi  x^2-2 x+\pi \right)^2+12\right]\right.
      	\nonumber \\
      &&+12 q^2\left(x^2+1\right)^2 \arctan(x)^2-4 \sec ^2\left(\frac{\arctan(x)}{\sqrt{3}}+\psi_0\right)\biggl\}\,.
\end{eqnarray}
\end{widetext}

Upon analysing this expression, as well as the remaining components, at the limit $x\to\infty$, we find it is null, which is consistent with our previous description of this spatial infinity. Conversely, at $x\to-\infty$, we find it may either vanish or approach a non-zero finite value, depending on whether $M$ equals $\pi q^2/3$ or not, respectively, supporting our previous discussion. 

Furthermore, the above expression, as well as the remaining $K$, diverges if and only if the conformal factor, $\cos ^2\left(\frac{\arctan(x)}{\sqrt{3}}+\psi_0\right)$, is zero at a finite $x$, which, if it occurs, only happens at a single point. From the brief analysis of the functions in the Einstein frame (\ref{phi ph})--(\ref{A(x) ph}), all of which are regular, it was already expected that any divergence in $K$ would only arise in the Jordan-frame solution, due to the vanishing of the conformal factor.

Whether or not it occurs, and the point at which it does, depends solely on the value of $\psi_0$. This is because of the function $\frac{\arctan(x)}{\sqrt{3}}$ that is part of the conformal factor, since the argument of the cosine function gets limited to the range $]\psi_0-\frac{\pi}{2\sqrt{3}}, \psi_0+\frac{\pi}{2\sqrt{3}}[$ (these endpoints are obtained by taking the limits $x\to-\infty$ and $x\to\infty$, respectively). This way, as the cosine function zeros occur where the argument is equal to $\frac{\pi}{2}+c_1\pi$, this range may at most contain one zero in it, depending on $\psi_0$, as its length is smaller than $\pi$.

By analysing the conformal factor, we identify two distinct types of critical values for $\psi_0$. The first type is given by $\psi_0 = \frac{\pi}{2} - \frac{\pi}{2\sqrt{3}} + c_1\pi$, which causes the conformal factor to vanish as $x \to \infty$. The second type is given by $\psi_0 = \frac{\pi}{2} + \frac{\pi}{2\sqrt{3}} + c_1\pi$, leading to the vanishing of the conformal factor as $x \to -\infty$. In both cases, however, the vanishing of the conformal factor does not result in a divergence of $K$, which remains finite and non-zero.
Moreover, for values of $\psi_0$ above the first critical value and below the second, the conformal factor vanishes at $x = \tan[\sqrt{3}(\pi/2 - \psi_0 + c_1\pi)]$, where $c_1$ must be consistent with the critical values, since otherwise this expression does not hold. At this zero, $K$ diverges. Notably, this zero can take any real value, and for a fixed $c_1$, increasing $\psi_0$ results in a smaller value of $x$. Conversely, for values of $\psi_0$ below the first critical value and above the second, the conformal factor has no zeros (remaining strictly positive), ensuring that $K$ does not diverge anywhere.
However, due to the constraint $m > 0$, this latter scenario is not possible for values of $M \lessapprox 0.14$, as such values permit only $\psi_0$ configurations that lead to a divergence of $K$. As $M$ increases beyond this threshold, $\psi_0$ values that do not lead to a divergence become allowed, with the allowed range expanding as discussed after Eq.\,(\ref{Mphant}). Furthermore, for $M \gtrapprox 0.14$, all $\psi_0$ values that cause divergences are consistent with the imposition $m > 0$.

Furthermore, as aforementioned, the only possibility for a second spatial infinity arises at the limits $x\to-\infty$ or $x\to\infty$, depending on which is the first one. Moreover, $R(x)$ is zero only when $K$ diverges (if and only if the conformal factor vanishes at a finite $x$), and so, regular centres cannot exist. Additionally, we also have that the vanishing of that factor always leads to $g_{00}=0$. Accordingly, we conclude that this divergence of $K$ necessarily corresponds to a light-like, central singularity, and there we define $x_\text{min}=x_s$ or $x_\text{max}=x_s$, according to where the first infinity is located.

\subsection{Characterization of the metric: horizons, throats and bounces} 

The remaining analysis of the metric, particularly regarding the characterization of the horizons, benefits from taking into account what limit is the (first) spatial infinity. Therefore, we will now split it according to the value of $\psi_0$ into two cases: $\psi_0 \neq \frac{\pi}{2} - \frac{\pi}{2\sqrt{3}} + c_1\pi$ and $\psi_0 = \frac{\pi}{2} - \frac{\pi}{2\sqrt{3}} + c_1\pi$.

\subsubsection{$\psi_0 \neq \frac{\pi}{2} - \frac{\pi}{2\sqrt{3}} + c_1\pi$}

As mentioned earlier, in this case, the limit $x\to\infty$ is always a flat spatial infinity, and is Minkowskian if $\psi_0=-\frac{\pi}{2\sqrt{3}}+c_1\pi$. Additionally, the limit $x\to-\infty$ is not an infinity when $\psi_0=\frac{\pi}{2}+\frac{\pi}{2\sqrt{3}}+c_1\pi$; otherwise, it is, being flat when $M=\pi q^2/3$ and non-flat (dS or AdS) for any other values. This way, as explained below, we will, once again, split the analysis of this subsection according to another condition regarding $\psi_0$.

As discussed before, when a flat spatial infinity is a possibility, which is always the case in this subsection, we will always consider it to be the (first) one. Therefore, when $M\neq \pi q^2/3$ we will only consider $x\to\infty$ as such, whereas, when $M= \pi q^2/3$ we may consider either limit (except when $x\to-\infty$ is not an infinity, in which case we have to consider $x\to\infty$). However, both of them lead to the same spacetime geometries, and so, even in this case, we will only consider the limit $x\to\infty$ to be the (first) spatial infinity, which is what characterizes this subsection. Accordingly, from this point onward, when imposing $m>0$, we will always be considering the mass determined at the limit $x\to\infty$ (see Eq. (\ref{m ex 2-1}) and the discussion following Eq. (\ref{Mphant})).

We can now proceed with the analysis of the metric. Firstly, we have that the spherical radius function and its derivative depend only on $\psi_0$. Thus, the existence and location of minima of $R(x)$ -- throats ($g_{00}>0$), extremal null throats ($g_{00}=0$), or bounces ($g_{00}<0$) -- hence zeros of its derivative, are only dependent on this constant, as we have verified for singularities. In fact, this type of structures (minima of $R(x)$) only exist if there is a second spatial infinity, which, as aforementioned, is only possible for $\psi_0\neq\frac{\pi}{2}+\frac{\pi}{2\sqrt{3}}+c_1\pi$ (and given the values of this subsection). Thus, from this point onward, we will split the analysis once again. 

Accordingly, from now on we will first consider only the values of $\psi_0$ of this subsection that satisfy $\psi_0\neq\frac{\pi}{2}+\frac{\pi}{2\sqrt{3}}+c_1\pi$, for which both $x\to\pm\infty$ are infinities. After that analysis, we will consider the values $\psi_0=\frac{\pi}{2}+\frac{\pi}{2\sqrt{3}}+c_1\pi$, for which only $x\to\infty$ is an infinity.

\begin{center}
    Case 1: $\psi_0\neq\frac{\pi}{2}+\frac{\pi}{2\sqrt{3}}+c_1\pi$
\end{center}

By analysing the $R(x)$ function and its derivative, we actually find that, in any case in which there is a singularity, minima of $R(x)$ do not exist; conversely, if there is no singularity, then $R(x)$ has exactly one minimum. As aforementioned, its type depends on the sign of $g_{00}$, and so, it depends on its position relative to the horizons -- a simple horizon inverts the sign of $g_{00}$, whereas an extremal horizon has no effects on it. This way, the minimum will be located at $x_T$ (throat), $x_T=x_H$ (extremal null throat), or $x_B$ (bounce). These values can be determined numerically and may take any real value, according to $\psi_0$. We find that when $\psi_0\in]\frac{\pi}{2}+\frac{\pi}{2\sqrt{3}}+c_1\pi,(c_1+1)\pi[$ the minimum of $R(x)$ lies at $x<0$, when $\psi_0=(c_1+1)\pi$ it is located at $x=0$ and when $\psi_0\in](c_1+1)\pi, \frac{\pi}{2}-\frac{\pi}{2\sqrt{3}}+(c_1+1)\pi[$ it lies at $x>0$. 

In cases where any one of these structures exists, there is always a second spatial infinity at $x\to-\infty$. 
Furthermore, in these cases, we can analyse the symmetry of the metric with respect to $x_T$, $x_{T}=x_H$, or $x_B$. In the range of no singularity, $R(x)$ is symmetric if and only if $\psi_0=c_1\pi$. This is because $r(x)$ is always symmetric (see Eq. (\ref{r pha})) and, in this range, the conformal factor is symmetric only for these values. Note that, regardless of $|q|$, $M\approx 0.73796$ corresponds to the value of $M$ for which $\psi_0=c_1\pi$ is the maximum value in $\psi_0\in]-\frac{\pi}{2\sqrt{3}}+c_1\pi,\frac{\pi}{2}-\frac{\pi}{2\sqrt{3}}+c_1\pi[$ such that $m>0$ -- as $M$ increases, this maximum value also increases. Thus, that is the minimum value of $M$ for which $R(x)$ is allowed to be symmetric. Additionally, the metric functions $g_{00}$ and $g_{11}$ exhibit symmetry under the same condition for $\psi_0$ and if and only if the spatial infinity at $x\to-\infty$ is also flat, hence when $M=\frac{\pi q^2}{3}$. Therefore, in these conditions, the metric -- and consequently the spacetime -- is symmetric relative to $x_T$, $x_T=x_H$, or $x_B$, otherwise it is asymmetric. 

Now, regarding the search for horizons, we will split $g_{00}$ in two terms, the conformal factor and $A(x)$, as shown in Eq. (\ref{A(x) ph}). When $g_{00}$ vanishes due to the former, there is a singularity, as already discussed. Thus, horizons may only emerge at the zeros of $A(x)$, which is independent of $\psi_0$. Whether or not each of them is a horizon depends on its location relative to the singularity. If they coincide, in which case it is still a singularity, we have
\begin{equation}\label{psi0-xs}
    \psi_0=\frac{\pi}{2}-\frac{\arctan(x_H)}{\sqrt{3}}+c_1\pi\;,
\end{equation}
where $x_H$ is the location of that zero. Thus, in this case, we have $x_s=x_H$. If $\psi_0$ is lower than this, then the singularity occurs first ($x_s>x_H$), whereas, if it is greater, then the zero occurs first ($x_s<x_H$), corresponding to a horizon. In both cases, we have to be careful to consider only $\psi_0$ inside the range for which there is a singularity. 

The number of zeros of the term $A(x)$ may range from 0 to 3, depending on the relationship between the electric charge, $q$, and $M$, similar to what was found in Example 1 (canonical sector). As before, this number is related to a critical behaviour, however, now, in general, there is not one but three critical values of $M$ or $q$, denoted as $M_{ci}$ and $q_{ci}$, respectively, where $i=\{1,2,3\}$. The values of all three are not unique, with each set being always associated with a specific fixed value of the other parameter. The values relative to $i=1$ and $i=3$ are determined in the same way as in Example 1, by using a numerical method that requires both $A(x)$ and its derivative to be null simultaneously. On the other hand, we have $M_{c2}=\pi q^2/3$ and $|q_{c2}|=\sqrt{3M/\pi}$, which are precisely the values that lead to $x\to-\infty$ being flat and also the only ones that may lead to a symmetric spacetime, as aforementioned. As before, both signs of $q$ always lead to the same results, as it appears as $q^2$ in the metric, so we can always use its absolute value. However, this symmetry does not hold for $M$, being its sign important. In fact, unlike what was verified for Example 1, now, the zeros of $A(x)$ may only exist for $M>0$ and all the $M_{ci}$ are always positive.

In the present case, the critical behaviour, including how each critical value affects the number of zeros, and the total number of such values, depends on the range within which the fixed constant lies, existing four different possibilities. Thus, we now present four examples for each constant, each with a different critical behaviour. 

In each example, we will present the corresponding horizon structure for different values of the non-fixed constant, considering that all of the existing zeros of $A(x)$ are horizons, in order to simplify the discussion. Nevertheless, this is only true when there is either no singularity or when it is located past all of the zeros. From Eq. (\ref{psi0-xs}), we find that this latter scenario, in which $x_s\leq x_{H\text{min}}$, where $x_{H\text{min}}$ is the location of the innermost zero (horizon), occurs if $\psi_0\geq\frac{\pi}{2}-\arctan(x_{H\text{min}})/\sqrt{3}+c_1\pi$. Note that, as aforementioned, due to the imposition $m>0$, the former scenario is not a possibility for values of $M\lessapprox 0.14$ (this includes $M<0$, as with any similar inequality in the remaining analysis), as they only allow values of $\psi_0$ for which a singularity exists. Thus, as always, in the following discussion, only the combinations of constants that are allowed by the imposition $m>0$ are of interest.

\begin{figure*}[ht!]
    \centering
    \makebox[0.994\textwidth][r]{%
    \includegraphics[width=0.515\linewidth]{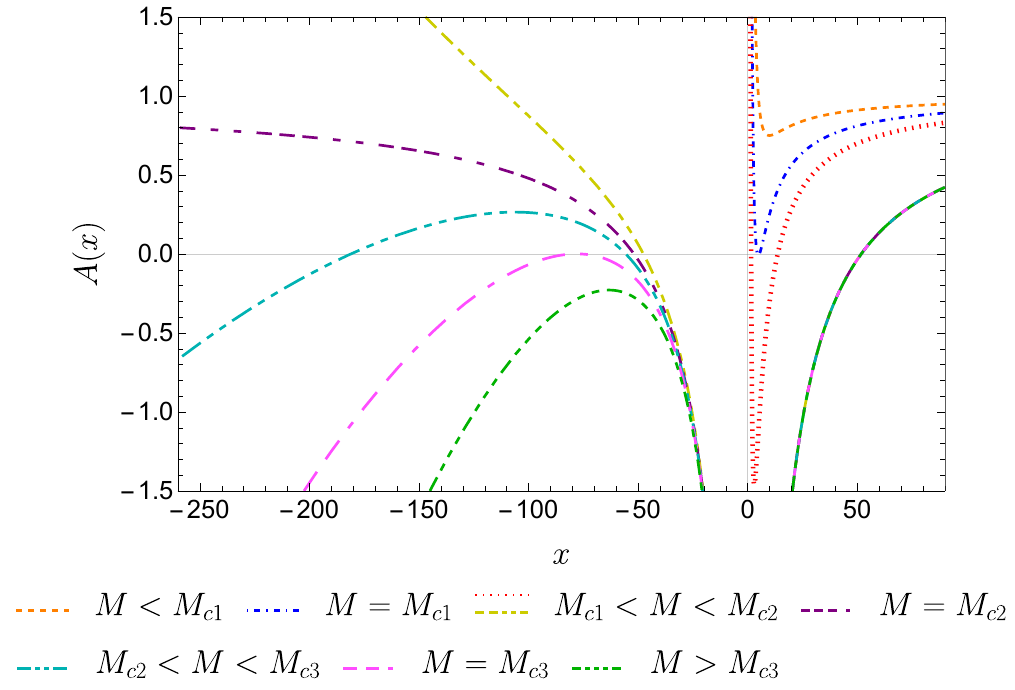}
    \hspace{0.13cm}
    \includegraphics[width=0.45\linewidth]{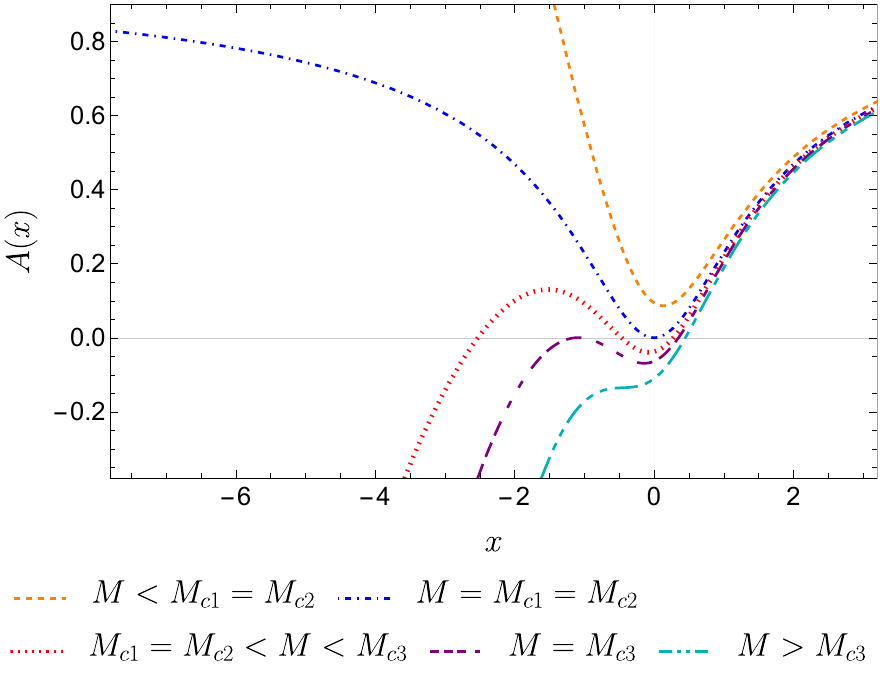}
    }
    \vspace{1em}
    \makebox[0.994\textwidth][r]{%
        \includegraphics[width=0.956\linewidth]{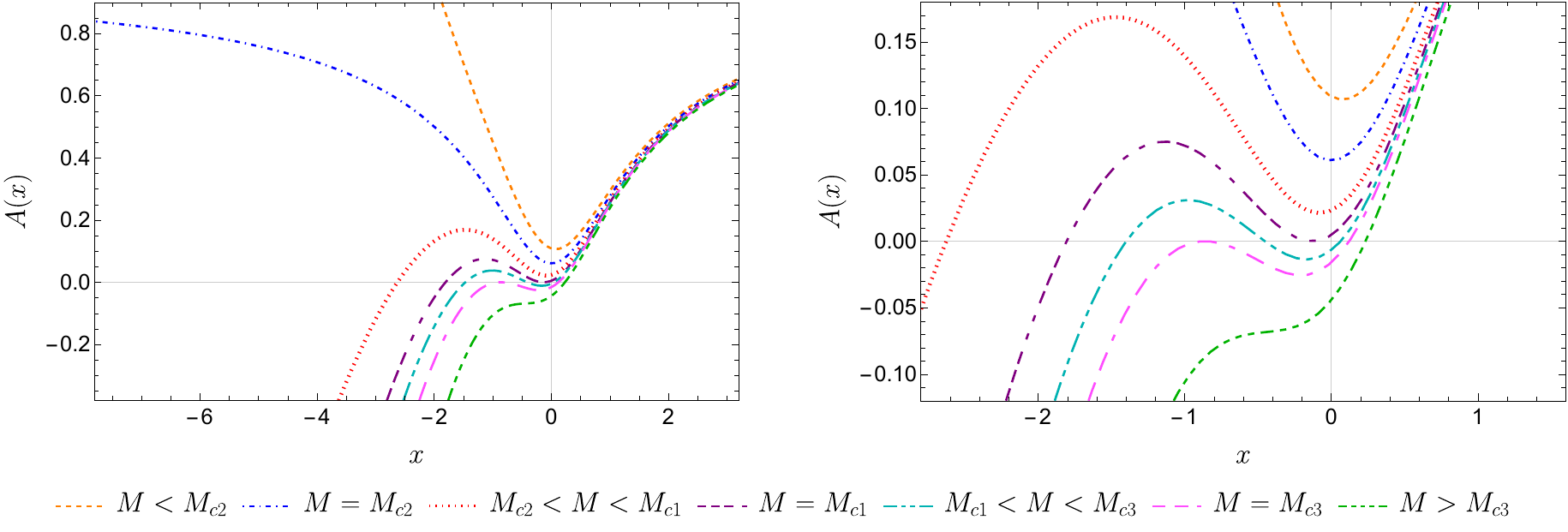}
    }
    \vspace{0.9em}
     \includegraphics[width=0.45\linewidth]{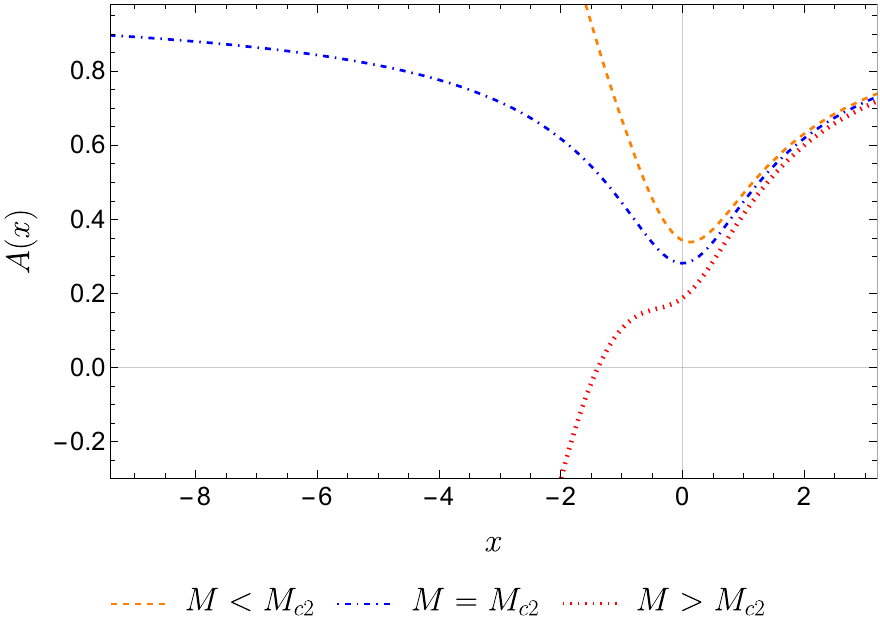}
    \caption{Plots of the function $A(x)$, given by Eq.~(\ref{A(x) ph}), obtained by fixing $|q|$. Top left plot: $|q|=5$, $M_{c1}\approx 4.98439$, $M_{c2}=25\pi/3$ and $M_{c3}\approx26.1799$. Seven different mass regimes, associated with these critical values, are shown, with the one corresponding to $M_{c1}<M<M_{c2}$ represented by two different curves to provide a clearer depiction of this range. The one in red (dotted) has $M=M_{c1}+2.8$, and the other one (yellow, dashed pattern) has $M=M_{c3}-1\times10^{-4}$. The cases $M<M_{c1}$, $M_{c2}<M< M_{c3}$ and $M>M_{c3}$ correspond, respectively, to $M=2.5$, $M=M_{c3}-3.5\times10^{-6}$ and $M=M_{c3}+5\times10^{-6}$. Top right plot: $|q|\approx0.825516$, $M_{c1}=M_{c2}=q^2\pi/3\,(\approx 0.713641)$ and $M_{c3}\approx0.727119$. Five different mass regimes are shown. The cases $M< M_{c1}=M_{c2}$, $M_{c1}=M_{c2}<M< M_{c3}$, and $M>M_{c3}$ correspond, respectively, to $M=M_{c2}-0.02$, $M=M_{c2}+0.008$ and $M=M_{c3}+0.01$. Middle row: The plot on the left serves to illustrate the asymptotic behaviour of $A(x)$, hence of $g_{00}$, as $x\to-\infty$, while the plot on the right provides a zoomed-in view of the zeros. Both consider $|q|=0.8$, $M_{c1}\approx0.68214$, $M_{c2}=0.64\pi/3$ and $M_{c3}\approx0.68653$. Seven different mass regimes are shown. The cases $M< M_{c2}$, $M_{c2}<M< M_{c1}$, $M_{c1}<M< M_{c3}$ and $M> M_{c3}$ correspond, respectively, to $M=0.66$, $M=M_{c2}+0.008$, $M=M_{c1}+0.002$ and $M=M_{c3}+0.006$. Bottom plot: $|q|=0.7$ and $M_{c2}=0.49\pi/3$. Three different mass regimes are shown. The cases $M<M_{c2}$ and $M>M_{c2}$ correspond, respectively, to $M=0.5$ and $M=M_{c2}+0.02$.}
    \label{criticalM}
\end{figure*}

First of all, Fig. \ref{criticalM} shows the behaviour of the term $A(x)$ for different mass regimes, taking into account the critical values of $M$. Each plot is associated with a different fixed value of $|q|$, except for the one on the right in the middle row, which is a zoom of the one on the left. Note that, as discussed before, when the conformal factor does not vanish, the asymptotic and general behaviours of $g_{00}$ are similar to those of $A(x)$.
Let us analyse the following cases:

\begin{itemize}
	\item
By fixing $|q|=5$, we obtain $M_{c1}\approx 4.98439$, $M_{c2}=25\pi/3$ and $M_{c3}\approx26.1799$. This case is represented by the left plot in the top row of Fig. \ref{criticalM}. There, seven different mass regimes are shown, with the one corresponding to $M_{c1}<M<M_{c2}$ represented by two different curves, with different values of $M$, to enhance clarity. By analysing this plot, assuming that all zeros are horizons (as will be done in the following analyses), we find that for $M<M_{c1}$, there are no horizons; for $M=M_{c1}$, there is an extremal event horizon, at $x_{eH}$; for $M_{c1}<M\leq M_{c2}$, there is an event horizon (EH), at $x_{EH}$, and a Cauchy horizon (CH), at $x_{CH}$, with $x_{CH}<x_{EH}$; for $M_{c2}<M< M_{c3}$, there is an EH, at $x_{EH}$, and two internal horizons (IH), at $x_{IH1}$ and $x_{IH2}$, with $x_{IH2}<x_{IH1}<x_{EH}$; for $M=M_{c3}$, there is an EH, at $x_{EH}$, and an extremal IH, at $x_{iH}$, with $x_{iH}<x_{EH}$; at last, for $M>M_{c3}$, there is an EH, at $x_{EH}$. 

\item
For the particular case of $|q|\approx0.825516$, obtained by requiring $M_{c1}=M_{c2}$, we have $M_{c1}=M_{c2}=q^2\pi/3\,(\approx 0.713641)$ and $M_{c3}\approx0.727119$. This case is represented by the right plot in the top row of Fig. \ref{criticalM}, where five different mass regimes are shown. By analysing this plot, we find that for $M< M_{c1}=M_{c2}$, there are no horizons; for $M=M_{c1}=M_{c2}$, there is an extremal EH, at $x_{eH}$; for $M_{c1}=M_{c2}<M< M_{c3}$, there is an EH, at $x_{EH}$, and two IHs, at $x_{IH1}$ and $x_{IH2}$, with $x_{IH2}<x_{IH1}<x_{EH}$. For the remaining values of $M$ ($M\geq M_{c3}$), we have the exact same regimes and corresponding horizon structure as in the previous example.

\item
By fixing $|q|=0.8$, we obtain $M_{c1}\approx0.68214$, $M_{c2}=0.64\pi/3$ and $M_{c3}\approx0.68653$. Note that, in this case, $M_{c2}<M_{c1}$, which, once again, leads to distinct results. This case is represented by the middle row of Fig. \ref{criticalM}, where seven different mass regimes are shown. The plot on the left is intended solely to illustrate the asymptotic behaviour of $A(x)$, and hence of $g_{00}$, as $x\to-\infty$, while the one on the right, as mentioned earlier, provides a zoomed-in view, making the zeros of $A(x)$ easier to identify. By analysing this plot, we find that for $M\leq M_{c2}$, there are no horizons; for $M_{c2}<M< M_{c1}$, there is an EH, at $x_{EH}$; for $M=M_{c1}$, there is an extremal EH, at $x_{eH}$, and an IH, at $x_{IH}$, with $x_{IH}<x_{eH}$; for $M_{c1}<M< M_{c3}$, there is an EH, at $x_{EH}$, and two IHs, at $x_{IH1}$ and $x_{IH2}$, with $x_{IH2}<x_{IH1}<x_{EH}$. Once again, for the remaining values of $M$ ($M\geq M_{c3}$), we have the exact same regimes and corresponding horizon structure as in the previous examples.

\item
Finally, by fixing $|q|=0.7$, we find that $M_{c1}$ and $M_{c3}$ do not exist, and obtain $M_{c2}=0.49\pi/3$. This case is represented by the bottom plot in Fig. \ref{criticalM}, where only three mass regimes are shown. By analysing this plot, we find that for $M\leq M_{c2}$, there are no horizons, whereas, for $M>M_{c2}$, there is an EH, at $x_{EH}$. 

\end{itemize}

These examples help illustrate the following findings: for $|q|\gtrsim 0.825516$, the relations shown in the first example are verified; for $|q|\approx 0.825516$ it is exactly what was shown in the second case; for $0.776\lesssim|q|\lesssim 0.825516$, the relations of the third case are verified; at last, for $|q|\lessapprox 0.776$, the relations of the fourth example are the ones that hold. Furthermore, note that any critical value of $M$ is positive, as can be seen in these examples, which is the reason why there may only be zeros of $A(x)$ for $M>0$, as aforementioned. This means that, considering the imposition $m>0$, any $M\leq 0$ always leads to a naked singularity solution.\\

Now, we can derive similar relations by fixing $M$. Figure \ref{criticalQ} shows the behaviour of the term $A(x)$, analogous to Fig. \ref{criticalM}, but for different electric charge regimes, taking into account the critical values of $q$. Each plot is associated with a different fixed value of $M$. In Fig. \ref{criticalQ}, we omit a zoomed-in view of the left plot in the bottom row because it is similar, apart from the differing interpretations of each curve, to the right plot in the middle row of Fig. \ref{criticalM}. Note that, as before, when the conformal factor does not vanish, the asymptotic and general behaviours of $g_{00}$ are similar to those of $A(x)$. Consider the following cases:

\begin{figure*}[th!]
    \centering
    \includegraphics[width=0.517\linewidth]{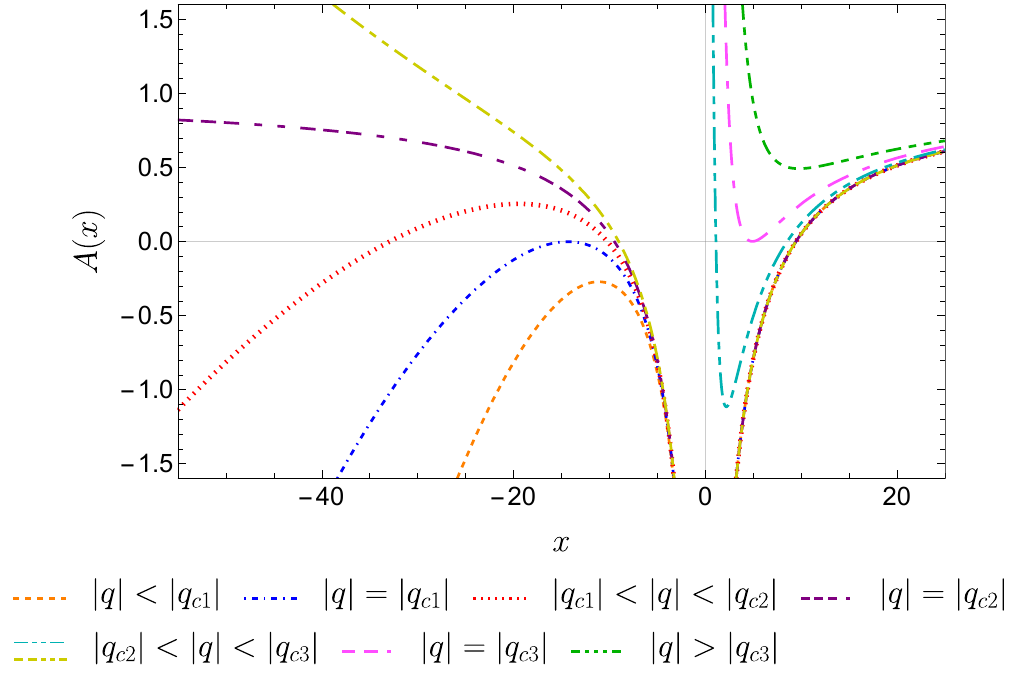}
    \includegraphics[width=0.463\linewidth]{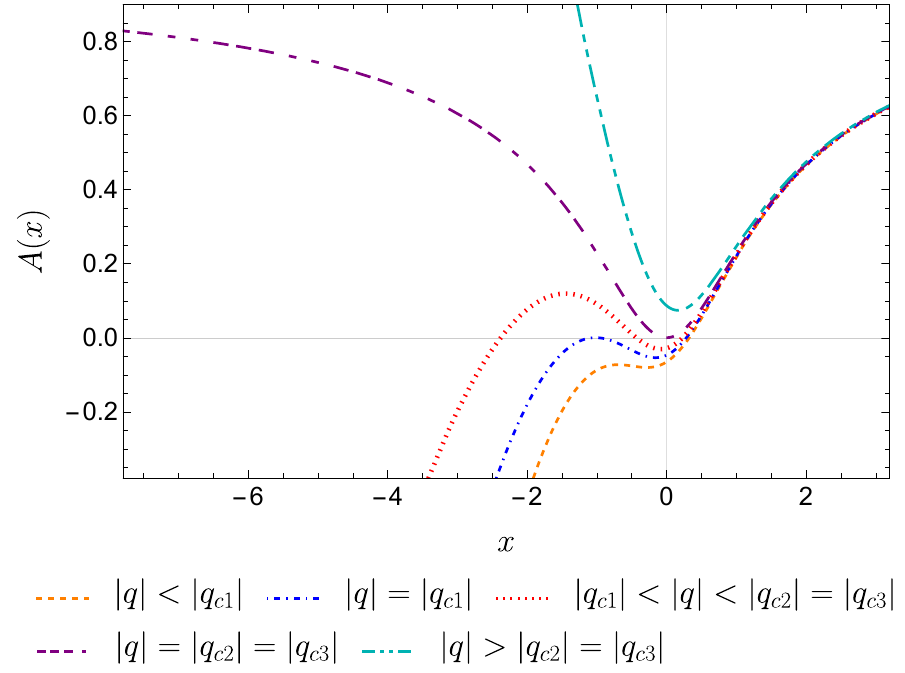}
    \vspace{1em} 
        \makebox[0.985\textwidth][r]{%
    \includegraphics[width=0.521\linewidth]{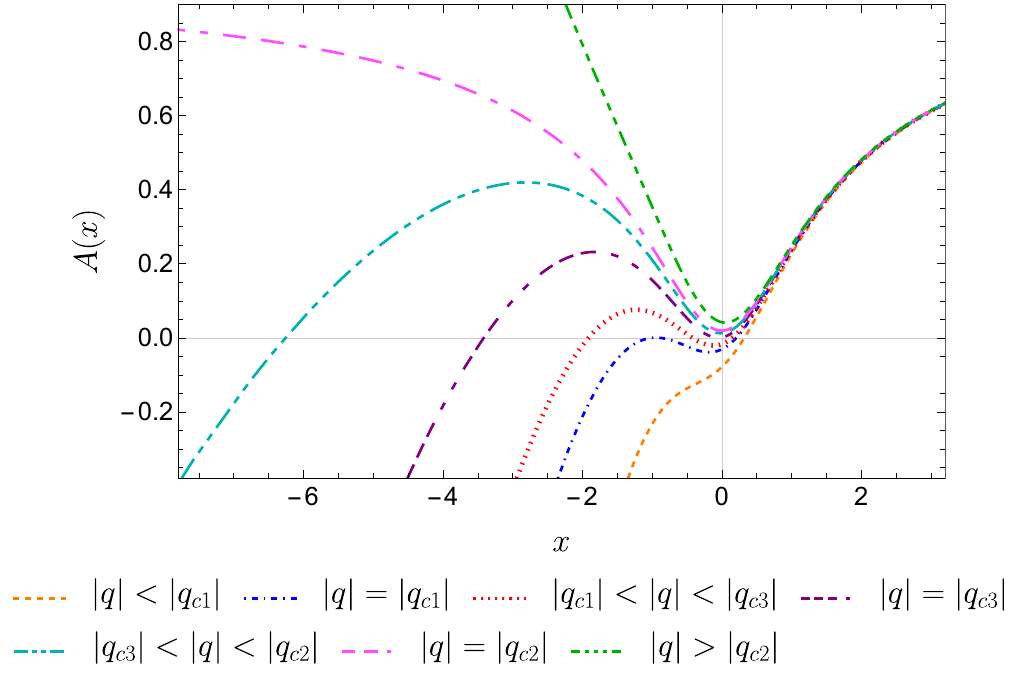}
	\hspace{0.01cm}
        \includegraphics[width=0.45\linewidth]{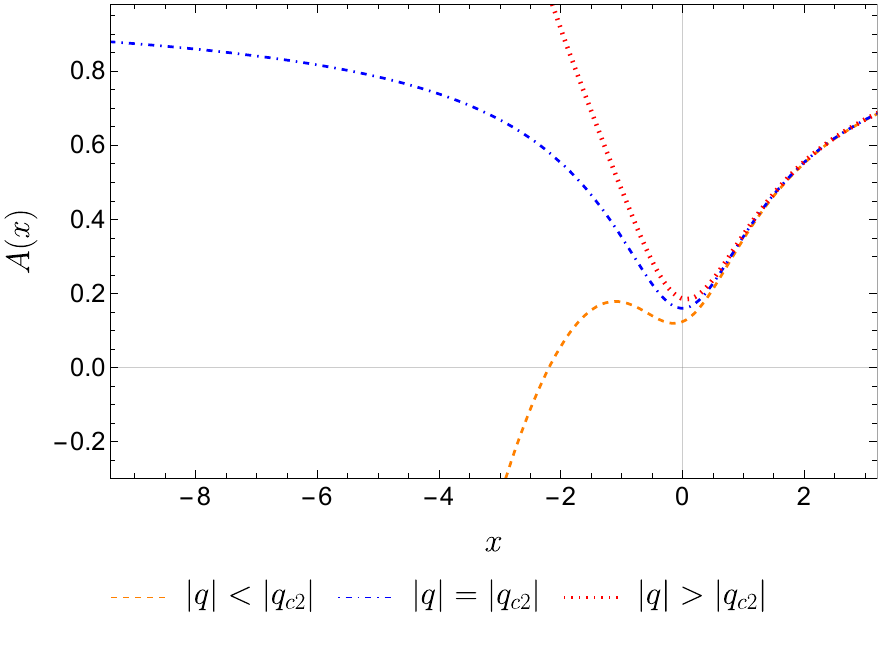}
    }
    \caption{Plots of the function $A(x)$, given by Eq.~(\ref{A(x) ph}), obtained by fixing $M$. Top left plot: $M=5$, $|q_{c1}|\approx2.18506$, $|q_{c2}|=\sqrt{15/\pi}$ and $|q_{c3}|\approx5.01556$. Seven different charge regimes, associated with these critical values, are shown, with the one corresponding to $|q_{c2}|<|q|<|q_{c3}|$ represented by two different curves to provide a clearer depiction of this range. The one that is entirely in the positive $x$-axis (cyan, dashed pattern) has $|q|=|q_{c2}|+1.3$, and the other one (yellow, dashed pattern) has $|q|=|q_{c2}|+1.3\times10^{-5}$. The cases $|q|<|q_{c1}|$, $|q_{c1}|<|q|< |q_{c2}|$ and $|q|>|q_{c3}|$ correspond, respectively, to $|q|=|q_{c1}|-4\times10^{-5}$, $|q|=|q_{c2}|-1.5\times10^{-5}$ and $|q|=|q_{c3}|+2$. 
    Top right plot: $M\approx0.713641$, $|q_{c1}|\approx0.817198$ and $|q_{c2}|=|q_{c3}|=\sqrt{3M/\pi}\,(\approx0.825516)$. Five different charge regimes are shown. The cases $|q|< |q_{c1}|$, $|q_{c1}|<|q|<|q_{c2}|=|q_{c3}|$, and $|q|>|q_{c2}|=|q_{c3}|$ correspond, respectively, to $|q|=0.814$, $|q|=|q_{c1}|+0.0033$ and $|q|=|q_{c2}|+0.015$. 
    Bottom left plot: $M=0.7$, $|q_{c1}|\approx0.80863$, $|q_{c2}|=\sqrt{2.1/\pi}$ and $|q_{c3}|\approx0.814359$. Seven different charge regimes are shown. The cases $|q|< |q_{c1}|$, $|q_{c1}|<|q|< |q_{c3}|$, $|q_{c2}|<|q|< |q_{c2}|$ and $|q|> |q_{c2}|$ correspond, respectively, to $|q|=0.8$, $|q|=|q_{c1}|+0.0025$, $|q|=|q_{c3}|+0.002$ and $|q|=|q_{c2}|+0.004$. 
    Bottom plot: $M=0.6$ and $|q_{c2}|=\sqrt{1.8/\pi}$. Three different charge regimes are shown. The cases $|q|<|q_{c2}|$ and $|q|>|q_{c2}|$ correspond, respectively, to $|q|=0.75$ and $|q|=|q_{c2}|+0.005$.}
    \label{criticalQ}
\end{figure*}

\begin{itemize}
	\item
Starting by fixing $M=5$, we obtain $|q_{c1}|\approx2.18506$, $|q_{c2}|=\sqrt{15/\pi}$ and $|q_{c3}|\approx5.01556$. This case is represented by the left plot in the top row of Fig. \ref{criticalQ}. There, seven different charge regimes are shown, with the one corresponding to $|q_{c2}|<|q|<|q_{c3}|$ represented by two different curves, with different values of $|q|$, to enhance clarity. By analysing this plot we find that for $|q|>|q_{c3}|$, there are no horizons; for $|q|=|q_{c3}|$, there is an extremal EH, at $x_{eH}$; for $|q_{c2}|\leq |q|< |q_{c3}|$, there is an EH, at $x_{EH}$, and a CH, at $x_{CH}$, with $x_{CH}<x_{EH}$; for $|q_{c1}|<|q|< |q_{c2}|$, there is an EH, at $x_{EH}$, and two IHs, at $x_{IH1}$ and $x_{IH2}$, with $x_{IH2}<x_{IH1}<x_{EH}$; for $|q|=|q_{c1}|$, there is an EH, at $x_{EH}$, and an extremal IH, at $x_{iH}$, with $x_{iH}<x_{EH}$; at last, for $|q|<|q_{c1}|$, there is an EH, at $x_{EH}$. 

\item
In the particular case $M\approx0.713641$, obtained by requiring $|q_{c2}|=|q_{c3}|$ (analogous to the previous particular case of $|q|$), we obtain $|q_{c1}|\approx0.817198$ and $|q_{c2}|=|q_{c3}|=\sqrt{3M/\pi}\,(\approx0.825516)$. This case is represented by the right plot in the top row of Fig. \ref{criticalQ}, where five different charge regimes are shown. For $|q|> |q_{c2}|=|q_{c3}|$, we find that no horizons exist; for $|q|=|q_{c2}|=|q_{c3}|$, there is an extremal EH, at $x_{eH}$; for $|q_{c1}|<|q|< |q_{c2}|=|q_{c3}|$, there is an EH, at $x_{EH}$, and two IHs, at $x_{IH1}$ and $x_{IH2}$, with $x_{IH2}<x_{IH1}<x_{EH}$. For the remaining values of $|q|$ ($|q|\leq |q_{c1}|$), we have the exact same regimes and corresponding horizon structure as in the previous example.

\item
By fixing $M=0.7$, we obtain $|q_{c1}|\approx0.80863$, $|q_{c2}|=\sqrt{2.1/\pi}$ and $|q_{c3}|\approx0.814359$. Note that, in this case, $|q_{c2}|>|q_{c3}|$, which, as before, lead to distinct results. This case is represented by the left plot in the bottom row of Fig. \ref{criticalQ}, where seven different charge regimes are shown. For $|q|\geq |q_{c2}|$, there are no horizons; for $|q_{c3}|<q< |q_{c2}|$, there is an EH, at $x_{EH}$; for $|q|=|q_{c3}|$, there is an extremal EH, at $x_{eH}$, and an IH, at $x_{IH}$, with $x_{IH}<x_{eH}$; for $|q_{c1}|<q< |q_{c3}|$, there is an EH, at $x_{EH}$, and two IHs, at $x_{IH1}$ and $x_{IH2}$, with $x_{IH2}<x_{IH1}<x_{EH}$. Once again, for the remaining values of $|q|$ ($|q|\leq |q_{c1}|$), we have the exact same regimes and corresponding horizon structure as in the previous examples.

\item
At last, by fixing $M=0.6$, we find that $|q_{c1}|$ and $|q_{c3}|$ do not exist, and obtain $|q_{c2}|=\sqrt{1.8/\pi}$. This case is represented by the right plot in the bottom row of Fig. \ref{criticalQ}, where three different charge regimes are shown. For $|q|\geq |q_{c2}|$, we find that no horizons exist, whereas for $|q|<|q_{c2}|$, there is an EH, at $x_{EH}$.

\end{itemize}

Once again, these examples help illustrate the following: for $M\gtrsim 0.713641$, the relations that are verified are the ones shown in the first example; for $M\approx 0.713641$ it is as shown in the second case; for $0.652\lesssim M\lesssim 0.713641$, the relations in the third case are the ones verified; at last, for $0<M\lessapprox 0.652$, the relations in the fourth example are the ones satisfied. Additionally, for $M\leq 0$, no critical values of $q$ exist, nor do zeros of $A(x)$, as aforementioned.

Furthermore, from all possible scenarios, we also find that when there is 1 horizon, it can either be an EH or an extremal EH; when there are 2 horizons, the configuration can be an EH and a CH, or an extremal EH and an IH, or an EH and an extremal IH; when there are 3 horizons, the only possible configuration is an EH and two IHs. Note that this is verified in all cases, independently of the location of the singularity.

Now, in general for any adequate values of the constants, we will present the solutions that arise based on the presence of either a singularity or a minimum of the radius function, the number of horizons, and the relative positions between these spacetime structures.
That being said, we will now start with the cases in which there is a singularity. 

Whenever there are no zeros of $A(x)$, such cases correspond to naked singularity solutions. In these cases, at $x_s$, we find that $g_{00}\to 0^+$, which can be deduced from Figs. \ref{criticalM} and \ref{criticalQ} -- bearing in mind that the conformal factor is always non-negative -- since $A(x)>0$ is verified. Thus, as discussed before, at this point, which is $x_\text{min}$, there is a light-like, naked, attractive central singularity. This occurs, for example, for $|q|\gtrsim 0.825516$ and $M<M_{c1}$, as long as $\psi_0$ is such that leads to a singularity, being allowed if it also ensures $m>0$. As aforementioned, as long as $M \gtrapprox 0.14$, any $\psi_0$ that leads to a singularity is allowed, and so, the requirement $m>0$ will no longer be mentioned in the following cases. 

\begin{figure}[ht!]
	\centering
	\includegraphics[scale=0.725]{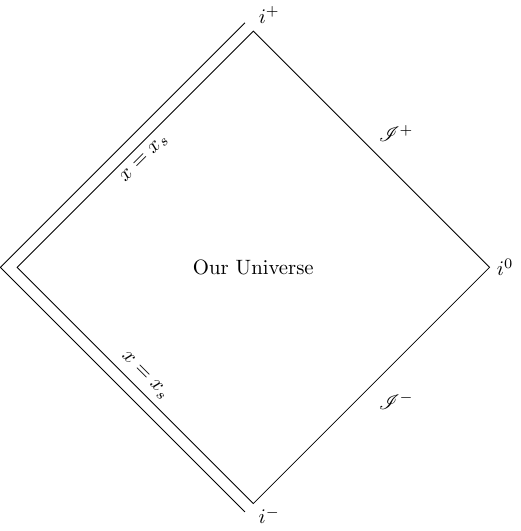}
	\caption{Penrose diagram for a naked, central, light-like singularity solution. On the right there is the asymptotically flat infinity, and on the left the upper and lower diagonal double lines at $x=x_s$ depict the future and past branches of the light-like singularity, respectively.}
	\label{PenrosePhant10}
\end{figure}

The Penrose diagram for this solution, shown in Fig. \ref{PenrosePhant10}, is similar to that of the left plot of Fig. \ref{PenroseCan}, but considering a light-like singularity, instead of a time-like one.

\begin{figure*}[ht!]
	\centering
	\includegraphics[scale=0.48]{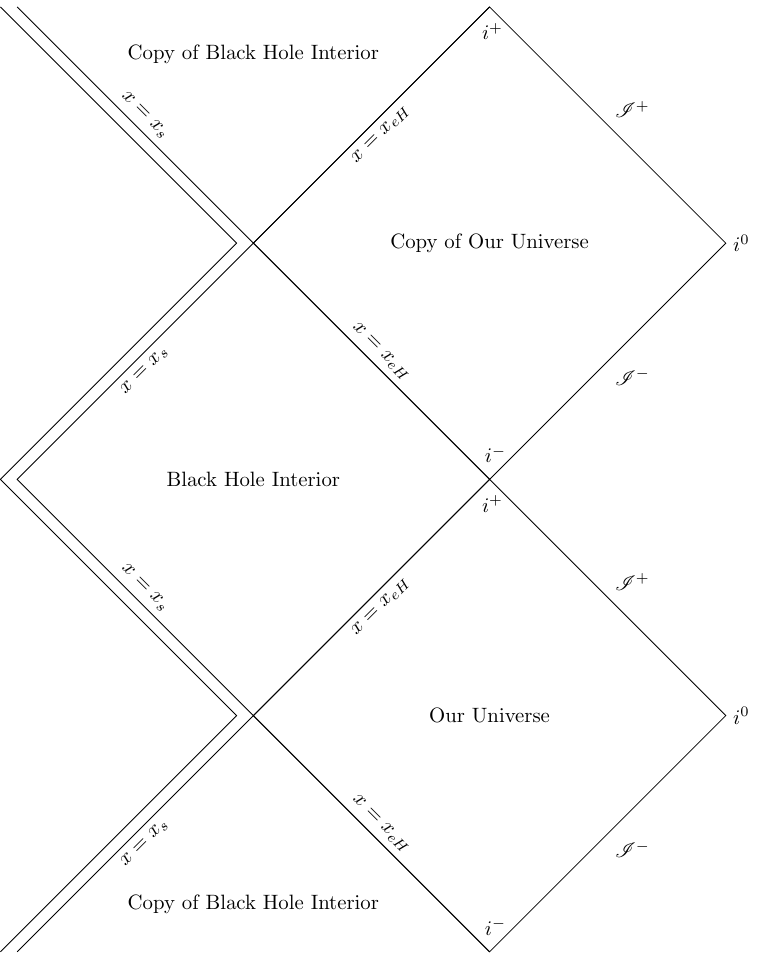}
	\hspace{1.1cm}
	\includegraphics[scale=0.48]{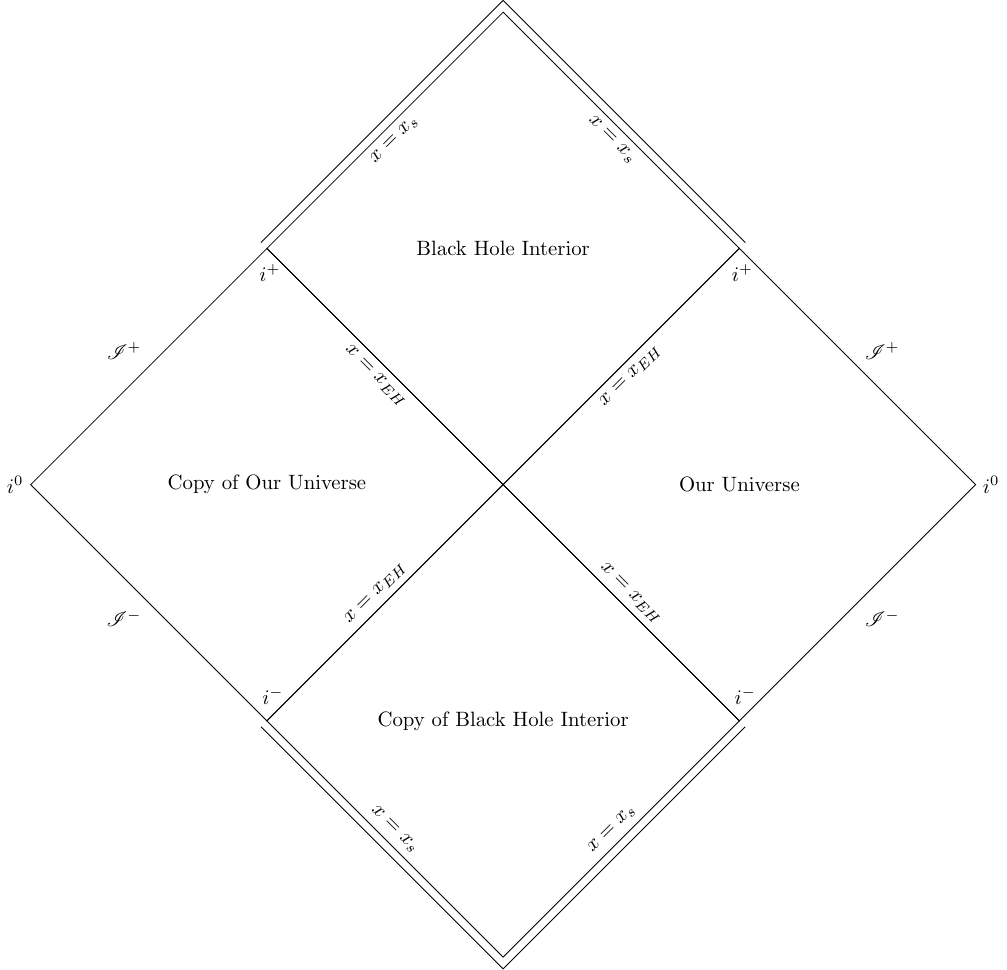}
	\caption{Left plot: Penrose diagram for a black hole solution with an extremal event horizon and a central, light-like singularity. In ``Our Universe", the infinity lies on the right, while on the left, at $x=x_{eH}$, there are future and past branches of the extremal event horizon. Within them there are future and past branches of this horizon, on the right, and of the light-like singularity, on the left. There are also copies of each region that can be accessed by traversing certain horizon branches. Moreover, the entire diagram actually extends infinitely upwards and downwards, from both regions labelled ``Copy of Black Hole Interior", repeating the same structure depicted here. Right plot: Penrose diagram for a black hole solution with an event horizon and a central, light-like singularity. Now, in ``Our Universe", at $x=x_{EH}$, there are future and past branches of the event horizon. Within the future one, there are two past branches of this horizon and two future branches of the light-like singularity. Moreover, there is a copy of this interior, as well as a ``Copy of Our Universe", the latter of which is connected to the original by an Einstein-Rosen bridge.}
	\label{PenrosePhant1}
\end{figure*}

When the term $A(x)$ has between 1 and 3 zeros, each of them is a horizon only if the singularity occurs after it. Consequently, in a given combination of $M$ and $|q|$, all of the zeros -- whether 1, 2, or 3 -- may correspond to horizons, or only some of them, or even none at all. As discussed earlier, the location of the singularity depends only on $\psi_0$ and is given by $x=\tan[\sqrt{3}  (\pi/2 - \psi_0 + c_1\pi)]$. Then, to determine whether $x_s$ occurs before, after or at a certain zero, given the value of $\psi_0$, we can simply use Eq. (\ref{psi0-xs}) (see the discussion following that equation).

In the case where none of the zeros corresponds to a horizon, we have $x_s\geq x_{H\text{max}}$, where $x_{H\text{max}}$ is the location of the outermost zero. From Eq. (\ref{psi0-xs}), we find that this occurs if $\psi_0\leq\frac{\pi}{2}-\arctan(x_{H\text{max}})/\sqrt{3}+c_1\pi$. Whenever this happens, it corresponds to a naked singularity solution, in full similarity to the one discussed above, that exists when $A(x)$ has no zeros. In this case, this can be verified for all combinations of $M$ and $|q|$, as long as $\psi_0$ has the adequate values. The Penrose diagram is the same as before.

Now, considering that the singularity occurs at least after one of the zeros, such that $x_s < x_{H\text{max}}$, being $\psi_0 >\frac{\pi}{2}-\arctan(x_{H\text{max}})/\sqrt{3}+c_1\pi$, such cases always correspond to black hole solutions. The particular spacetime geometry, hence the corresponding type of black hole, which is related to the number and type of horizons, and also the type of singularity, depends on where $x_s$ is located relative to the zeros of $A(x)$.

Figures \ref{criticalM} and \ref{criticalQ} can be used, when a singularity exists, to analyse its nature. A singularity that occurs at a region where $A(x)>0$ is attractive, since, as $x\to x_s^+$, we have $g_{00}\to 0^+$. 
Conversely, when $A(x)<0$, the singularity is repulsive, because, at that same limit, $g_{00}\to 0^-$. 
If the singularity occurs at a horizon ($A(x)=0$), its nature depends on the sign of $g_{00}$, in particular of $A(x)$, in the region outside of it.  

In the remaining discussion, when we mention the interior of a given horizon, and there are more at lower values of $x$, we are referring to the spacetime between that and the next one (at lower $x$). Whereas, if that is the last or only one, we are referring to the remaining spacetime until the singularity (or second spatial infinity, as will be seen later). A similar reasoning applies when we talk about the outside of a certain horizon.

That being said, based on the previous analysis of the horizon structure (see Figs. \ref{criticalM} and \ref{criticalQ}), there is a light-like, attractive central singularity, whenever $x_s$ is located inside an extremal EH, or a CH, or the outermost IH (relative to $x\to\infty$) when three horizons exist (in which case this horizon is actually reclassified as a CH). On the other hand, there is a light-like, repulsive central singularity, whenever $x_s$ lies inside an EH, or an extremal IH 
, or an IH located inside an extremal EH, or the innermost IH when three horizons exist. 

For instance, consider the case where the singularity is located beyond all the $A(x)$ zeros, that is, when $x_s<x_{H\text{min}}$. In this situation, it is of the former type for all cases in which $M\leq M_{c2}$, or $|q|\geq |q_{c2}|$, and of the latter type in all the cases in which $M>M_{c2}$, or $|q| < |q_{c2}|$.

In addition to the singularity classification for each scenario, the different possible horizon configurations, given the number of horizons that exist, have already been discussed following the analysis of the critical values. There are black hole solutions with 1, 2 or 3 horizons, and we now present an example of each possible configuration.

Solutions with one horizon occur, for example, for $|q|\approx 0.825516$ and $M=M_{c1}=M_{c2}$, and also $M_{c1}=M_{c2}<M<M_{c3}$, as long as $\psi_0 >\frac{\pi}{2}-\arctan(x_{eH})/\sqrt{3}+c_1\pi$ and $\frac{\pi}{2}-\arctan(x_{EH})/\sqrt{3}+c_1\pi<\psi_0 <\frac{\pi}{2}-\arctan(x_{CH})/\sqrt{3}+c_1\pi$, respectively. In the former case, it is an extremal EH and the singularity is attractive, and in the latter, it is an EH (we are considering only the first horizon for this solution) and the singularity is repulsive. 

The Penrose diagram for the former case, shown in the left plot of Fig. \ref{PenrosePhant1}, is similar to that in the middle of Fig. \ref{PenroseCan}, but considering a light-like singularity instead. On the other hand, the diagram of the latter case, even if it also has one horizon, is shown in the right plot of Fig. \ref{PenrosePhant1}, which is similar to that of the Schwarzchild solution, but with a light-like singularity, instead of a space-like one.

The difference between these two diagrams arises due to the type of the horizons. In fact, an extremal EH does not change the metric signature, and thus, in the left plot, in the region labelled ``Black Hole Interior", the light-like singularity appears on the left, featuring both future and past branches, and the horizon appears on the right, also featuring future and past branches (once one enters this interior, one can either approach the singularity or traverse the horizon into a future copy of our universe). Furthermore, the entire diagram of this solution actually extends infinitely upwards and downwards, from both regions labelled ``Copy of Black Hole Interior", repeating the same structure depicted in this portion. In fact, this applies to any diagram in this work that presents truncated regions, with the infinite extension oriented in the direction of those truncations (it can also extend sideways or in all directions).

Conversely, an EH changes the metric signature, in this case to $(-+- ~-)$, and so, in the right plot, in the ``Black Hole Interior", the singularity appears upwards, with both branches lying in the future (once one enters this horizon, falling into the singularity becomes inevitable), and the horizon appears downwards, with both branches in the past. A feature that appears in this diagram is the Einstein-Rosen bridge, connecting ``Our Universe" and its copy (to the left). 

Note that, in the former diagram, all the regions that are copies retain the original orientation, whereas, in the latter, they are reversed.

Black holes with two horizons exist, for example, for $|q|\gtrsim 0.825516$ and $M_{c1}<M\leq M_{c2}$, for $|q|\gtrsim 0.825516$ and $M= M_{c3}$, and also for $0.776\lesssim|q|\lesssim 0.825516$ and $M=M_{c1}$, given that $\psi_0 >\frac{\pi}{2}-\arctan(x_{CH})/\sqrt{3}+c_1\pi$, $\psi_0 >\frac{\pi}{2}-\arctan(x_{iH})/\sqrt{3}+c_1\pi$ and $\psi_0 >\frac{\pi}{2}-\arctan(x_{IH})/\sqrt{3}+c_1\pi$, respectively. 
In the first case, there are an EH and a CH, and the singularity is attractive. In the second case, there are an EH and an extremal IH, being the singularity repulsive. In the last case, there are an extremal EH and an IH, and the singularity is also repulsive.

Even if all three cases present two horizons, their respective Penrose diagrams are different from each other, similar to when there is only one horizon, which is due to the different horizon structures. As seen before, any simple horizon changes the metric signature, whereas any extremal horizon does not. What was explained before, regarding the different positions of the branches of the light-like singularity in the two diagrams of Fig. \ref{PenrosePhant1}, which depended on whether the event horizon it was in was simple or extremal, actually applies to the branches of any spacetime structure and depends solely on the metric signature of the region they are in. In fact, in the diamond of any region with metric signature $(-+-\,-)$, the branches of any internal horizon or singularity (or second spatial infinity, as will be seen later) that follows that region (lower $x$) appear upwards, both lying in the future, and the branches that are related to that region appear downwards, in the past; whereas, in the diamond of any region with metric signature $(+--~-)$, the branches of those structures appear to the left, lying in the future and past, and the branches linked to that region appear to the right, also in the future and past. Note that this does not apply directly to all of their copies, since some of them retain the original orientation, but others are reversed, depending on their position relative to the original. 

\begin{figure*}[ht!]
	\centering
	\includegraphics[scale=0.47]{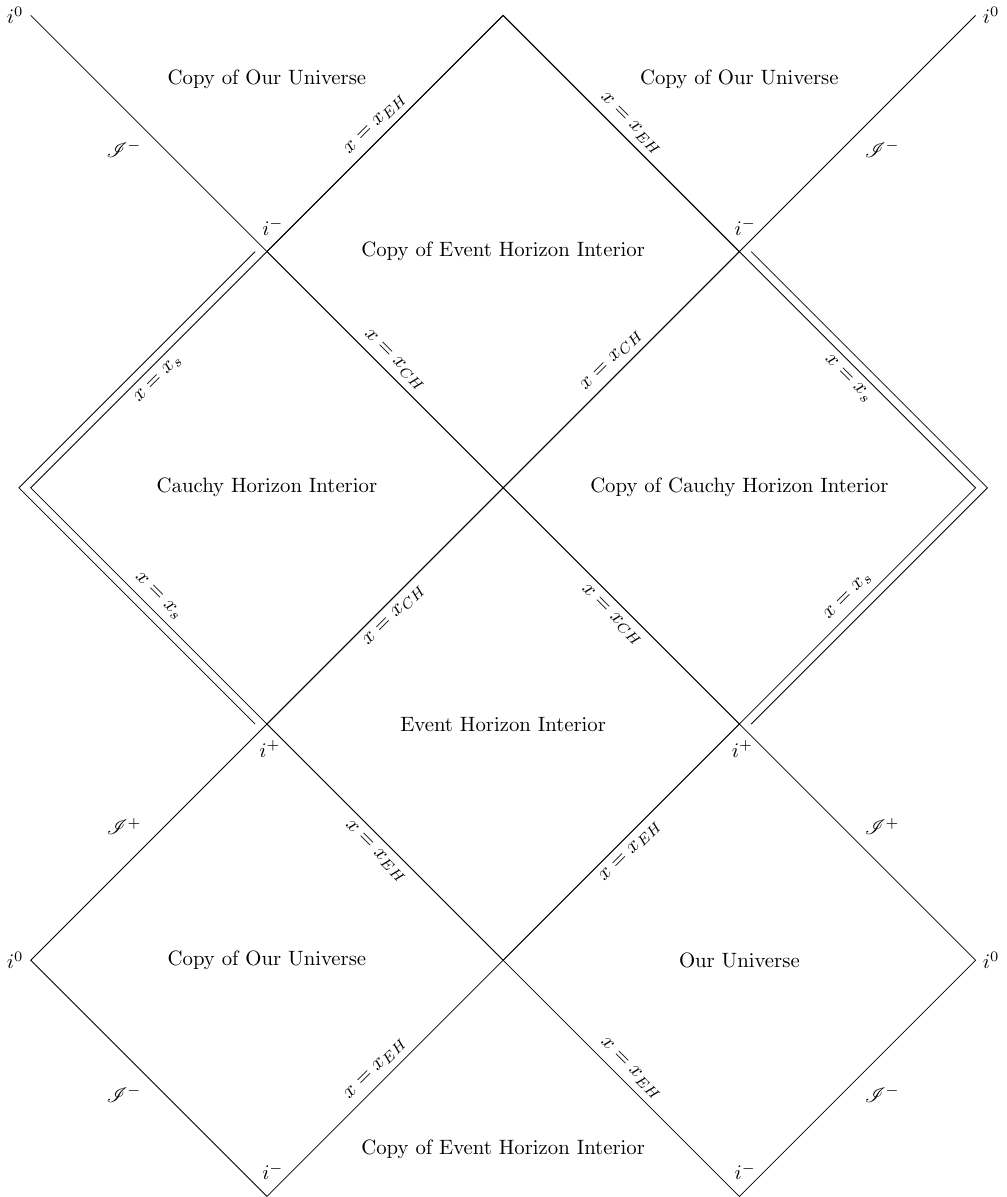}
	\hspace{1.25cm}
	\includegraphics[scale=0.412]{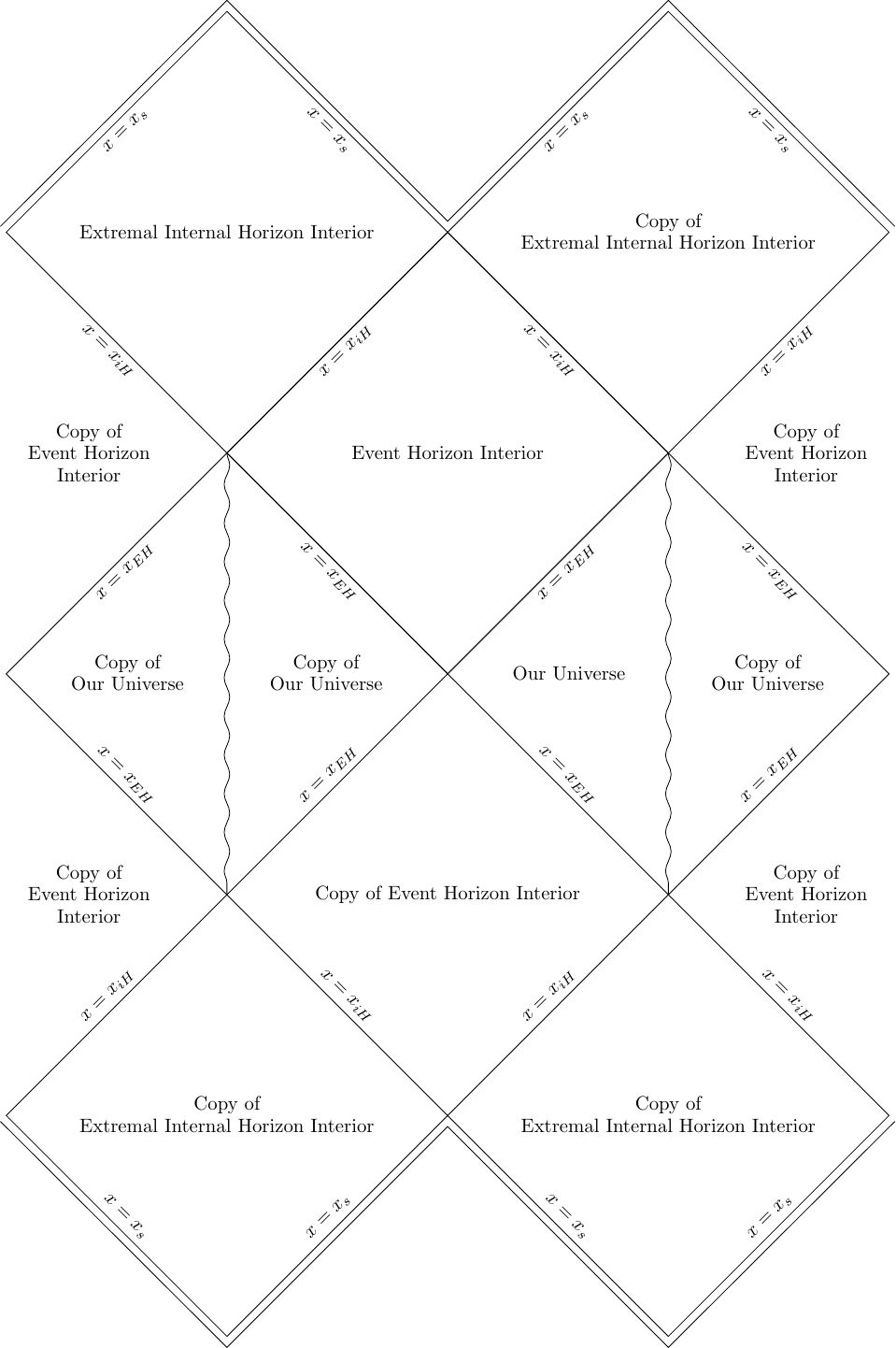}
	\caption{Left plot: Penrose diagram for a black hole solution with an event horizon, a Cauchy horizon and a central, light-like singularity.  
		Right Plot: Penrose diagram for a black hole solution with an event horizon, an extremal internal horizon and a central, light-like singularity. We refer the reader to the text for more details.}
	\label{PenrosePhant2}
\end{figure*}

\begin{figure*}[ht!]
   \centering
      \includegraphics[scale=0.58]{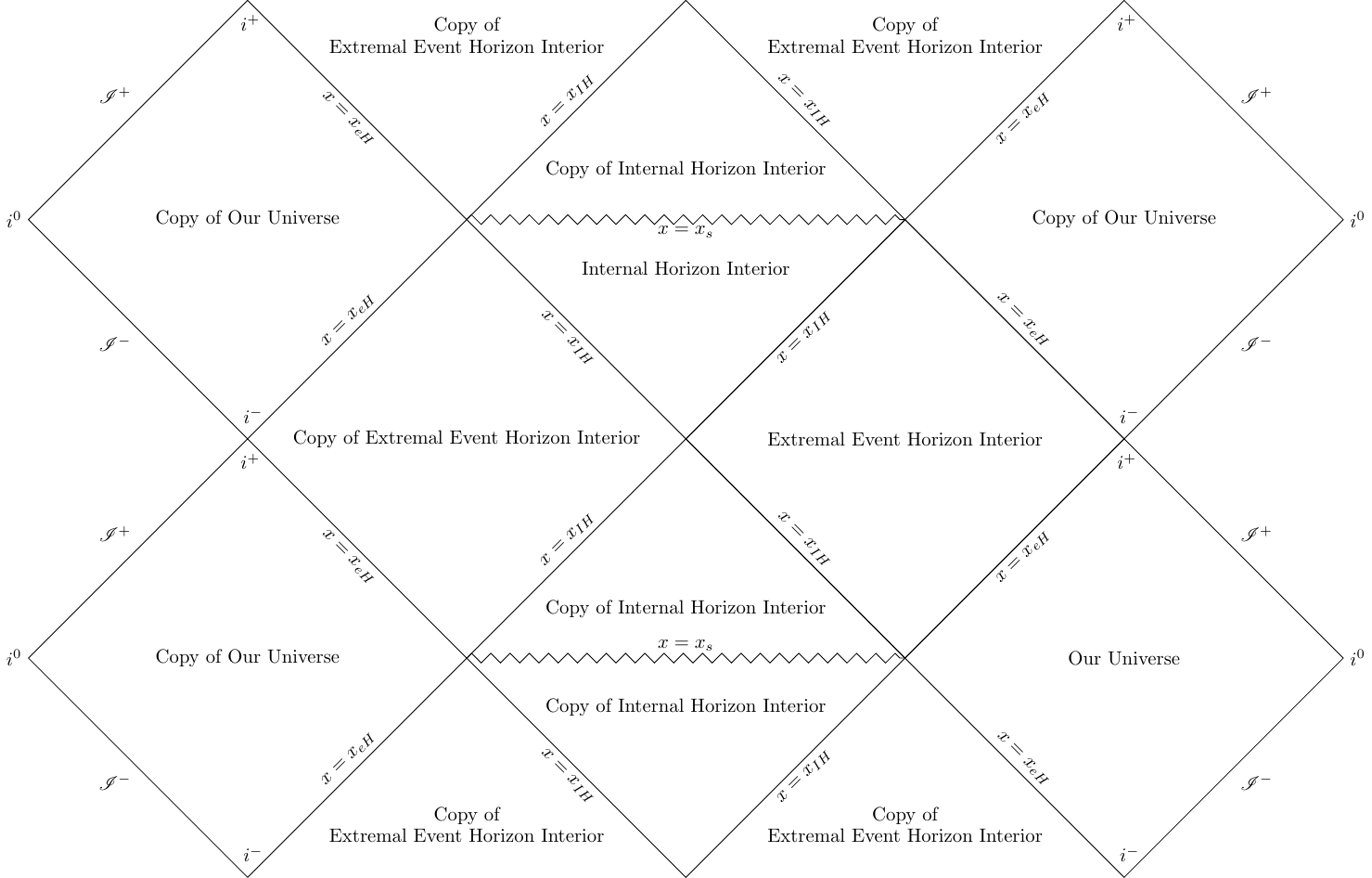}
    \caption{Penrose diagram for a black hole solution with an extremal event horizon, an internal horizon and a central, light-like singularity.  We refer the reader to the text for more details.}
    \label{PenrosePhant2.1}
\end{figure*}

This being said, we are able to construct the three diagrams.
They are shown, respectively, from the first case to the third, in the left and right plots of Fig. \ref{PenrosePhant2} and in Fig. \ref{PenrosePhant2.1}. The first plot is similar to that in the right plot of Fig. \ref{PenroseCan}, but considering a light-like singularity instead of a time-like one. In the second and third plots there are regions, not directly related in any way, that overlap due to the way the diagram is constructed, namely ``Our Universe" and a copy, and ``Internal Horizon Interior" and a copy, respectively. To show and distinguish the two different regions, we used wavy and zigzag lines, respectively, to separate them. Note that these lines do not represent any type of finite spacetime structure and cannot be traversed. \\

More specifically, the left plot of Fig. \ref{PenrosePhant2} depicts a Penrose diagram for a black hole solution with an event horizon, a Cauchy horizon and a central, light-like singularity. 

\begin{itemize}

\item In ``Our Universe" the asymptotically flat infinity lies on the right, while on the left, at $x_{EH}$, there are future and past branches of the event horizon.

\item Within the future one, there are two past branches of this horizon, and two future branches of the Cauchy horizon, at $x=x_{CH}$. 

\item Within the left one of these, there are future and past branches of this horizon, on the right, and of the light-like singularity, on the left. 

\item Moreover, copies of each region exist, accessed by traversing certain horizon branches. 
The ``Copy of Our Universe" lying to the left of the original is connected to it by an Einstein-Rosen bridge. 

\end{itemize}

The right plot of Fig. \ref{PenrosePhant2} depicts a Penrose diagram for a black hole solution with an event horizon, an extremal internal horizon and a central, light-like singularity. 

\begin{itemize}

\item 
Now, unlike before, within the future branch of the event horizon that lies in ``Our Universe" there are two future branches of the extremal internal horizon, at $x=x_{iH}$. 

\item 
Within the left one, in the ``Extremal Internal Horizon Interior", there are two past branches of this horizon and two future branches of the light-like singularity. 

\item 
By traversing the past horizon branch on the left, there is a ``Copy of Event Horizon Interior". From here, by traversing the past branch of the event horizon, we can reach a ``Copy of Our Universe", which, however, is not the same as the one that lies immediately to the left of ``Our Universe" (connected to it by an Einstein-Rosen bridge). 

\item 
These regions are not directly related in any way, but, due to how the diagram is constructed, the respective diamonds completely overlap. Thus, to show and distinguish the two different regions, we drew a wavy line separating them, which does not represent any type of finite spacetime structure and cannot be traversed. 

\item 
A similar situation occurs in the diamond where ``Our Universe" lies. Each region’s asymptotically flat infinity is represented by two diagonal lines; however, as they overlap with the horizon branches of the adjacent region, they are not shown explicitly in this diagram. Moreover, as before, more copies of each region exist.

\end{itemize}

Figure \ref{PenrosePhant2.1} depicts a Penrose diagram for a black hole solution with an extremal event horizon, an internal horizon and a central, light-like singularity.

\begin{itemize}

\item In ``Our Universe" the asymptotically flat infinity lies on the right, and future and past branches of the extremal event horizon lie on the left. 

\item Within the future one, there are future and past branches of this horizon, on the right, and of the internal horizon, on the left, at $x=x_{IH}$. 

\item Within the future one of these, in the ``Internal Horizon Interior", there are two past branches of this horizon, and two future branches of the light-like singularity, which are not explicitly represented. This interior is separated from the ``Copy of Internal Horizon Interior", which lies above it, by a zigzag line; however, they are neither the same nor are they directly related in any way. The use of this line and the relation between these two regions are similar to the wavy line and the relation between the two ``Copy of Our Universe'' in Fig. \ref{PenrosePhant2}: the diamonds of the two regions completely overlap because of the way the diagram is constructed, so, in order to show and distinguish both of them, we drew the zigzag line.

\item Similar to the infinities in that figure, now, the two branches of the light-like singularity overlap with the horizon branches of the adjacent region, thus, they are not explicitly represented in this diagram. Moreover, as before, more copies of each region exist, and another zigzag line, with the same meaning, appears in the region following the past branch of the internal horizon aforementioned.

\end{itemize}

At last, black hole solutions with three horizons occur, for example, for $|q|\gtrsim 0.825516$ and $M_{c2}<M< M_{c3}$, as long as $\psi_0 >\frac{\pi}{2}-\arctan(x_{IH2})/\sqrt{3}+c_1\pi$. In this case, as in all cases in which three horizons exist, there is an EH and two IHs, and the singularity is repulsive. 

\begin{figure*}[ht!]
   \centering
      \includegraphics[scale=0.5]{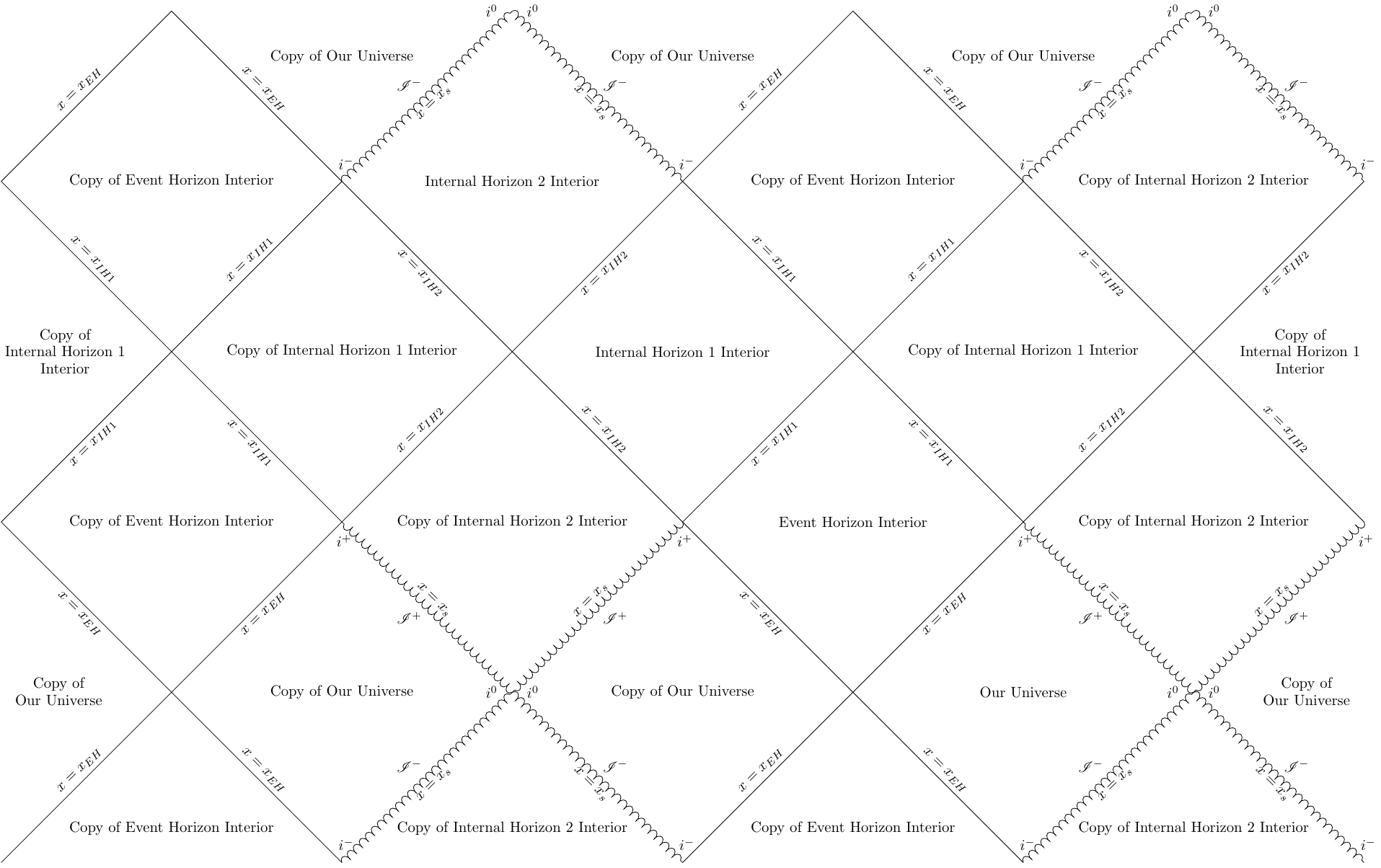}
    \caption{Penrose diagram for a black hole solution with an event horizon, two internal horizons and a central, light-like singularity. We refer the reader to the text for more details.}
    \label{PenrosePhant3}
\end{figure*}

The arrangement of the different branches in a Penrose diagram, according to the metric signature of the region they are in, was already explained in the discussion about the plots of Figs. \ref{PenrosePhant2} and \ref{PenrosePhant2.1}. With all of that taken into account, the Penrose diagram for this case was constructed and is shown in Fig. \ref{PenrosePhant3}. In this diagram there are coiled lines used to separate the ``Universe" regions from the interiors of the internal horizon 2, which are drawn adjacent to one another due to how the diagram is constructed, not being directly related. These lines cannot be traversed, and its smoother side represents a branch of the asymptotically flat infinity, while its sharper side represents a branch of the light-like singularity.

More specifically, Fig. \ref{PenrosePhant3} depicts a Penrose diagram for a black hole solution with an event horizon, two internal horizons and a central, light-like singularity. 

\begin{itemize}

\item In ``Our Universe" the asymptotically flat infinity is on the right, represented by the smoother (and, from this region's perspective, inner) side of the coiled line (the use of this line will be further explained later), and on the left there are future and past branches of the event horizon. 

\item Within the future one there are two past branches of this horizon, and two future branches of internal horizon 1 (the first, relative to infinity, of two internal horizons), at $x=x_{IH1}$. 

\item Within the left one of these there are future and past branches of this horizon, on the right, and of internal horizon 2 (the second of the two internal horizons), on the left, at $x=x_{IH2}$. 

\item Within the future one of these there are two past branches of this horizon and two future branches of the light-like singularity, represented by the sharper (and, from this region's perspective, inner) side of the coiled line. Note that this line, as any other coiled line in this diagram, cannot be traversed. They are used to separate the exterior regions (``Our Universe" and its copies) from the interiors of internal horizon 2 (``Internal Horizon 2 Interior" and its copies), which are drawn adjacent to each other solely due to the way the diagram is constructed but are not directly related. 

\item Unlike in the previous figures, this distinct line is not used to distinguish regions that overlap, but to separate two very distinct behaviours: the smoother side represents a branch of the asymptotically flat infinity and the sharper one represents a branch of the light-like singularity. 

\item Moreover, copies of each region exist, accessed by traversing certain horizon branches. Note that both the top-left and top-right corners, which are left blank due to the lack of space in the diagram, should be labelled ``Copy of Our Universe".

\end{itemize}

We will now explore the cases in which there is no singularity (and $x\to-\infty$ is an infinity). Note that the horizon configuration, given the combination of constants, is exactly the same as described in the analysis of the critical values. Furthermore, as can be seen from Figs. \ref{criticalM} and \ref{criticalQ}, and given that the asymptotic behaviours of $A(x)$ and $g_{00}$ are similar, as discussed before, in all cases in which $M< M_{c2}$, or $|q|> |q_{c2}|$, the second spatial infinity is AdS, when $M=M_{c2}$, or $|q| = |q_{c2}|$, it is flat, and in all cases in which $M>M_{c2}$, or $|q| < |q_{c2}|$, it is dS. 

Now, whenever there are no horizons, two-way traversable wormhole solutions emerge. The minimum of $R(x)$ is always a throat in these cases, since $g_{00}$ remains always positive -- as can be seen in Figs. \ref{criticalM} and \ref{criticalQ}, and bearing in mind that the conformal factor is positive. Note that, as discussed before, the location of the minimum of $R(x)$ may take any real value in $x$, depending only on $\psi_0$. Furthermore, these solutions may be asymmetric or symmetric, if they attend the conditions previously discussed for a symmetric spacetime. This latter case is verified for $|q|\lesssim 0.825516$, $M=M_{c2}$ and $\psi_0=c_1 \pi$. However, as discussed earlier, $M\approx 0.73796$ is the minimum value of $M$ for which $\psi_0=c_1\pi$ is allowed ($m>0$), and, in this particular case, it is not possible to have $M=M_{c2} \geq 0.73796$. Thus, any one of these combinations leads to $m<0$, and so, we discard them. This way, only asymmetric ones are allowed. 

These wormholes occur, for example, for $|q|\gtrsim 0.825516$ and $M<M_{c1}$, or $|q|\lesssim 0.825516$, $M\leq M_{c2}$, with $\psi_0$ lying inside the range of no singularity. These are allowed as long as $\psi_0$ also ensures $m>0$. In the first example, there are combinations for which $\psi_0=c_1\pi$ -- hence $x_T=0$ -- is allowed, and so, even though fully symmetric solutions do not exist, the radius function is symmetric in these cases. On the other hand, in the second example, it is not possible to have $M\approx 0.73796$, thus, $\psi_0=c_1\pi$ -- hence $x_T=0$ and a symmetric $R(x)$ -- is not allowed in any case. In fact, in this example, only $\psi_0 \in ~]\frac{\pi}{2}+\frac{\pi}{2\sqrt{3}}+c_1\pi,(c_1+1)\pi[$ -- hence $x_T<0$ -- are allowed.

Additionally, we verify that in the first example all solutions feature an AdS second spatial infinity. In contrast, in the second example some solutions have an AdS second spatial infinity, while others have a flat one (see the previous discussion on Fig. \ref{criticalM}).

\begin{figure*}[ht!]
   \centering
      \includegraphics[scale=0.75]{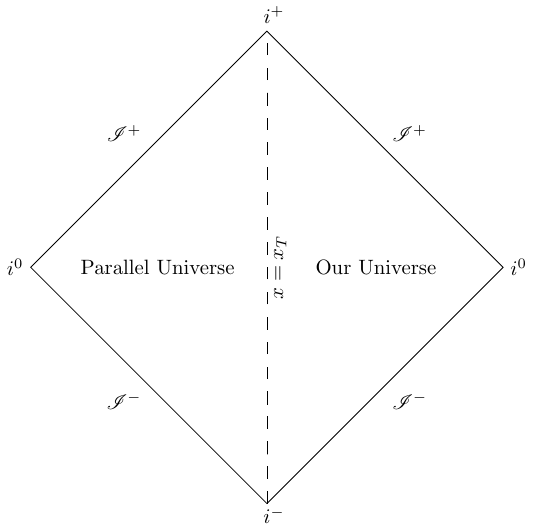}
      \hspace{1.5cm}
      \includegraphics[scale=0.75]{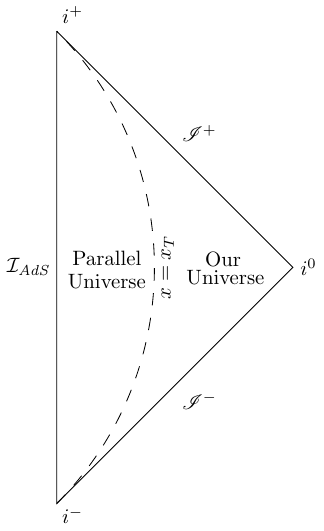}
    \caption{Left plot: Penrose diagram for a two-way traversable wormhole solution featuring two asymptotically flat regions. The throat is depicted as the dashed line at $x=x_T$. Even though $R(x)$ is not symmetric in the solution analysed, $x_T$ is equidistant from both infinities, which is why it is drawn in the middle. To the right of the throat lies ``Our Universe", with an asymptotically flat infinity, and to its left lies a ``Parallel Universe" with a similar infinity. Right plot: Penrose diagram for an asymmetric two-way traversable wormhole solution, asymptotically flat on one side and AdS on the other. The throat is tilted to the right because that is the only possible representation in this diagram, regardless of the symmetry of $R(x)$. Now, to the left of the throat lies a ``Parallel Universe" with an AdS infinity, depicted as the time-like (vertical) conformal boundary $\mathcal{I}_{AdS}$.}
    \label{PenrosePhant4}
\end{figure*}

The Penrose diagrams of these solutions are shown in Fig. \ref{PenrosePhant4}. The left plot corresponds to the case where both spatial infinities are flat, for which we have verified that $R(x)$ is asymmetric, and that the throat is only allowed at $x_T<0$. Despite this, since both infinities lie at $x\to\pm\infty$, we, in fact, consider $x_T$ to be equidistant from them, and so, it is drawn in the middle of the diamond. Thus, this diagram is similar to that of the well known Morris-Thorne wormhole \cite{Morris:1988cz}. 
Now, about the right plot, it corresponds to the cases where the second infinity is AdS, for which a symmetric $R(x)$, with $x_T=0$, may be allowed. The AdS infinity is depicted as the time-like (vertical) conformal boundary $\mathcal{I}_{AdS}$. Thus, regardless of the symmetry of $R(x)$, and the distance between the throat and the infinities, the throat is drawn tilted to the right because that is the only possible representation in this diagram. In both cases, by traversing the throat, one reaches a parallel universe. The ``second spatial infinity" we have been discussing corresponds to this universe’s spatial infinity.

In cases in which there are 1 to 3 horizons, the particular spacetime geometry will depend on the location of the minimum of $R(x)$ relative to each of them. In fact, it may be located before, between, or after the entire set, or even at the same point as only one of them. This location determines if it is a throat, an extremal null throat or a bounce. As discussed before, the emergence of each of these depends on whether $g_{00}>0$, $g_{00} = 0$, or $g_{00} < 0$, respectively, at the corresponding point. 

That being said, and based on the previous analysis of the horizon structure (see Figs. \ref{criticalM} and \ref{criticalQ}), the minimum of $R(x)$ is a throat whenever it is located outside the first or only horizon, or inside an extremal EH, or a CH, or the outermost IH (relative to $x\to\infty$) when three horizons exist. Furthermore, it is an extremal null throat whenever it coincides with any one of the horizons. At last, it is a bounce whenever it is located inside an EH, or an extremal IH inside an EH, or an IH located inside an extremal EH, or the innermost IH when three horizons exist.

In all of these cases, the second spatial infinity is either flat, dS, or AdS, according to what was previously discussed. Furthermore, all solutions that have $M=M_{c2}$, or $|q|=|q_{c2}|$, may be asymmetric or symmetric, as aforementioned, being allowed as long as $m$ remains positive. Any other values of those constants always lead to asymmetric solutions.

In the following analysis, we present examples of one-way traversable wormhole solutions, solutions in which a throat precedes any horizon, including when it is an extremal null throat, and black bounce solutions.

The former type of solutions emerges when exactly one horizon exists -- either an EH or an extremal EH -- and it is also a throat ($x_T=x_H$), corresponding to an extremal null throat. Similarly to the case of the two-way traversable wormholes, the one-way traversable ones can be either symmetric or asymmetric. The former occurs when $|q|\approx 0.825516$, $M=M_{c2}$ and $\psi_0=c_1 \pi$, which leads to $x_T=x_{eH}=0$. However, once again, since $M=M_{c2} \gtrapprox 0.73796$ is not verified, this combination leads to $m<0$, thus, only asymmetric ones are allowed. In fact, $x_T=x_{H}=0$ always leads to $m<0$, not being allowed in any case because the values of $M$ that yield $x_H=0$ never verify $M\gtrapprox 0.73796$ (see the examples provided below). This means that $\psi_0=c_1\pi$, which leads to $x_T=0$, is not allowed; hence, the radius function is also not allowed to be symmetric in any case now.

Furthermore, we find that the asymmetric geometries are only allowed if the horizon is an EH located at some $x_T=x_{EH}<0$. This is because when an extremal EH is the only horizon ($M=M_{c1}$ and $|q|\gtrapprox 0.825516$) it always occurs at some $x\geq 0$, and any $x_T=x_{H}\geq 0$ when there is only one horizon always leads to $m<0$ (the combinations of $M$ and $\psi_0$ that lead to that intersection, lead to $m<0$; the particular case $x_T=x_{H}=0$ was discussed above). 

Thus, this type of wormhole is only allowed for $0.776\lesssim|q|\lesssim 0.825516$ and $M_{c2}<M<M_{c1}$, or $|q|\lessapprox 0.776$ and $M>M_{c2}$, with $\psi_0$ such that $x_T=x_{EH}<0$, which is a necessary, but not sufficient, condition to have $m>0$, thus, $\psi_0$ must also ensure $m>0$. The condition $x_T=x_{EH}<0$ implies that $\psi_0$ must lie in the interval $]\frac{\pi}{2}+\frac{\pi}{2\sqrt{3}}+c_1\pi,(c_1+1)\pi[$, as discussed earlier. We find that in neither example do the values of $M$ that yield $x_H=0$ satisfy $M\gtrapprox 0.73796$, thereby supporting our previous discussion. Furthermore, we verify that in all of the allowed solutions, the second spatial infinity is dS (see Fig. \ref{criticalM}). 

\begin{figure*}[ht!]
   \centering
      \includegraphics[scale=0.5]{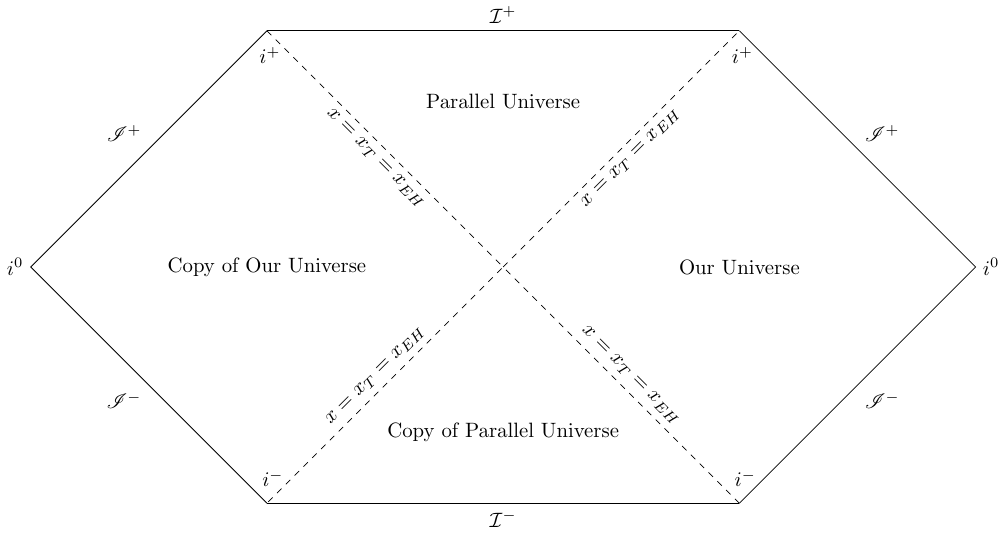}
      \hspace{0.5cm}
      \includegraphics[scale=0.5]{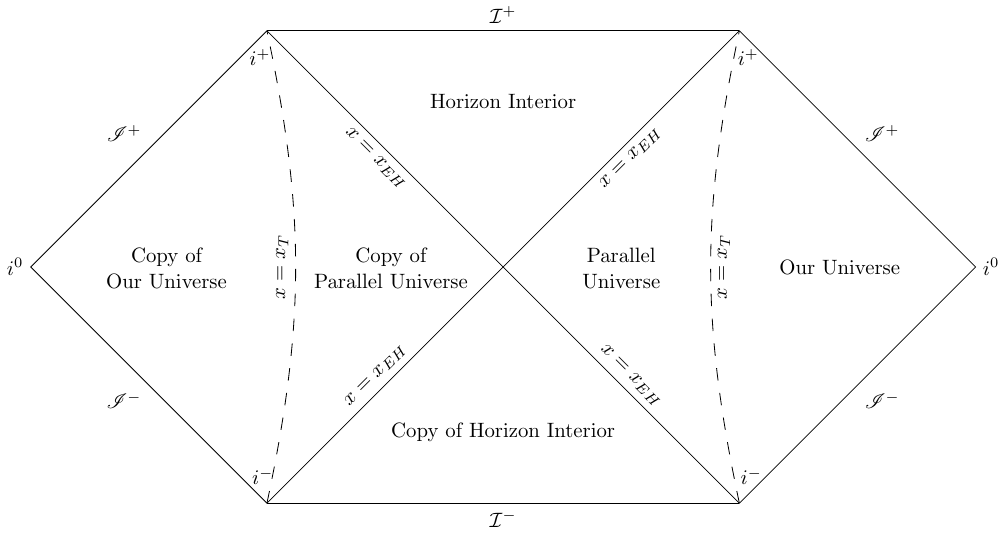}
      \vspace{0.1cm}
      \includegraphics[scale=0.5]{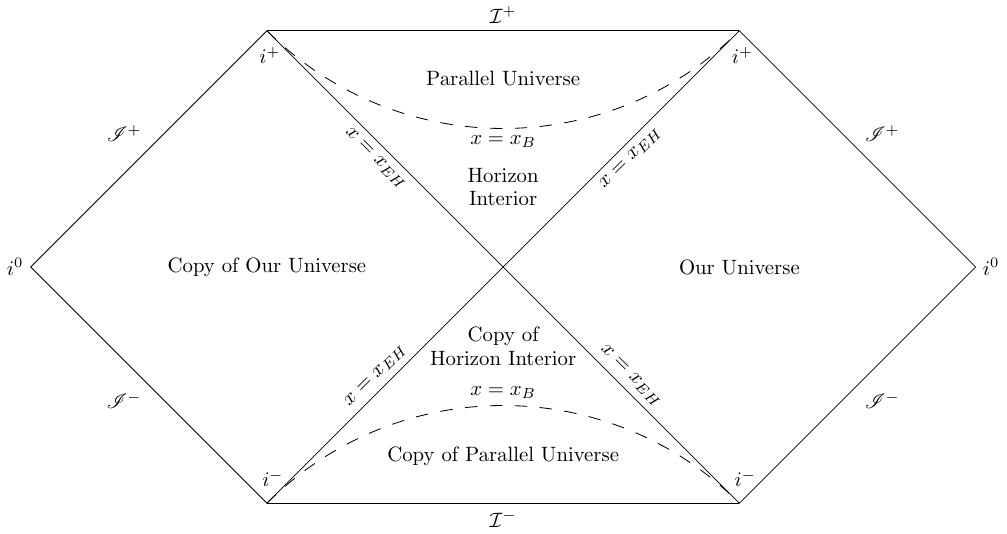}
    \caption{Top left plot: Penrose diagram for an asymmetric one-way traversable wormhole solution, asymptotically flat on one side of the throat and dS on the other.    	
    Top right plot: Penrose diagram for an asymmetric regular spacetime featuring a throat beyond which lies an event horizon, with one side approaching an asymptotically flat infinity and the other a dS infinity.
    Bottom plot: Penrose diagram for an asymmetric black bounce solution featuring a bounce located inside an event horizon, with one side leading to an asymptotically flat infinity and the other to a dS infinity. We refer the reader to the text for further details.}
    \label{PenrosePhant5}
\end{figure*}

The Penrose diagram for this solution is shown in the top left plot of Fig. \ref{PenrosePhant5}. This diagram is similar to the one shown in the right plot of Fig. \ref{PenrosePhant1}, and to that of the Schwarzchild solution, but with a dS infinity instead of the singularity, and an extremal null throat instead of the EH. Future and past dS infinities are depicted in the diagram as the space-like (horizontal) conformal boundaries $\mathcal{I}^+$ and $\mathcal{I}^-$, respectively.

More specifically, the top left plot of Fig. \ref{PenrosePhant5} depicts a Penrose diagram for an asymmetric one-way traversable wormhole solution, asymptotically flat on one side of the throat and dS on the other.

\begin{itemize}

\item In ``Our universe", there are an asymptotically flat infinity, on the right, and future and past branches of the extremal null throat on the left. This structure, depicted as diagonal dashed lines at $x=x_T=x_H$, emerges due to the intersection of the throat with the event horizon, being only one-way traversable. This is the only way to represent it in this diagram, since a horizon is always a line at a $45^\circ$ angle. 

\item Beyond the future one (the metric signature is $(-+-\,-)$, just as with an event horizon) lies a ``Parallel Universe", with two past branches of the extremal null throat, and a future dS infinity, depicted as the space-like (horizontal) conformal boundary $\mathcal{I}^+$. 

\item Moreover, there is a ``Copy of Our Universe", connected to the original by an Einstein-Rosen bridge, as well as a ``Copy of Parallel Universe", featuring a past dS infinity.

\end{itemize}

The second type of solution mentioned above occurs whenever the minimum of $R(x)$, in this case a throat, is located outside the outermost or only horizon, whether it is an EH or an extremal EH ($x_{H\text{max}}<x_T$). Thus, now, only asymmetric solutions may emerge. Similar to the one-way wormhole geometry, given the requirement $m>0$, we find that this new type of geometry is only allowed if $x_{H\text{max}}<x_T<0$. This is because any $x_T\geq 0$ that satisfies $x_{H\text{max}}<x_T$ leads to $m<0$. The particular case where $x_T=0$ is never allowed because any $M$ that yields $x_{H\text{max}}<0$ (ensuring $x_{H\text{max}}<x_T$) does not satisfy $M\gtrapprox 0.73796$ ($\psi_0=c_1\pi$ is not allowed). This way, as before, not even the radius function is allowed to be symmetric.

Thus, this type of geometry is only allowed for $0.776\lesssim|q|\lesssim 0.825516$ and $M>M_{c2}$ (which includes the mass scenarios $M_{c2}<M<M_{c1}$, $M=M_{c1}$, $M_{c1}<M<M_{c3}$, $M=M_{c3}$ and $M>M_{c3}$), or $|q|\lessapprox 0.776$ and $M>M_{c2}$, with $\psi_0$ such that $x_{H\text{max}}<x_T<0$, while also ensuring $m>0$. The condition $x_{H\text{max}}<x_T<0$ implies that $\psi_0$, once again, must lie in the interval $]\frac{\pi}{2}+\frac{\pi}{2\sqrt{3}}+c_1\pi,(c_1+1)\pi[$. Additionally, once again, we verify that in all of the allowed solutions, the second spatial infinity is dS (see Fig. \ref{criticalM}).

The Penrose diagram for the solution that features only one EH, which occurs, for example, for $0.776\lesssim|q|\lesssim 0.825516$ and $M_{c2}<M<M_{c1}$, is shown in the top right plot of Fig. \ref{PenrosePhant5}. It is similar to the top left plot, but with the throat located outside the horizon. In $x$, $x_T$ lies closer to $x_{EH}$ than to the infinity at $x\to\infty$, so the throat is drawn tilted towards $x_H$. Although $x_T$ is equidistant from both infinities (as always, since they are at $x\to\pm\infty$), it has to be drawn as is, in the diamond that includes ``Our Universe".

More specifically, the top right plot of Fig. \ref{PenrosePhant5} depicts the Penrose diagram of an asymmetric regular spacetime featuring a throat beyond which lies an event horizon, with one side approaching an asymptotically flat infinity and the other a dS infinity.

\begin{itemize}

\item To the left of the throat lying in ``Our Universe", there is a ``Parallel Universe" with future and past branches of the event horizon. Beyond the future one (``Horizon Interior"), there are past branches of the event horizon and a future dS infinity. 

\item Moreover, copies of each region exist, accessed by traversing certain horizon branches or throats.  In this case, the Einstein-Rosen bridge connects the ``Parallel Universe" and its copy.

\end{itemize}

For simplicity, we only show this example, with one EH, since the diagrams of the remaining cases can now be easily derived from the previously shown diagrams of black hole solutions that share the same horizon structure (Fig. \ref{PenrosePhant2.1}, Fig. \ref{PenrosePhant3} and the right plot of Fig. \ref{PenrosePhant2}, in the order of the mass scenarios ranging from $M=M_{c1}$ to $M=M_{c3}$ for $0.776\lesssim|q|\lesssim 0.825516$). In fact, one essentially needs to replace the singularity (that is located in a $(- + -\,-)$ region) by a dS infinity (space-like conformal boundary), and draw the throat as is illustrated in the shown diagram, with a parallel universe lying beyond it. In the adapted version of Fig. \ref{PenrosePhant2.1}, the zigzag line would not be used to distinguish two overlapped regions, but to separate two regions that are not directly related, as done with the coiled line in Fig. \ref{PenrosePhant3}. In this latter figure's adapted version, the coiled line would not even be necessary, because the dS infinity would not be in contact with the flat infinity, unlike the case with a light-like singularity.

Another type of geometry, related to the previous two, occurs when the minimum of $R(x)$ coincides with the outermost horizon (among several), whether it is an EH or an extremal EH ($x_T=x_{H\text{max}}$), thereby forming an extremal null throat that is located outside all the other horizons, analogous to the previous case (a throat also preceded all horizons). Thus, once again, only asymmetric solutions may emerge. Similar to the two previous cases, due to the requirement $m>0$, we find that this type of geometry is only allowed if $x_T=x_{H\text{max}}<0$, since any non-negative $x_T=x_{H\text{max}}$ leads to $m<0$. The particular case $x_T=x_{H\text{max}}=0$ is never allowed because any $M$ that yields $x_{H\text{max}}=0$ does not satisfy $M\gtrapprox 0.73796$ ($\psi_0=c_1\pi$ is not allowed). Thus, as before, not even the radius function is allowed to be symmetric.

Therefore, we find that this type of geometry is only allowed for $0.776\lesssim|q|\lesssim 0.825516$ and $M_{c1}\leq M\leq M_{c3}$ (which includes the mass scenarios $M=M_{c1}$, $M_{c1}<M<M_{c3}$, $M=M_{c3}$), with $\psi_0$ such that $x_T=x_{H\text{max}}<0$, while also ensuring $m>0$. The condition $x_T=x_{H\text{max}}<0$ implies that $\psi_0$ must lie in the interval $]\frac{\pi}{2}+\frac{\pi}{2\sqrt{3}}+c_1\pi,(c_1+1)\pi[$. Furthermore, as before, in all of the allowed solutions, the second spatial infinity is dS (see Fig. \ref{criticalM}).

The Penrose diagrams for these solutions can be adapted, respectively -- following the order of the mass scenarios -- from the diagrams shown in Fig. \ref{PenrosePhant2.1}, Fig. \ref{PenrosePhant3} and the right plot of Fig. \ref{PenrosePhant2}. To perform the adaptation one needs to replace the singularity by a dS infinity, as explained before, and draw an extremal null throat in place of the outermost horizon.

A black bounce solution emerges whenever the minimum of $R(x)$, independently of the sign of $g_{00}$, is located inside a horizon, whichever it is. Thus, in our case, we find three different types of black bounce, according to whether the minimum that exists is a bounce, a throat or an extremal null throat. When one or more horizons exist past one of these structures, their classification is ``reset" because traversing such a structure leads to a different universe. This means that the horizon following the minimum of $R(x)$, when it exists, is reclassified as an EH or an extremal EH, depending on whether it is simple or extremal, respectively. Additionally, when two horizons exist after that structure, which in our case is only possible in the case where three simple horizons exist, the innermost is reclassified as a CH. Note that in the previous geometries, although a different universe also arises, the classification remains unchanged as no horizons appear before the minimum of $R(x)$ while others are present beyond it.

Furthermore, unlike the previous geometries, where $x_T>x_{H\text{max}}$ or $x_T=x_{H\text{max}}$ were verified, now, $x_{R_\text{min}}$ (location of the minimum of $R(x)$, regardless of its type) is lower than some $x_H$. Thus, at least $x_{R_\text{min}}<x_{H\text{max}}$ is verified. In this case, the requirement $m>0$ does not always force this minimum to be at some $x_{R_\text{min}}<0$, thus it may be located at $x_{R_\text{min}}\geq0$. The particular case where $x_{R_\text{min}}=0$ may now be allowed, depending on the combination of constants, because a value of $M$ that yields $x_{H\text{max}}>0$ may satisfy $M\gtrapprox 0.73796$. Thus, symmetric radius functions may be allowed, as well as symmetric solutions, which are a possibility in certain cases. Nevertheless, $m>0$ still imposes great constraints on the allowed range of $x_{R_\text{min}}\geq 0$; however, as $M$ increases, that range expands, since the allowed range of $\psi_0$ also expands.

All of this also means that more scenarios of $M$ and $|q|$ are allowed, and so, for each of the three types of black bounce, we will only present one example for each different horizon structure possible.

The solutions in which there is a bounce occur, for example, for $|q|\gtrsim 0.825516$ and $M_{c1}<M\leq M_{c2}$, and also $M>M_{c3}$, or $0.776\lesssim|q|\lesssim 0.825516$ and $M_{c1}\leq M\leq M_{c3}$, with $\psi_0$ such that the minimum of $R(x)$ occurs in a region inside a horizon where $g_{00}<0$ (see Fig. \ref{criticalM}). These solutions are allowed as long as the value of $\psi_0$ also ensures $m>0$. In these solutions, the second spatial infinity may be AdS, flat or dS, as previously discussed (see Figs. \ref{criticalM} and \ref{criticalQ}). In particular, in the examples associated with $|q|\gtrsim 0.825516$ it is either AdS or flat and in the ones associated with $0.776\lesssim|q|\lesssim 0.825516$ it is dS.

When in the first example ($|q|\gtrsim 0.825516$ and $M_{c1}<M\leq M_{c2}$) we have $M=M_{c2}$ and $\psi_0=c_1\pi$, a symmetric black bounce emerges. In this case, the bounce is located at $x_B=0$, and beyond it there is an EH -- following our previous discussion -- located at $x_{EH2}=-x_{EH1}$ (there are two EHs in this solution, one before and one after the bounce). This configuration is allowed ($m>0$) when $|q|$ is such that it yields $M_{c2}\gtrapprox 0.73796$. Apart from this case, the remaining examples always correspond to asymmetric solutions, though symmetric radius functions are possible if $\psi_0=c_1\pi$, which is allowed when $M\gtrapprox 0.73796$. We find that these values of $M$ may occur only in the examples associated with $|q|\gtrsim 0.825516$.

In Appendix~\ref{apB}, we present the detailed derivation of the $f(\mathcal{R})$ function from action (\ref{2}) -- which is ultimately obtained in parametric form -- associated with a particular instance within the family of symmetric black bounce solutions described above, namely with $q=5$, $M=25\pi/3$ ($M=M_{c2}$) and $\psi_0=0$.

The Penrose diagram of the case where the bounce lies inside an EH and there is no other horizon, which may occur, for example, for $|q|\gtrsim 0.825516$ and $M>M_{c3}$, is shown in the bottom plot of Fig. \ref{PenrosePhant5}. This is similar to the other diagrams shown in that figure, which were previously discussed, but with the minimum of $R(x)$ -- a bounce -- located inside the horizon. Note that the bounce is equidistant from both infinities, as discussed before for the throats, and as verified for any minimum of $R(x)$ in this work, however, it has to be drawn after the horizon, in the half-diamond where the boundary of the dS infinity lies -- in any case, the bounce or throat is always positioned according to the horizon it precedes or follows. Furthermore, $x_B$ lies closer to $x_{EH}$ than to $x\to-\infty$, and so, in any case, it must be tilted as shown. However, in this diagram, regardless of that proximity, that is actually the only possible way to represent it, due to how the dS infinity is depicted.

More specifically, this is shown in the bottom plot of Fig. \ref{PenrosePhant5}, which depicts a Penrose diagram for an asymmetric black bounce solution featuring a bounce located inside an event horizon, with one side leading to an asymptotically flat infinity and the other to a dS infinity. 

\begin{itemize}

	\item Within the future branch of the event horizon that lies in ``Our Universe", in the ``Horizon Interior", there are past branches of the event horizon and a bounce, depicted as the dashed line at $x=x_B$. 
	
	\item Beyond it there is a ``Parallel Universe" with a future dS infinity. Note that, in $x$, $x_B$ lies closer to $x_{EH}$ than to infinity, which means, in any case, it must be tilted as is. 
	
	\item However, in this diagram, regardless of that proximity, that is actually the only possible way to represent it. This is also the reason it lies closer to the dS infinity and not equidistant from both infinities as it is in $x$. Moreover, copies of each region exist.

\end{itemize}

\begin{figure*}[ht!]
   \centering
      \includegraphics[scale=0.342]{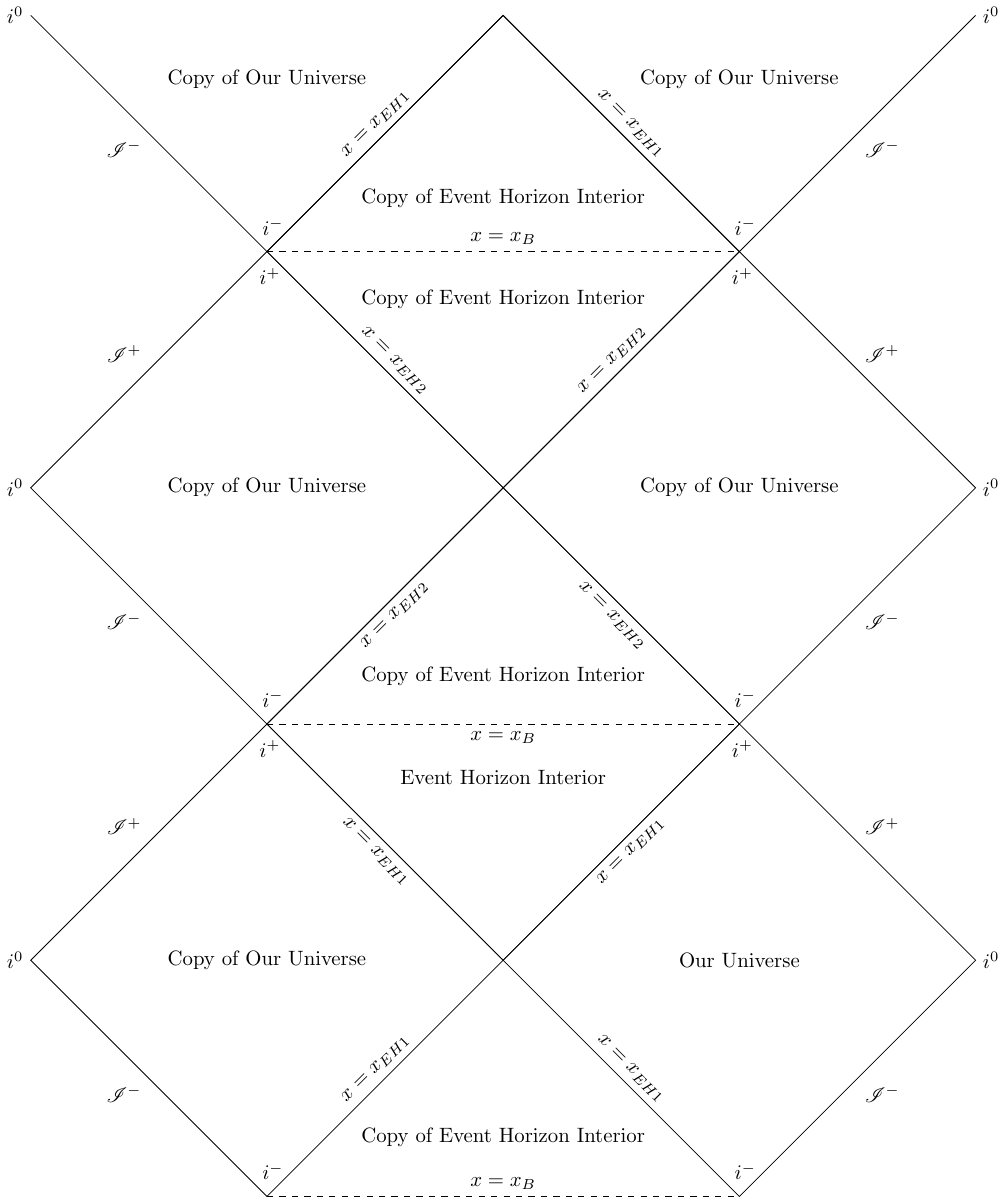}
      \hspace{0.01cm}
      \includegraphics[scale=0.342]{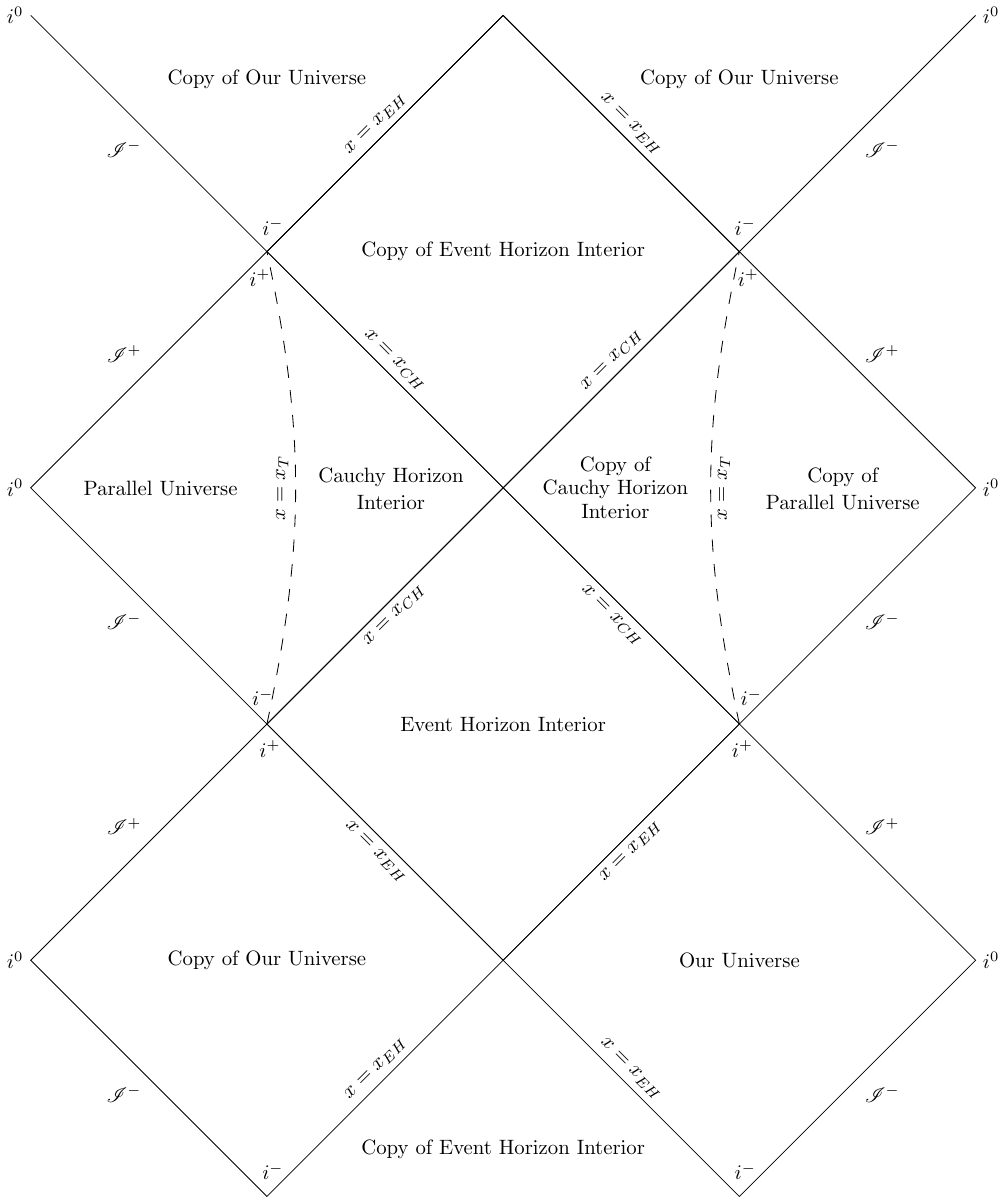}
      \hspace{0.01cm}
      \includegraphics[scale=0.342]{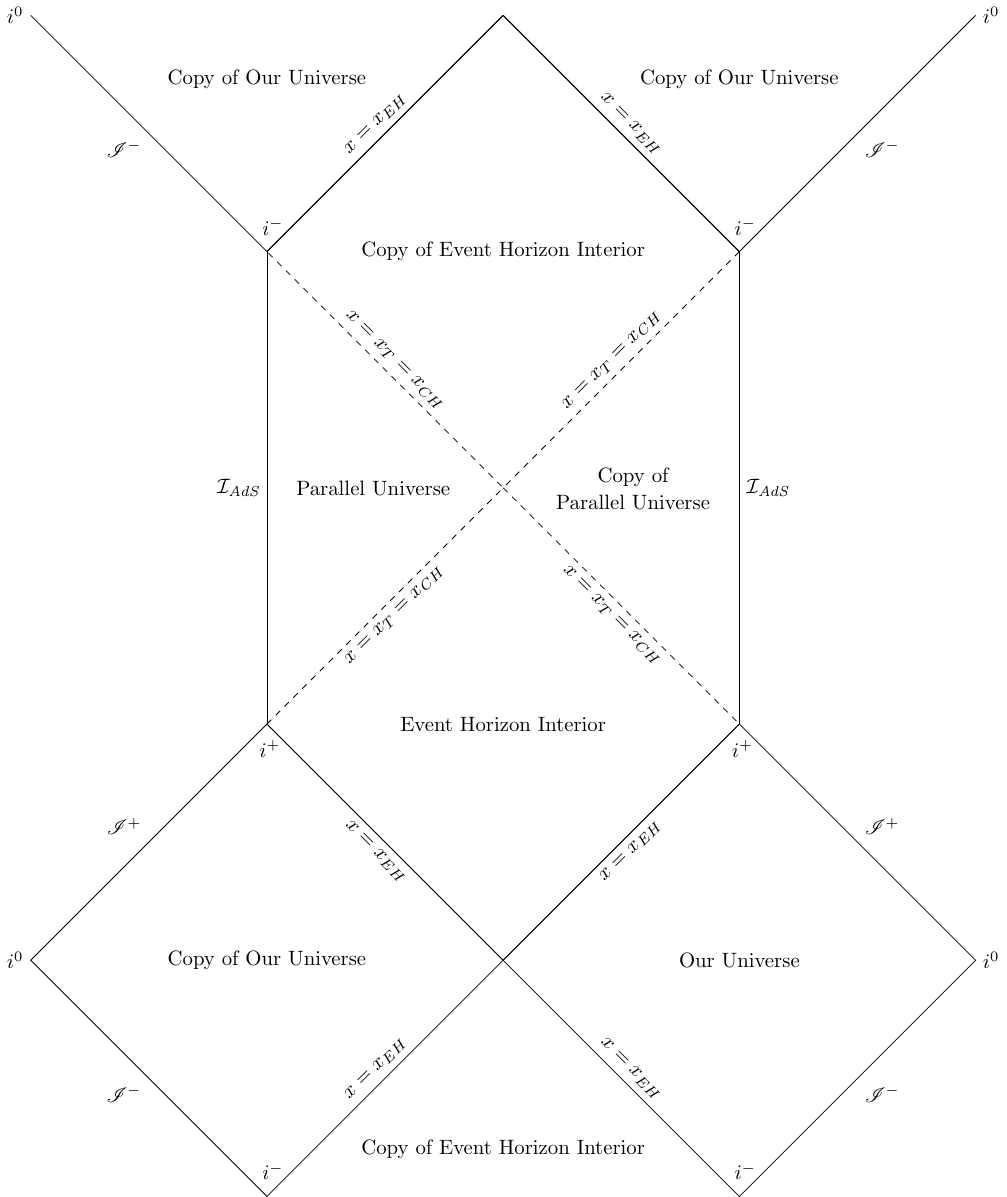}
    \caption{Left plot: Penrose diagram for a symmetric black bounce solution featuring a bounce located between an event horizon and a Cauchy horizon, with both sides leading to asymptotically flat infinities.
    Middle plot: Penrose diagram for an asymmetric black bounce solution featuring a throat located inside a Cauchy horizon, which lies inside an event horizon, with both sides leading to asymptotically flat infinities.
    Right plot: Penrose diagram for a black bounce solution featuring an extremal null throat -- arising from the intersection of the throat with the Cauchy horizon -- located inside an event horizon, with one side leading to an asymptotically flat infinity and the other to an AdS infinity. We refer the reader to the text for further details.}
    \label{PenrosePhant6}
\end{figure*}

The Penrose diagram of the symmetric black bounce discussed above is shown in the left plot of Fig. \ref{PenrosePhant6}. It is similar to the left plot of Fig. \ref{PenrosePhant2}, but with a flat infinity, instead of a singularity, and a bounce in the middle of both horizons (not tilted because $x_B$ is equidistant from both), which are now both EHs, as explained before. In this case, beyond the bounce there is a copy of our universe, rather than a parallel one, as in other diagrams, because the solution is symmetric. In the case of an asymmetric solution this diagram must be adapted: the bounce has to be tilted towards the horizon to which $x_B$ is closer to (it may still be in the middle), and beyond the bounce lies a parallel universe, rather than a copy, with an AdS or flat infinity.

More specifically, the Penrose diagram in the left plot of Fig.
\ref{PenrosePhant6}, depicts a symmetric black bounce solution featuring a bounce located between two event horizons, with both sides leading to asymptotically flat infinities. 

\begin{itemize}

\item On the left of ``Our Universe", which is asymptotically flat, there are future and past branches of the event horizon, at $x_{EH1}$ (the $1$ denotes the first of two event horizons). 

\item Within the future one, in the ``Event Horizon Interior", there are past branches of the event horizon and a bounce, beyond which lies a ``Copy of Event Horizon Interior", which is part of a ``Copy of Our Universe". 

\item In this interior, there are future branches of the second event horizon, at $x=x_{EH2}$. 

\item Beyond the left one lies a ``Copy of Our Universe", with an asymptotically flat infinity on the left, and future and past branches of the event horizon on the right. 

\item Note that in asymmetric solutions the bounce could also be tilted either upwards or downwards, depending on which horizon it is closer to, in terms of $x$. 

\item Furthermore, in these solutions, the copies mentioned are replaced by parallels (``Parallel of Our Universe", for instance). Moreover, in both symmetric and asymmetric diagrams, there are copies of each region. 

\end{itemize}

For simplicity, we only show these examples, since the diagrams of the remaining cases can now be easily derived from the previously shown diagrams of black hole solutions that share the same horizon structure. All the remaining cases present a dS second infinity, that is part of a parallel universe. Thus, in all cases, to perform the adaptation, one needs to replace the singularity by a dS infinity, as explained before, and draw a bounce, tilted or not, in the diamond that corresponds to the interior $x_B$ is in (always in a region with a metric signature $(-+-~-)$).

Furthermore, the solutions in which a throat is present occur, for example, for $|q|\gtrsim 0.825516$ and $M=M_{c1}$, and also $M_{c1}<M\leq M_{c2}$, or $0.776\lesssim|q|\lesssim 0.825516$ and $M_{c1}\leq M < M_{c3}$, with $\psi_0$ such that the minimum of $R(x)$ occurs in a region inside a horizon where $g_{00}>0$ (see Fig. \ref{criticalM}). These solutions are allowed as long as the value of $\psi_0$ also ensures $m>0$. 

At last, the solutions in which there is an extremal null throat occur, for example, for $|q|\gtrsim 0.825516$ and $M_{c1}<M\leq M_{c2}$, or $0.776\lesssim|q|\lesssim 0.825516$ and $M_{c1}\leq M\leq M_{c3}$, with $\psi_0$ such that the minimum of $R(x)$ coincides with an inner horizon ($g_{00}=0$) (see Fig. \ref{criticalM}). These solutions are allowed as long as the value of $\psi_0$ also ensures $m>0$. 

Whenever either a throat or an extremal null throat is present, the resulting black bounce solutions are asymmetric. However, a symmetric radius function is possible if $\psi_0=c_1\pi$, which is allowed when $M\gtrapprox 0.73796$. In both types of black bounce, we find that these values of $M$ may occur only in the examples associated with $|q|\gtrsim 0.825516$. Furthermore, as before, in the examples associated with $|q|\gtrsim 0.825516$ the second spatial infinity is either AdS or flat and in the ones associated with $0.776\lesssim|q|\lesssim 0.825516$ it is dS.

The Penrose diagram of the solution with a throat inside a CH, which may occur for $|q|\gtrsim 0.825516$ and $M_{c1}<M\leq M_{c2}$, is shown in the middle plot of Fig. \ref{PenrosePhant6}. This is similar to the left plot of the same figure, but with the minimum of $R(x)$, which is a throat now, inside the CH, and with a parallel universe beyond it. Here we show the case where the second spatial infinity is flat, but there is also the possibility of an AdS infinity. 
Moreover, the throat has to be tilted as is, since $x_T$ lies closer to the horizon than to $x\to-\infty$. 

More specifically, the middle plot of Fig. \ref{PenrosePhant6} depicts a Penrose diagram for an asymmetric black bounce solution featuring a throat located inside a Cauchy horizon, which lies inside an event horizon, with both sides leading to asymptotically flat infinities. 

\begin{itemize}

	\item Now, unlike before, in the ``Event Horizon Interior" there are future branches of the Cauchy horizon. 
	
	\item Within the left one, in the ``Cauchy Horizon Interior", there are future and past branches of this horizon, on the right, and a throat, beyond which lies a ``Parallel Universe" with an asymptotically flat infinity. 
	
	\item The throat is tilted towards $x_{CH}$ because, in $x$, it lies closer to it than to infinity. 
	
	\item Furthermore, in the diagram, it lies closer to the second infinity than to the first one, even if, in $x$, it is equidistant from both, because that is the only way to represent it. As before, copies of each region exist.

\end{itemize}

The Penrose diagram of the solution with an extremal null throat -- which is the result of the intersection of a throat and a CH -- inside an EH, which may occur for $|q|\gtrsim 0.825516$ and $M_{c1}<M\leq M_{c2}$, is shown in the right plot of Fig. \ref{PenrosePhant6}. This is similar to the middle plot of the same figure (just discussed), but with the extremal null throat and an AdS infinity instead. Here we show the case of an AdS second infinity, but there is also the possibility of a flat spatial infinity. As explained before, the extremal null throat has to be drawn as is, regardless of its proximity to other structures. 

The right plot of Fig. \ref{PenrosePhant6} depicts a Penrose diagram for a black bounce solution featuring an extremal null throat -- arising from the intersection of the throat with the Cauchy horizon -- located inside an event horizon, with one side leading to an asymptotically flat infinity and the other to an AdS infinity. 

\begin{itemize}
	
	\item First, unlike the previous diagram, in the ``Event Horizon Interior" there are future branches of the extremal null throat, at $x=x_T=x_{CH}$. 
	
	\item Beyond the left one (the metric signature changes back to $(+--\,-)$, as with a Cauchy horizon) lies a ``Parallel Universe" with future and past branches of the extremal null throat, on the right, and an AdS infinity, on the left. 
	
	\item Once again, in $x$, the throat is equidistant from both infinities, however, in the diagram it has be drawn as it is, closer to the AdS infinity. As before, copies of each region exist.

\end{itemize}

As before, for simplicity, we only show these examples, since the diagrams of the remaining cases can now be easily derived from the previously shown diagrams of black hole solutions that share the same horizon structure. All the remaining cases present a dS second infinity, that is part of a parallel universe. This way, to perform the adaptation, in any case, the singularity must be replaced by a dS infinity, as explained before, and the throat or extremal null throat must be drawn (the throat must be tilted towards the branches of the structure it lies closer to) in the diamond that corresponds to where $x_T$ is in (always in a region with a metric signature $(+--~-)$ or coinciding with an inner horizon, respectively).

\begin{center}
    Case 2: $\psi_0=\frac{\pi}{2}+\frac{\pi}{2\sqrt{3}}+c_1\pi$
\end{center}

In what follows, we will analyse this particular case of $\psi_0$, for which only $x\to\infty$ is an infinity. In fact, as discussed earlier, in this case, as $x\to-\infty$, the radius converges to a non-zero finite value, namely, $R(x)\to 1/\sqrt{3}$, regardless of $M$ or $|q|$, and so, it is not an infinity. 

Furthermore, the metric function $g_{00}$ also converges to finite values at that limit and $g_{11}\to 0$ (for any $M$ or $|q|$). Regarding $K$, we find it converges to non-zero finite values. In the particular case of $M = \pi q^2/3$, we find that $g_{00}\to 0$ and $K_1\to 3$. This means that $x\to -\infty$ is a horizon in this case and an ordinary regular surface, with a non-zero finite radius, otherwise. 

This way, we still need, if possible, to analyse the metric beyond that surface, until a singularity, second spatial infinity, or regular centre emerges. However, that means we have to analyse values of $x$ smaller than $x\to -\infty$, which is, apparently, an absurd. Nevertheless, that might be possible by transforming the coordinate $x$ into a coordinate in which $-\infty$ is transformed into a finite value. By doing so, we might be able to extend the solution beyond that surface \cite{Bronnikov:2006bv}. This is known as analytical continuation. 

Let us define a new, adequate coordinate as $y=\tanh(x)$, for which the endpoints $x\to \pm\infty$ are transformed into $x\to\pm1$. Then, by rewriting the line element in Eq. (\ref{dsJ pha}) in terms of $y$, we obtain the new line element:
\begin{eqnarray}\label{dsJ pha_y}
    &&ds_J^2=\cos^2\left(\frac{\arctan(x(y))}{\sqrt{3}}+\psi_0\right)\times
    	\nonumber\\	
    &&\hspace{-0.6cm}\left[A(x(y))dt^2-\frac{1/(1-y^2)^2}{A(x(y))}dx^2-(x(y)^2+1)d\Omega^2\right],
\end{eqnarray}
where $x(y)=\text{arctanh}(y)$. The coordinate $y$ is then restricted to the range of values that result in a real-valued metric. If that range goes beyond $y=-1$, it means it is possible to extend, via an analytical continuation, the solution past that regular surface. However, by analysing this line element, we find that continuation is not possible, with the range of $y$ remaining $y\in]-1,1[$.

This way, this scenario is non-physical, because it features a non-analytical region where particles are forbidden. Consequently, we will discard this case and not pursue further analysis.

\subsubsection{$\psi_0 = \frac{\pi}{2} - \frac{\pi}{2\sqrt{3}} + c_1\pi$}

As discussed earlier, when $\psi_0$ satisfies this condition, we find that at the limit $x\to \infty$, $R(x)\to 1/\sqrt{3}$, regardless of $M$ or $|q|$, thus, it is not an infinity. Nevertheless, the limit $x\to -\infty$ is, which means we have to consider it as the spatial infinity, even in scenarios it is AdS or dS.
Furthermore, as $x\to\infty$, $K$ approaches a non-zero finite value (in particular, $K_1\to3$) and $g_{00}\to 0$, as well as $g_{11}$, and so, this is a horizon. Note that this is analogous to the Case 2 analysed, and discarded, above, the main difference being that now the regular surface occurs as $x\to\infty$.

This way, as before, we still need, if possible, to analyse the metric beyond that surface. However, since that means we have to analyse values of $x$ higher than $x\to \infty$, this requires an analytical continuation -- via an appropriate coordinate transformation -- as described above. For this purpose, we can use the same coordinate $y$ as before, for which the endpoints $x\to \pm\infty$ are transformed into $x\to\pm1$, as aforementioned, and which leads to the same transformed line element (\ref{dsJ pha_y}).

From the previous analysis of that line element we derived that the range of values that result in a real-valued metric is $y\in]-1,1[$. This way, once again, the analytical continuation turns out to be impossible. As a result, this case is also non-physical, and so, we discard it and do not analyse it any further.

\section{Conclusion}\label{section:conclusion}

This paper builds upon previous investigations of hybrid metric-Palatini gravity (HMPG) by deriving and thoroughly analysing exact electrically charged solutions in the presence of a non-zero scalar potential. By integrating key aspects of both the metric and Palatini formalisms, HMPG establishes a versatile theoretical framework capable of addressing persistent challenges in General Relativity, including the nature of cosmic acceleration and the dark matter problem.
Focusing on spherically symmetric spacetimes, we systematically explored solutions in both the Jordan and Einstein conformal frames. To construct these solutions, we employed an inverse problem approach, wherein specific forms of the radius function are initially proposed, allowing for the subsequent derivation of the associated metric functions, scalar field configurations, and the corresponding scalar potential. This procedure was applied separately in two cases: one yielding a canonical scalar field and the other a phantom scalar field. For computational convenience, the solutions were first obtained within the Einstein frame, where the field equations take a simpler form, before being transformed back into the Jordan frame. This dual-frame analysis provides deeper insight into the physical interpretation of the solutions, highlighting their causal structures, horizon properties, and potential astrophysical implications.

A comprehensive analysis was conducted on the horizon and throat structures, asymptotic behaviours, and the existence of singularities. A key result, observed in both the canonical and phantom sectors, is the identification of ``critical values'' for the charge and the Einstein-frame mass parameters, which determine the number and types of horizons present. Specifically, depending on the relative magnitudes of charge and mass, the theory allows for configurations with no horizons, as well as solutions with up to two or even three horizons.
Furthermore, the choice of the scalar field, canonical or phantom, leads to fundamentally different causal structures. In the canonical case, the solutions resemble those of the Reissner–Nordström spacetime, featuring naked singularities and black holes with either a single extremal horizon or two distinct simple horizons. In contrast, the phantom sector gives rise to a much broader range of solutions, including naked singularities, traversable wormholes (both one-way and two-way), black holes, and black bounces, where a classical singularity is replaced by a throat or a bounce. These configurations exhibit a variety of horizon structures, such as simple event, Cauchy, or internal horizons, as well as extremal horizons of different types. Additionally, certain regular solutions emerge, extending beyond the throat or bounce, with asymptotic behaviours that can be flat, de Sitter (dS), or anti-de Sitter (AdS). To further elucidate the causal properties of these geometries, Penrose diagrams were constructed, offering a visual representation of their distinct causal structures.

It is worth noting that the Einstein-frame solution we obtain in the phantom sector is the same as the one derived in Ref.~\cite{Bolokhov:2012kn}, also under the assumption (\ref{r pha}). In that work a detailed analysis of the solution was carried out, including the classification of all possible geometries -- based on the number and nature of the horizons and the asymptotic behaviours -- as well as the construction of the corresponding Penrose diagrams. In our work, however, the main focus is not on that solution, but rather on the Jordan-frame version, where, in some cases, the conformal factor introduces significant modifications to the geometry and the global structure of the solutions. In particular, when it vanishes at finite values of the radial coordinate, a singularity emerges, which does not occur in any case in the analysis performed in Ref.~\cite{Bolokhov:2012kn}. On the other hand, when it does not vanish, leading to entirely regular solutions, the results are indeed similar to those of that work, and several geometries are identical. Nevertheless, even in this case, novel geometries emerge due to the conformal factor, which allows the local minimum of $R(x)$ to occur at any real value of $x$, rather than always at $x=0$ as in Ref.~\cite{Bolokhov:2012kn}. In any case, note that the introduction of the conformal factor leads to physically inequivalent solutions, even when the spacetime geometry is identical. In particular, thermodynamic properties such as temperature and entropy differ, implying that the underlying thermodynamic system is fundamentally distinct.

Our findings emphasize that, with suitable parameter choices, the inclusion of a non-zero scalar potential significantly extends the variety of geometric and causal structures beyond those found in earlier HMPG studies that assumed a vanishing potential. This enrichment manifests in the emergence of novel horizon configurations, throat structures, and regular black hole alternatives, demonstrating the profound impact of the scalar sector on the theory’s solution space.
By broadening the range of possible solutions, this work not only enhances the astrophysical relevance of HMPG but also uncovers its potential to describe entirely new families of black hole and wormhole geometries. These findings could have important implications for gravitational physics, guiding future research into the stability properties of these solutions, their observational signatures, such as gravitational lensing and gravitational wave emission, and their connections to quantum gravity and beyond-standard-model extensions of General Relativity.

Indeed, recent advancements in observational techniques have significantly enhanced our ability to test and refine theoretical models of black holes. The detection of gravitational waves by LIGO and Virgo has opened new observational frontiers, allowing for the direct study of black hole mergers and compact binary coalescences. Simultaneously, high-resolution imaging of black hole shadows by the EHT has provided unprecedented empirical validation of black hole existence and behaviour in strong-field regimes. These observational breakthroughs serve as crucial tests of GR while also offering opportunities to probe potential deviations that could indicate new physics beyond Einstein’s theory.
In fact, despite its success in describing large-scale gravitational phenomena, GR faces significant theoretical challenges in extreme curvature regimes, particularly near black hole singularities. Resolving these issues necessitates incorporating quantum gravity effects, which remain largely elusive but are essential for a more complete understanding of spacetime. Furthermore, higher-dimensional theories inspired by string theory present promising avenues for addressing these inconsistencies, yielding testable predictions for black hole behaviour in modified gravitational frameworks.

In this broader context, modified theories of gravity, such as HMPG, have emerged as compelling alternatives to address some of GR’s foundational limitations, including the resolution of the singularity problem, stability issues, and the complexities of the causal structure. HMPG and related models provide a rich theoretical setting to investigate these fundamental questions, offering novel approaches to unify gravitational phenomena across different regimes.
Among the most intriguing developments in black hole physics are the concepts of regular black holes and black bounces. Unlike traditional GR solutions, which lead to singularities where classical physics breaks down, these alternative models propose geometries that remain regular throughout. Regular black holes provide singularity-free descriptions of gravitational collapse, while black bounces introduce distinct topological features, permitting smooth transitions between black holes and expanding cosmological solutions.

In conclusion, the study of black holes remains a pivotal intersection between observation and theory, offering a unique framework to test GR in extreme conditions, explore singularity-free solutions, and incorporate quantum corrections. As observational capabilities continue to advance and theoretical models evolve, black hole research stands poised to deliver transformative insights into the nature of gravity and the underlying structure of the universe.

\acknowledgments{
The authors thank Kirill A. Bronnikov for the helpful insights and guidance, which significantly contributed to the development of this work.
FSNL acknowledges support from the Funda\c{c}\~{a}o para a Ci\^{e}ncia e a Tecnologia (FCT) Scientific Employment Stimulus contract with reference CEECINST/00032/2018, and funding through the research grants UIDB/04434/2020, UIDP/04434/2020 and PTDC/FIS-AST/0054/2021.
MER thanks Conselho Nacional de Desenvolvimento Cient\'ifico e Tecnol\'ogico - CNPq, Brazil, for partial financial support. This study was financed in part by the Coordena\c{c}\~{a}o de Aperfei\c{c}oamento de Pessoal de N\'{i}vel Superior - Brasil (CAPES) - Finance Code 001.
}


\appendix
\section{Derivation of $f(\mathcal{R})$ -- Example 1}\label{apA}

In this appendix, we present in detail the derivation of the $f(\mathcal{R})$ function from action (\ref{2}) associated with a solution obtained in Example 1 (see Section \ref{example1_section}).

We begin by noting that, by setting $E = \mathcal{R}$ in Eq.~(\ref{3}), the scalar field $\phi$ and the potential $V(\phi)$ reduce to:
\begin{equation}\label{Ap_1}
    \phi\equiv f^\prime(\mathcal{R})\,, \qquad  V(\phi)=\mathcal{R}f^\prime(\mathcal{R})-f(\mathcal{R})\,\,.
\end{equation}

Then, by relating these, we have
\begin{equation}\label{Ap_2}
    f(\mathcal{R})=\mathcal{R}\phi-V(\phi)\,\,.
\end{equation}

In Eq.~(\ref{phi exp}) there is the solution for $\bar{\phi}(x)$ in Example 1, which can be inverted to obtain $x$ in terms of $\bar{\phi}$. Furthermore, from the canonical case ($-1<\phi<0$) of Eq.~(\ref{8}), which provides $\phi(\bar{\phi})$, we get the relation
\begin{equation}\label{Ap_3}
    \bar{\phi}=\sqrt{6}\text{ arctanh}(\sqrt{-\phi})\;.
\end{equation}

By expressing $x$ in terms of $\bar{\phi}$ in Eq.~(\ref{potential}), which is the solution for $U(\bar{\phi}(x))$ in Example 1, and then replacing it by Eq.~(\ref{Ap_3}), we get $U(\phi)$. Finally, by using Eq.~(\ref{V_canonical}), we obtain an expression for $V(\phi)$:
\begin{widetext}
\begin{eqnarray}
V(\phi)&=&-4\,\bigl(1+\phi\bigr)^{2}\Bigl\{-9q^{2}+q^{2}\cosh\Bigl(2\sqrt{2}\,\bar{\phi}_0-4\sqrt{3}\,\arctanh\sqrt{-\phi}\Bigr)-12\,\Bigl(\sqrt{2}\,\bar{\phi}_0-2\sqrt{3}\arctanh\sqrt{-\phi}\Bigr)\notag\\
&&\times\Bigl[M -q^{2}\Bigl(\sqrt{2}\,\bar{\phi}_0
      -2\sqrt{3}\arctanh\sqrt{-\phi}\Bigr)\Bigr]+\cosh\Bigl(\sqrt{2}\,\bar{\phi}_0-2\sqrt{3}\arctanh\sqrt{-\phi}\Bigr)\notag\\
&& \times \Bigl[8q^{2}-6\,\Bigl(\sqrt{2}\,\bar{\phi}_0 -2\sqrt{3}\arctanh\sqrt{-\phi}\Bigr)\Bigl[M -q^{2}\Bigl(\sqrt{2}\,\bar{\phi}_0-2\sqrt{3}\arctanh\sqrt{-\phi}\Bigr)\Bigr]\Bigr]\label{Ap_4}\\
&&+6\,\Bigl[3M-2q^2\Bigl(\sqrt{2}\,\bar{\phi}_0-2\sqrt{3}\arctanh\sqrt{-\phi}\Bigr)\Bigr]\sinh\Bigl(\sqrt{2}\,\bar{\phi}_0-2\sqrt{3}\arctanh\sqrt{-\phi}\Bigr)\Bigr\}\notag\;.
\end{eqnarray}
\end{widetext}

Using this together with Eq.~(\ref{phi exp}), we can express both $\phi$ and $V(\phi)$ as functions of $x$. 
However, our goal is to obtain $f(\mathcal{R})$ as an explicit analytic function of $\mathcal{R}$ only. To achieve this, we need to express $\phi$ solely as a function of $\mathcal{R}$ -- since $V(\mathcal{R})$ then follows directly. The relation between them is encoded in Eq.~(\ref{RVphi}), which may be written as
\begin{equation}\label{Ap_5}
    \mathcal{R}=V^{\prime}(\phi)\,\,.
\end{equation}
Therefore, if this equation is invertible, it yields an explicit analytic expression for $\phi(\mathcal{R})$. By taking the derivative of $V(\phi)$, we have
\begin{widetext}
\begin{eqnarray}
V^{\prime}(\phi)&=&-\frac{8\bigl(1+\phi\bigr)}{\sqrt{-\phi}}\,\Bigl\{
-6\sqrt{3}M - 9q^2\sqrt{-\phi} + q^2\sqrt{-\phi} \cosh\Bigl(2\sqrt{2}\,\bar{\phi}_0 - 4\sqrt{3}\arctanh\sqrt{-\phi}\Bigr)\notag\\
&& -\,12\,\Bigl(\sqrt{2}\,\bar{\phi}_0 - 2\sqrt{3}\arctanh\sqrt{-\phi}\Bigr)\Bigl[-\sqrt{3}q^2 + M\sqrt{-\phi} - q^2\sqrt{-\phi} \,\Bigl(\sqrt{2}\,\bar{\phi}_0 - 2\sqrt{3}\arctanh\sqrt{-\phi}\Bigr)\Bigr] \notag\\
&& +\Bigl[-2\sqrt{3}q^2 + 18M\sqrt{-\phi} - 3\Bigl(\sqrt{2}\,\bar{\phi}_0 - 2\sqrt{3}\arctanh\sqrt{-\phi}\Bigr)\Bigl[\sqrt{3}M + 4q^2\sqrt{-\phi} \label{Ap_6}\\
&& -\,\sqrt{3}q^2 \Bigl(\sqrt{2}\,\bar{\phi}_0 - 2\sqrt{3}\arctanh\sqrt{-\phi}\Bigr)\Bigr]\Bigr]\sinh\Bigl(\sqrt{2}\,\bar{\phi}_0 - 2\sqrt{3}\arctanh\sqrt{-\phi}\Bigr) \notag \\
&& +\, 2\cosh\Bigl(\sqrt{2}\,\bar{\phi}_0 - 2\sqrt{3}\arctanh\sqrt{-\phi}\Bigr)\Bigl[3\sqrt{3}M + 4q^2\sqrt{-\phi} - 3\sqrt{-\phi}\Bigl(\sqrt{2}\,\bar{\phi}_0 - 2\sqrt{3}\arctanh\sqrt{-\phi}\Bigr)\notag\\
&&\times \Bigl[M - q^2 \Bigl(\sqrt{2}\,\bar{\phi}_0 - 2\sqrt{3}\arctanh\sqrt{-\phi}\Bigr)\Bigr] + \sqrt{3}q^2 \sinh\Bigl(\sqrt{2}\,\bar{\phi}_0 - 2\sqrt{3}\arctanh\sqrt{-\phi}\Bigr)\Bigr]\Bigr\}\notag\;.
\end{eqnarray}
\end{widetext}
Given the complexity of this expression, it is not possible to analytically invert Eq.~(\ref{Ap_5}) to obtain $\phi(\mathcal{R})$. As a result, we must rely on a numerical approach, representing $f(\mathcal{R})$ as a parametric curve.

For this purpose, we take $\phi$ as the free parameter and use Eq.~(\ref{Ap_5}) to express $\mathcal{R}$, as well as $f(\mathcal{R})$, in Eq.~(\ref{Ap_2}), as a function of it. Then, after fixing $q$, $M$ and $\bar{\phi}_0$, both $\mathcal{R}$ and $f(\mathcal{R})$ are computed for a range of $\phi$, from which a parametric representation of $f(\mathcal{R})$ is obtained.

\begin{figure}[ht]
	\centering
	\includegraphics[scale=0.57]{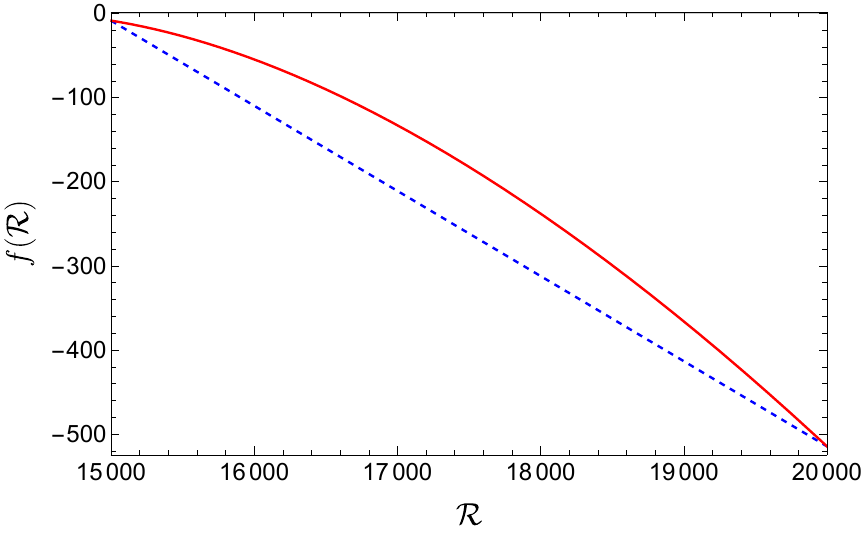}
	\caption{Parametric plot of $f(\mathcal{R})$ as a function of $\mathcal{R}$, associated with a solution of Example 1 with $q = 5$, $M=5.50389$ and $\bar{\phi}_0 = 0$. The solid red curve shows the numerical result for $\phi \in [-0.158069,\, -0.027436]$. The dashed blue straight line is a hypothetical linear case, included for reference, highlighting the non-linear nature of $f(\mathcal{R})$ in this case.}
	\label{fR_canonical}
\end{figure}

Accordingly, by fixing $q=5$, $M=5.50389$ ($M=|M_c|+0.5$), $\bar{\phi}_0=0$ (which is equivalent to $\psi_0=0$ in our analysis of the solutions) -- corresponding to a black hole solution with both an EH and a CH -- and varying $\phi$ in the range $[-0.158069,\,-0.027436]$, we obtain the solid red curve in Fig.~\ref{fR_canonical} by plotting $f(\mathcal{R})$ as a function of $\mathcal{R}$. For comparison, a dashed blue straight line is also plotted, corresponding to a hypothetical linear case constructed only to serve as a reference. This clearly illustrates that the $f(\mathcal{R})$ function obtained here is non-linear.

\vspace{1\baselineskip}

\section{Derivation of $f(\mathcal{R})$ -- Example 2}\label{apB}

In this appendix, we present in detail the derivation of the $f(\mathcal{R})$ function from action (\ref{2}) associated with a solution obtained in Example 2 (see Section \ref{example2_section}). We follow the same approach as in Appendix~\ref{apA}.

The $f(\mathcal{R})$ function is given by Eq.~(\ref{Ap_2}). Furthermore, in Eq.~(\ref{phi ph}) there is the solution for $\bar{\phi}(x)$ in Example 2, which can be inverted to obtain $x$ in terms of $\bar{\phi}$. Moreover, from the phantom case ($\phi>0$) of Eq.~(\ref{8}), which provides $\phi(\bar{\phi})$, we get the relation
\begin{equation}\label{Ap_B1}
    \bar{\phi}=\sqrt{6}\arctan(\sqrt{\phi})\;.
\end{equation}

By expressing $x$ in terms of $\bar{\phi}$ in Eq.~(\ref{potential ph}), which is the solution for $U(\bar{\phi}(x))$ in Example 2, and then replacing it by Eq.~(\ref{Ap_B1}), we get $U(\phi)$. Then, by using Eq.~(\ref{V_phantom}), we obtain an expression for $V(\phi)$:
\begin{widetext}
\begin{eqnarray}
V(\phi)&=&\frac{1}{4}(1 + \phi)^2 \Bigl\{-24 M \pi + \bigl(9 + 4\pi^2\bigr) q^2- q^2 \cos\Bigl(2\sqrt{2}\,\bar{\phi}_0-4\sqrt{3}\arctan\sqrt{\phi}\Bigr) \notag \\
&&- 8q^2 \arctan\left(\tan\left[\bar{\phi}_0/\sqrt{2} - \sqrt{3}\arctan\sqrt{\phi}\right]\right)^2
        \Bigl[-2 + \cos\Bigl(\sqrt{2}\,\bar{\phi}_0 - 2\sqrt{3}\arctan\sqrt{\phi}\Bigr)\Big] \notag\\
&& + 2\,\Bigl[6M\pi + \bigl(4 - \pi^2\bigr)\,q^2\Bigr]
        \cos\Bigl(\sqrt{2}\,\bar{\phi}_0 - 2\sqrt{3}\arctan\sqrt{\phi}\Bigr) \label{Ap_B2} \\
&& + 12\,\Bigl(-3M + \pi q^2\Bigr)
        \sin\Bigl(\sqrt{2}\,\bar{\phi}_0 - 2\sqrt{3}\arctan\sqrt{\phi}\Bigr) + 8\,\arctan\left(\tan\left[\bar{\phi}_0/\sqrt{2} - \sqrt{3}\arctan\sqrt{\phi}\right]\right) \notag\\
&& \times\Bigl[
            \Bigl(3M - \pi q^2\Bigr)
                \Bigl[-2 + \cos\Bigl(\sqrt{2}\,\bar{\phi}_0 - 2\sqrt{3}\arctan\sqrt{\phi}\Bigr)\Big]
            + 3q^2\,\sin\Bigl(\sqrt{2}\,\bar{\phi}_0 - 2\sqrt{3}\arctan\sqrt{\phi}\Bigr)
        \Bigr]\Bigr\}\notag \;,
\end{eqnarray}
\end{widetext}

Now, as before, to obtain $f(\mathcal{R})$ as an explicit analytic function of $\mathcal{R}$ only, we need to express $\phi$ solely as a function of $\mathcal{R}$. These are related by Eq.~(\ref{Ap_5}). By taking the derivative of $V(\phi)$, we have
\begin{widetext}
\begin{eqnarray}
V^{\prime}(\phi) &=& \frac{1+\phi}{2\sqrt{\phi}}\Bigl\{-2\sqrt{3} \cos\Bigl(\bar{\phi}_0/\sqrt{2} - \sqrt{3} \arctan\sqrt{\phi}\Bigr)
\Bigl\{4\bigl(-3M + \pi q^2\bigr) \cos\Bigl(\bar{\phi}_0/\sqrt{2} - \sqrt{3} \arctan\sqrt{\phi}\Bigr)  \notag\\
&&+ q^2 \Bigl[8\,\arctan\left(\tan\left[\bar{\phi}_0/\sqrt{2} - \sqrt{3}\arctan\sqrt{\phi}\right]\right)
  \cos\Bigl(\bar{\phi}_0/\sqrt{2} - \sqrt{3} \arctan\sqrt{\phi}\Bigr)\notag \\
&&+ \sin\Bigl(3\bar{\phi}_0/\sqrt{2} - 3\sqrt{3} \arctan\sqrt{\phi}\Bigr) \Bigr] - 6M\pi \sin\Bigl(\bar{\phi}_0/\sqrt{2} - \sqrt{3} \arctan\sqrt{\phi}\Bigr)\\
&& + \Bigl[\bigl(1 + \pi^2\bigr) q^2 + 4\,\arctan\left(\tan\left[\bar{\phi}_0/\sqrt{2} - \sqrt{3}\arctan\sqrt{\phi}\right]\right)
\Bigl[-3M + \pi q^2 \notag\\
&& + q^2\,\arctan\left(\tan\left[\bar{\phi}_0/\sqrt{2} - \sqrt{3}\arctan\sqrt{\phi}\right]\right)\Bigr]\Bigr] \sin\Bigl(\bar{\phi}_0/\sqrt{2} - \sqrt{3} \arctan\sqrt{\phi}\Bigr) \Bigr\} +\sqrt{\phi}\frac{4V(\phi)}{(1 + \phi)^2}\Bigr\}\;. \notag
\end{eqnarray}
\end{widetext}

As in the previous appendix, the complexity of this expression prevents an analytical inversion of Eq.~(\ref{Ap_5}) to obtain $\phi(\mathcal{R})$. We therefore adopt the same numerical approach: we fix $q$, $M$ and $\bar{\phi}_0$, take $\phi$ as a free parameter and vary it over a suitable range, using Eqs.~(\ref{Ap_5}) and (\ref{Ap_2}) to compute $\mathcal{R}$ and $f(\mathcal{R})$, respectively, so that a parametric representation of $f(\mathcal{R})$ is obtained.

\begin{figure}[t]
	\centering
	\includegraphics[scale=0.55]{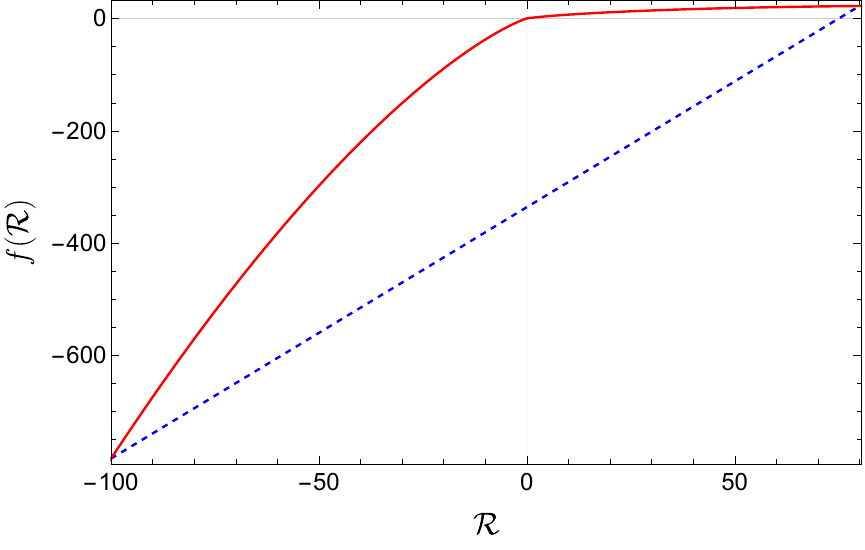}
	\caption{Parametric plot of $f(\mathcal{R})$ as a function of $\mathcal{R}$, associated with a solution of Example 2 with $q = 5$, $M=25\pi/3$ and $\bar{\phi}_0 = 0$. The solid red curve shows the numerical result for $\phi \in [0.072709,\,11.251057]$. The dashed blue straight line is a hypothetical linear case, included for reference, highlighting the non-linear nature of $f(\mathcal{R})$ in this case.}
	\label{fR_phantom}
\end{figure}

Accordingly, by fixing $q=5$, $M=25\pi/3$ ($M=M_{c2}$), $\bar{\phi}_0=0$ (equivalent to $\psi_0=0$) -- corresponding to a symmetric black bounce solution with an EH preceding the bounce and another following it -- and by varying $\phi$ in the range $[0.072709,\,11.251057]$, we can plot $f(\mathcal{R})$ as a function of $\mathcal{R}$, obtaining the solid red curve in Fig.~\ref{fR_phantom}. As before, a dashed blue straight line is included to represent a hypothetical linear case used solely as a reference. This once again illustrates that the $f(\mathcal{R})$ function obtained here is non-linear.

\end{document}